\documentclass[aps,prb,reprint,showpacs,floatfix]{revtex4-1}  
\usepackage{amsmath,amssymb,xcolor}
\usepackage[colorlinks,bookmarks=false,citecolor=red,linkcolor=blue,urlcolor=blue]{hyperref}
\usepackage{graphicx,epstopdf,pdfpages,xcolor,tabularx,fancyhdr}
 
\usepackage[caption=false]{subfig} 
\begin{document}
\title{Topological transitions in Ising models}
\author{Somenath Jalal} 
\email{somenath.jalal@gmail.com}
\author{Rishabh Khare}
\author{Siddhartha Lal}
\email{slal@iiserkol.ac.in}
\affiliation{Department of Physical Sciences, Indian Institute of Science Education and Research-Kolkata, W.B. 741246, India}


%
\def\c{\hat{c}}
\def\cdag{{\hat{c}}^{\dagger}}
\def\d{\hat{d}}
\def\ddag{{\hat{d}}^\dagger} 
\def\sbar{\bar{\sigma}}
\def\taubar{\bar{\tau}}

\begin{abstract}
The thermal dynamics of the two-dimensional Ising model and quantum dynamics of the one-dimensional transverse-field Ising model (TFIM) are mapped to one another through the transfer-matrix formalism. We show that the fermionised TFIM undergoes a Fermi-surface topology-changing Lifshitz transition at its critical point. We identify the degree of freedom which tracks the Lifshitz transition via changes in topological quantum numbers (e.g., Chern number, Berry phase etc.). An emergent $SU(2)$ symmetry at criticality is observed to lead to a topological quantum number different from that which characterises the ordered phase. The topological transition is also understood via a spectral flow thought-experiment in a Thouless charge pump, revealing the bulk-boundary correspondence across the transition. The duality property of the phases and their entanglement content are studied, revealing a holographic relation with the entanglement at criticality. The effects of a non-zero longitudinal field and interactions that scatter across the singular Fermi surface are treated within the renormalisation group (RG) formalism. The analysis reveals that the critical point of the 1D TFIM and the 1D spin-1/2 Heisenberg chain are connected via a line of $SU(2)$-symmetric theories. We extend our analysis to show that the classical to quantum correspondence links the critical theories of Ising models in various dimensions holographically through the universal effective Hamiltonian that describes the Lifshitz transition of the 1D TFIM. We obtain in this way a unified perspective of transitions in Ising models that lie beyond the traditional Ginzburg-Landau-Wilson paradigm. We discuss the consequences of our results for similar topological transitions observed in classical spin models, topological insulators, superconductors and lattice gauge-field theories which are related to the Ising universality class.
\end{abstract}

\maketitle  

\tableofcontents

\section{{\label{sec:intro}}Introduction}

Phase transitions are the genesis of all emergent phenomena in many-body systems, and are characterised by discontinuous (and often singular) behaviour in various measurable quantities. Landau's remarkable insight was to define the notion of an order parameter~\cite{landaulifshitzstatphys}, a measurable quantity in terms of which the transition could be observed. A discontinuity in the order parameter across the transition is called a first-order transition, while discontinuities in its derivatives are collectively called continuous transitions (as the order parameter grows continuously from zero at the critical point towards a finite value in the ordered phase). The order parameter in a continuous transition arises from the spontaneous breaking of a symmetry enjoyed by the Hamiltonian describing the finite-temperature ($T>0$) dynamics of the system (e.g., a spontaneous magnetisation below the critical temperature $T=T_{c}$ in a magnet).
\par
While it is easy to understand the appearance of a finite magnetisation in the presence of a non-zero polarising magnetic field $h$ (thus breaking explicitly the symmetry of the Hamiltonian), the case of a spontaneous magnetisation is subtle. In order to see how magnetisation can arise in the absence of a field, we employ a stratagem commonly used in examples displaying spontaneous symmetry breaking (SSB). This involves taking the thermodynamic limit while maintaining a small but finite field $h$, and taking the limit of a vanishing field ($h\to 0\pm$) only after the thermodynamic limit has been reached. As noted in Ref.(\cite{goldenfeld-book}), this is an example of ergodicity breaking. In the prevalent Ginzburg-Landau-Wilson paradigm of continuous transitions, such order parameters are spatially local in nature, i.e., they are collective macroscopic orders that arise from coarse-graining over microscopic degrees of freedom, and are expected to have global consequences in the thermodynamic limit. For instance, along with a smoothly vanishing magnetisation as temperature $T\to T_{c}-$, divergences are observed in the magnetic susceptibility, the correlation length and various inter-spin correlation functions. These divergences can be studied via scaling forms, i.e., functions of dimensionless quantities (e.g., $T/T_{c}$ and $h/h_{c}$) which help obtain the leading divergences in terms of various critical exponents. A set of critical exponents helps define the notion of universality: a grouping of systems seemingly different at the microscopic level, but all of whom possess the same type of continuous phase transition. The scaling functions themselves arise from a careful treatment of the divergent fluctuations present at the transition. These lead to anomalous dimensions being acquired by various couplings (i.e., parameters of the Hamiltonian) of the original theory as fluctuations at short distance (e.g., the lattice scale) are integrated out and a theory at the largest lengthscales is obtained~\cite{goldenfeld-book}. This is the essence of the renormalisation group (RG) formalism~\cite{wilson-1975}, and leads to the surprising conclusion that lengthscales different from the correlation length (e.g., microscopic lengthscales like the lattice scale) are also important at the transition.
\par
Progress in our understanding of phase transitions has been aided greatly by the study of the Ising model (IM)~\cite{ising-1925} in various guises (i.e., spins with $Z_{2}$-symmetry placed on an ordered lattice and interacting typically only with nearest neighbours). The Ising universality class was the first to be understood, and is also the best studied. Importantly, the exact solution of the two-dimensional (2D) Ising model on a square lattice in the absence of external fields is one of the few exactly solvable models ~\cite{onsager-1944,kaufman-1949}, and is a landmark achievement that spurred the growth of interest in the study of phase transitions. While the original works are technically involved, considerable insight can be gained more readily from the classical to quantum correspondence~\cite{suzuki-1976,fradkin-susskind-1978}: this method involves a mapping between the finite-$T$ transfer-matrix of the classical two-dimensional Ising model and a $T=0$ quantum Hamiltonian of the 1D transverse-field Ising model (TFIM or quantum Ising (QI)  model)~\cite{schultz-1964,elliot-1970}. The nearest-neighbor 1D TFIM also happens to be an exactly solvable model~\cite{pfeuty-1970} through an exact fermionic representation of the spin degree of freedom (called the real-space non-local Jordan-Wigner (JW) transformation). This mapping between the finite-$T$ partition function of the classical 2D Ising model and the path integral of the $T=0$ 1D TFIM equates the thermal fluctuation-driven transition in the former to the quantum fluctuation-driven transition in the latter; both transitions, therefore, belong to the same universality class. The mapping also offers an equivalence between the famous Kramers-Wannier duality of the 2D Ising model~\cite{kramers-wannier-1941} and the quantum order-disorder duality of the 1D TFIM~\cite{kadanoffceva-1971,fradkin-susskind-1978}. It is this equivalence that forms the bedrock of much of our understanding of $T=0$ quantum criticality and its implications at finite-$T$~\cite{sachdev-book}.
\par
Physical insight into the nature of the order-disorder transition of the 2D Ising model can also be gained from the early works of Peierls~\cite{peierls-1936} and Landau~\cite{landaulifshitzstatphys}. These authors showed that the fluctuations that lead to the disordering of the ordered equilibrium state are domain walls, i.e., regions that separate domains with different types of order (i.e., alignment of the spins). Indeed, a lucid heuristic argument was offered by Peierls and Landau on the phase transition in the Ising model as arising from the balance between the internal energy cost for, and the entropic gain from, the generation of domain walls. This argument shows unequivocally why the 1D Ising model cannot have ordering at any non-zero temperature. It also offers a value for the critical temperature of the 2D Ising model which is reasonably close to that obtained from the exact solution and the Kramers-Wannier duality relation~\cite{kramers-wannier-1941}. While rigorous proof of these heuristic results for the energetics have been obtained in later works, a question remains over the origin of the sensitivity of these domain wall fluctuations to boundary conditions in the vicinity of the transition. Evidence for this can be found in Onsager's display of the fact that domain walls are created by changes in boundary condition~\cite{onsager-1944}, and that the free energy cost for their generation vanishes linearly as the temperature approaches its critical value. Further, it was shown by M{\"u}ller-Hartmann and Zittartz~\cite{mullerhartmannzittartz-1977} that the vanishing interfacial free energy cost/surface tension at the transition leads to the Kramers-Wannier condition for self-duality~\cite{kramers-wannier-1941}. Another hint that boundary conditions are important to the exact solution can be found in Kasteleyn's finding~\cite{kasteleyn1961} that the partition function of the 2D Ising model, when placed on a spatial manifold whose topology is characterised by a genus $m$, is given by the sum of $4^{m}$ equal terms. Thus, only one term exists for open boundary conditions (OBC) in both the x and y directions, while 4 terms are needed for the torus (PBC in both the x and y directions). A related question deals with the subtlety of the SSB stratagem: the spontaneous magnetisation achieved via SSB involves avoiding the singularity, i.e., taking $(T\to T_{c}+, h\to 0\pm)$~\cite{yang-1952}, while the Onsager-Kaufman exactly solution~\cite{onsager-1944,kaufman-1949} is strictly at $h=0$. What, then, is the precise nature of the order parameter at $h=0$~\cite{wu-1966,mccoy-1968,pfeuty-1970}, and how is it related to that obtained for $h\to 0\pm$~\cite{yang-1952}?
\par
Various related issues arise in the exact solution of the 1D TFIM as well~\cite{pfeuty-1970,schultz-1964}. The JW fermions mentioned above correspond to the domain wall excitations created in the 1D TFIM via quantum fluctuations of the spins (i.e., the effect of the transverse field). Via the classical to quantum correspondence, they also correspond to the thermal-fluctuation induced domain walls of the 2D Ising model~\cite{fradkin-susskind-1978}. As we will see in the next section, the continuous phase transition of the 2D Ising model/1D TFIM is tracked via the appearance of a singular Fermi surface of these JW fermions at the Brillouin zone edge $k=\pm\pi$ with a Dirac-like dispersion (see Fig.(\ref{fig:dispersion-QI})), suggesting a change in Fermi-surface topology across the transition. Remarkably, this is precisely the hallmark of a Lifshitz transition for a fermionic system~\cite{lifshitz-1960}. Which degree(s) of freedom best characterise the transition? Given that the local magnetisation in the bulk vanishes in the ordered phase ($m(\vec{r})=0$) due to the $Z_{2}$ symmetry of the Ising Hamiltonian for longitudinal field $h=0$, we will instead adopt the philosophy of Ruelle~\cite{ruelle-book}, Kohn~\cite{kohn-1964} and Thouless~\cite{thoulessedwards-1972} (among others) that a phase transition should be signalled by a change in the sensitivity upon varying boundary conditions. Thus, we ask whether non-local order parameters can identify the transition? How is this related to the emergent $SU(2)$ symmetry of the massless Dirac spectrum at criticality~\cite{fradkin-book}? What is the connection between this singular Fermi surface and various physical properties of the original Ising spins, e.g., specific heat, spontaneous magnetisation, correlation functions, the duality property etc.? How is the thermodynamic limit to be taken? What are the effects of a non-zero explicit symmetry-breaking longitudinal field $h$? Which critical exponents can be computed in this way, and which scaling relations do they satisfy? Beginning with a discussion of the transfer matrix mapping between the 2D Ising model and the 1D TFIM in section \ref{sec:2d-classical-1d-QI}, we will attempt to answer the questions listed above in sections \ref{sec:lt}, \ref{sec:electric-field}, \ref{sec:dual} and \ref{sec:rg-2d-1d-ising}. In section \ref{sec:univ}, we will similarly analyse, and present the results obtained for, various other Ising and related models. Finally, we will conclude in section \ref{sec:summary} by considering the implications of our findings for finite-temperature transitions in classical Ising models, propose an experimental realisation as well as consider the nature of universality in the phase transitions we study. 
\section{\label{sec:2d-classical-1d-QI} Preliminaries: $T$-matrix mapping between 2D IM and 1D TFIM}
\begin{figure}[htp] 
   \centering
   \includegraphics[width=0.45\textwidth]{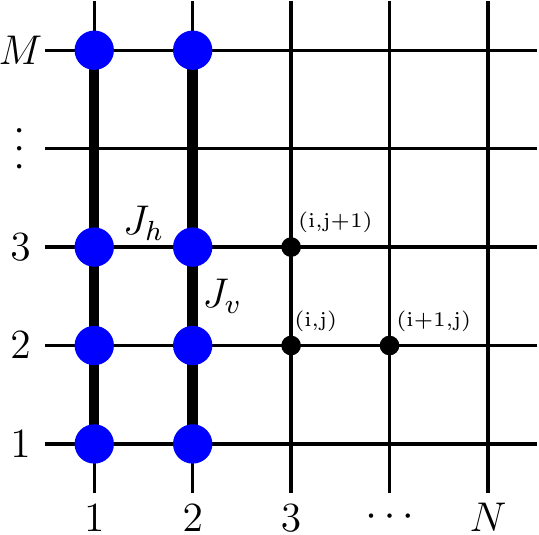}
   \caption{(Color online.)~A schematic diagram of the 2D lattice. Spin on each site is labelled as $\sigma_{i,j}$. The horizontal and vertical couplings are $J_h$ and $J_v$ respectively. $i$ and $j$ denotes row and column index ($i=1,2, \cdots N$ and $j=1,2, \cdots M$) respectively. The spins along a single column are fused to make a single M component spin vector $\bar{\sigma}^{i}$ and the effective quantum Ising model has N such spins.}
   \label{fig:2d-lattice}
\end{figure}

The two-dimensional (2D) Ising model is written as 
\begin{eqnarray}
H_{\text{2D}} = - \sum_{i,j} \left[ J_h  \sigma_{i,j} \sigma_{i+1,j} + J_v  \sigma_{i,j} \sigma_{i,j+1} \right]
\end{eqnarray}
where $J_h$ and $J_v$ are horizontal and vertical Ising interaction strength respectively. They are taken to be positive numbers and the Ising phase possesses ferromagnetic ordering of the spins $\sigma_{i,j}$ on a two-dimensional lattice point denoted by label $i$ and $j$ (see Fig.~\ref{fig:2d-lattice}). In attempting an exact solution, the goal is to find a suitable form of the transfer matrix ($T$) and solve for its largest eigenvalue. The partition function is defined as
\begin{eqnarray}
Z = \sum_{\text{configurations}} e^{-\beta H_{\text{2D}}} = {\text{Tr}} ~ {T^N}~,
\end{eqnarray}
where $\beta=1/\textrm{k}_{\textrm{B}}$T is inverse temperature and $N$ is the total number of sites on the lattice. Tr is the trace over all the spin configurations. We will now derive the transfer matrix $T$ for the 2D Ising model and show that it leads to the Hamiltonian for the transverse field Ising model (TFIM) in 1D with nearest neighbor interactions (also known as 1D quantum Ising model).~\cite{kogut-1979, emch-book, schultz-1964} 

\par
Let us consider a particular column (say $i^{th}$ column) on the 2D lattice. The spins are denoted as $\sigma_{(i,1)}, \sigma_{(i,2)} \cdots \sigma_{(i,M)}$ on that $i^{th}$ column. It is convenient to consider them all at a time by a single $M$ component vector parameter as ${\bar{\sigma}}^{(i)}$. In this language the 2D classical ising model can be written as 
\begin{eqnarray}
H_{\text{2D}} &=& - \sum_{i=1}^N H^h (\sbar^{(i)}, \sbar^{(i+1)}) + H^v (\sbar^{(i)}) \nonumber \\
&=& \sum_i \left[ J_h  \sbar^{(i)} \sbar^{(i+1)} + J_v \sbar^{(i)} \taubar^{(i)} \right]~,
\end{eqnarray}
where the Ising interaction along the horizontal direction maintains its nature in terms of the column variable (i.e., the $M$ component vector parameter ${\bar{\sigma}}^{(i)}$). On the other hand, the Ising interaction in the vertical direction becomes a local (intra-spin) interaction in the column variables: $\sbar^{(i)} \taubar^{(i)}$ , with $\taubar^{(i)}$ being the nearest-neighbour spin along the vertical. Now, the transfer matrix for this Hamiltonian can be written as 
\begin{eqnarray}
Z &=& {\text{Tr}} ~ e^{\beta J_h \sbar^{(i)} \sbar^{(i+1)} }  e^{\beta J_v \sbar^{(i)} \taubar^{(i)}} \nonumber \\
&=& {\text{Tr}} ~ (V_1 V_2)^N = \text{Tr} ~ V^N = \text{Tr} ~ (V_2^{1/2} V_1 V_2^{1/2})^N~,
\end{eqnarray}
where $V_1$ and $V_2$ are given by
\begin{eqnarray}
V_1 &=& e^{\beta J_h \sbar^{(i)} \sbar^{(i+1)}} = e^{\beta J_h \tau^z_i  \tau^z_{i+1}}\\
V_2 &=& e^{\beta J_v \sbar^{(i)} \taubar^{(i)}} =  e^{\beta J_v} I + e^{-\beta J_v} \tau^x \nonumber\\
&=& e^{\beta J_v} (1 + \tanh (\beta J_v^{\star}) \tau^x)\nonumber\\
&=& \sqrt{2 \sinh (2 \beta J_v)} e^{\beta J_v^\star \tau^x}~, 
\end{eqnarray}
and we have used the relations
\begin{eqnarray}
\tanh (\beta J_v^{\star}) =  e^{-2 \beta J_v}\\
\sinh {(2 \beta J_v)} \sinh {(2 \beta J_v^\star)} = 1~.
\end{eqnarray}
The full transfer matrix is then found to be 
\begin{eqnarray}
Z = \text{Tr} ~ e^{-\beta \sum_{i=1}^{N} \left[ J \tau_i^z \tau_{i+1}^z + h \tau_i^x\right]}~,
\end{eqnarray}
where we have defined $J = 2 \beta J_v$ and $h = 2 \beta J_h^{\star} \sim 2 \beta e^{- 2 \beta J_h}$~{\cite{fradkin-susskind-1978}}. 
\par
As shown in the next subsection, this transfer matrix for a translational invariant system with periodic boundary conditions is solved using the Jordan-Wigner transformation, and then diagonalised using a Bogoliubov transformation. The diagonalised transfer matrix has a Bogoliubov-de Gennes quasiparticle-like spectrum~{\cite{schultz-1964}}
\begin{eqnarray}
\cosh (\epsilon_q) &=& \cosh {(2 \beta J_h)} \cosh (2 \beta J_h^\star) \nonumber \\
&-& \sinh (2 \beta J_h) \sinh (2 \beta J_h^\star) \cos (q)~,
\end{eqnarray}
where the positive solution of $\epsilon_q$ in the above equation is the eigenspectrum. The $\pi$-momentum mode of $\epsilon_q$ i.e., $\epsilon_{q=\pi}$ gives rise to a special condition, $J_h^\star = J_v$, leading to the critical curve for the phase transition of the 2D classical Ising model denoted by~\cite{schultz-1964}  
\begin{eqnarray}
\sinh {(2 \beta_c J_h)} \sinh {(2 \beta_c J_v)} = 1 {\label{eq:kw}}~,
\end{eqnarray}
with $\beta_c = 1/T_c$. This relation reveals that the critical temperature is a function of $J_v$ and $J_h$, and is the Kramers-Wannier {\cite{kramers-wannier-1941}} relation for the anisotropic classical Ising model. 
At this point, an important connection can be made. The authors of Ref.(\cite{mullerhartmannzittartz-1977}) showed that this Kramers-Wannier duality relation for the anisotropic 2D Ising model for the square lattice with periodic boundary conditions in one of the spatial directions could also be obtained directly from the condition that the free energy cost for an interfacial domain wall that spans the periodic direction should vanish at the critical temperature. From the discussion above for the TFIM, we can now see that the classical to quantum correspondence for the transfer matrix maps the thermal dynamics of such an interfacial domain wall for $T\sim T_{c}$~\cite{mullerhartmannzittartz-1977} onto the quantum dynamics of the 1D TFIM for $h\sim J$. In sections \ref{sec:lt}-\ref{sec:rg-2d-1d-ising}, the topological consequences of such spanning domain walls in the 2D Ising model will be clarified.
\subsection{TFIM mapped to theory of 1D massive Dirac fermions}
As shown above, the transfer matrix for the finite-temperature 2D Ising model can be mapped to the one dimensional transverse field Ising model (TFIM)
\begin{equation}
H = {h}{\sum_{i=1}^{N}}\sigma_{i}^z + J {\sum_{i=1}^{N}}{\sigma_{i}^x} {\sigma_{i+1}^{x}}~,
\end{equation}
where $J$ is the Ising interaction strength, $h$ is the transverse field with respect to the Ising direction and $ \sigma_{N+1}^{x} = \sigma_{1}^{x}$ (periodic boundary conditions). It is important to note that this Hamiltonian has a $Z_{2}$ symmetry given by the non-local string operator which spans the system
\begin{equation}
Z = \Pi_{i=1}^{N} \sigma_{i}^{z} ~~,~~\textrm{such~that}~~ \left[H,Z\right] = 0~.
\label{z2topinv}
\end{equation}
Further, we rewrite the Hamiltonian as
\begin{eqnarray}
H &=& {h}{\sum_{i}}\sigma_{i}^z + {\sum_{i}} \big[ J ( \sigma_{i}^{+} \sigma_{i+1}^{+} + \sigma_{i}^{-} \sigma_{i+1}^{-}) \nonumber \\
&+& J ( \sigma_{i}^{-} \sigma_{i+1}^{+} + \sigma_{i}^{+} \sigma_{i+1}^{-} )\big]
\end{eqnarray}
The spin model can be transformed to bilinear fermionic problem by means of Jordan-Wigner transformation
\begin{eqnarray}
\sigma_{n}^{z} &=& c_{n}^{\dagger}c_{n} - \frac{1}{2}\nonumber\\
\sigma_{n}^{\dagger} &=& c_{n}^{\dagger}~e^{i\pi\sum_{j=1}^{n-1}c_{j}^{\dagger}c_{j}}\nonumber\\
\sigma_{n}^{\dagger} &=& e^{-i\pi\sum_{j=1}^{n-1}c_{j}^{\dagger}c_{j}}~c_{n}~,
\end{eqnarray}
and the model Hamiltonian is written as 
\begin{eqnarray}
H &=& - h N +2 h {\sum_{i=1}^{N}} {c_{i}^{\dagger} c_{i}}  + {\sum_{i}}  \big[ J ( c_{i}^{\dagger} c_{i+1}^{\dagger} + c_{i+1} c_{i} ) \nonumber \\
&& + J ( c_{i}^{\dagger} c_{i+1} +  c^{\dagger}_{i+1} c_{i} )\big]~. 
\end{eqnarray}
This model is exactly solvable and one solves this by Fourier transforming to momentum space 
\begin{eqnarray}
H &=& (h + J \cos (k)) (\cdag_{k} \c_{k} + \cdag_{-k} \c_{-k} ) \nonumber \\
&+& i J \sin (k) ( \cdag_{k} \cdag_{-k} - \c_{-k} \c_{k})
\end{eqnarray}
We can then diagonalise this Hamiltonian via a Bogoliubov transformation
\begin{eqnarray}
d_{k} &=& \cos(\theta_{k})~c_{k} + i~\sin(\theta_{k})~c^{\dagger}_{-k}\nonumber\\
d^{\dagger}_{-k} &=& i~\sin(\theta_{k})~c_{k} + \cos(\theta_{k})~c^{\dagger}_{-k}~,
\end{eqnarray}
with $\tan(2\theta_{k}) = J\sin (k)/(J\cos (k) + h)$~. This amounts to finding a suitable particle-hole symmetric Nambu basis~{\cite{pfeuty-1970}}.
It is worth noting that the TFIM problem conserves only fermion parity (being equivalent to the global $Z_2$ symmetry of the original spin problem) but does not conserve fermion number (evident from the presence of the pairing term). However, the $k=0$ and $k=\pm\pi$ points are special as the Hamiltonian for these modes is diagonal in the occupation number basis (i.e., they conserve fermion number) and thus need to be taken account of separately~\cite{schultz-1964}. Indeed, these three $k$-modes diverge under the Bogoliubov transformation, i.e., making it ill-defined for $k=0,\pm\pi$~. We shall soon see that the \textit{singular} $k=\pm\pi$ modes track the phase transition. 
\par
\begin{figure}[htp]
\centering
\includegraphics[width=0.45\textwidth]{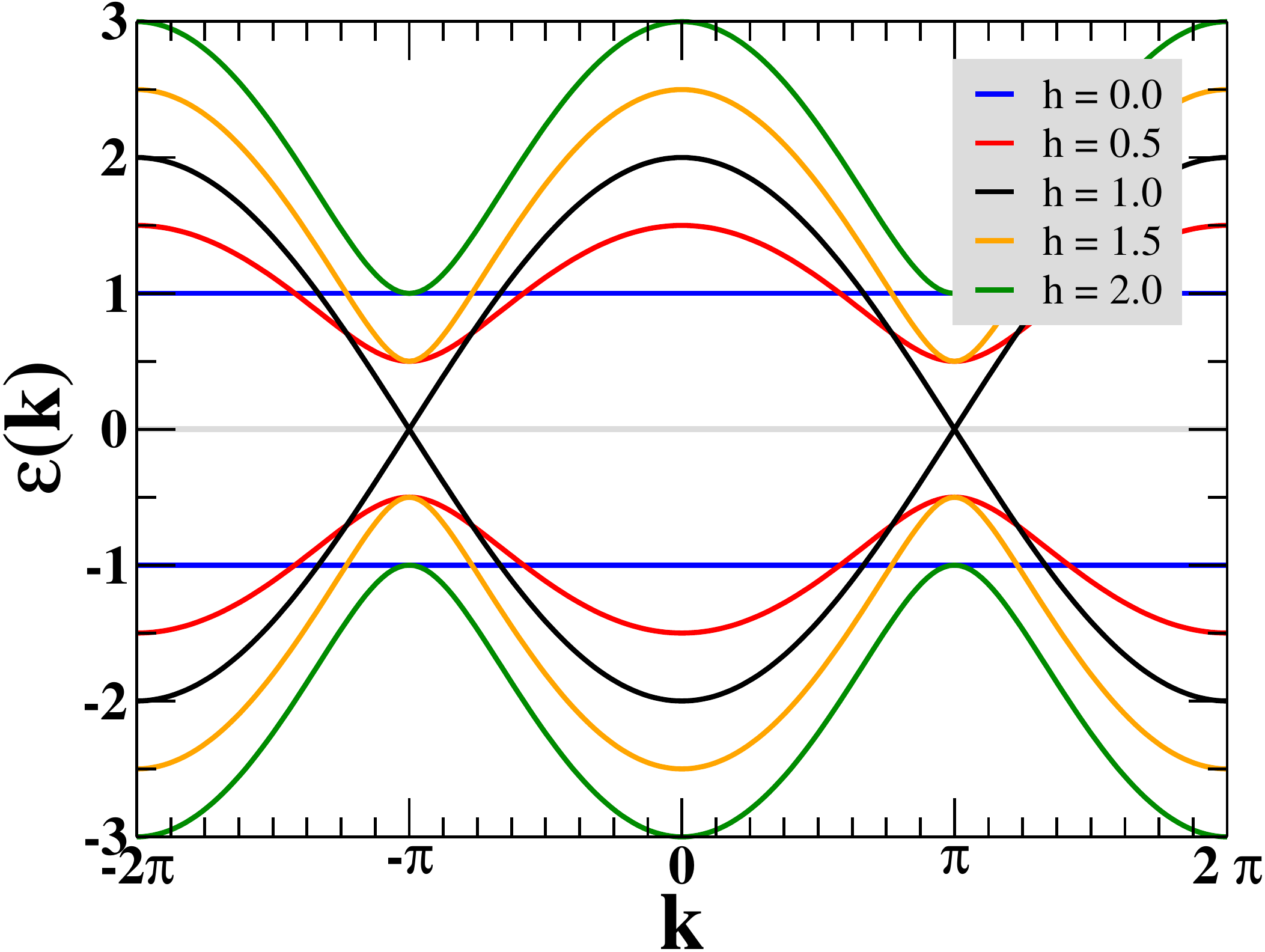}
\caption{(Color online.)~
Dispersion plot for the 1D TFIM eq.(\ref{eq:1dtfim}) for different values of $h$ and fixed $J = 1$~. This is a signature of a singular (Dirac-like) Fermi surface where the gap closes linearly at the critical point.
}
\label{fig:dispersion-QI}
\end{figure}

The dispersion relation for this (p-wave) superconductor of spinless electrons in 1D is given as 
\begin{eqnarray}
\epsilon_k = \pm \sqrt{(J \sin k)^2 + (J \cos {k} + h)^2},
\label{eq:1dtfim}
\end{eqnarray}
and the spectrum is plotted for different $h/J$ values in Fig.~{\ref{fig:dispersion-QI}}. The effective Dirac spectrum is seen near the $k = \pm \pi$ points for $J=h$, or the critical point.
The gap closes at $k=\pm \pi$ points and near those points it closes linearly as $\pm (J-h)$. This is the special feature (Dirac like) of 1D TFIM which is due to the nature of fermionic quasiparticles, and thus the massless fermionic Dirac spectrum is visible at the critical point. Indeed, the massless Dirac points at $k=\pm\pi$ form a Kramers doublet, and signal an emergent time reversal symmetry at criticality.
Near these points any small field (h) will open up a mass gap of the form $\pm \sqrt{k^2 + (1-J/h)^2}$. Thus near the critical point we can write the effective Hamiltonian of the 1D massless Dirac Hamiltonian as~\cite{kitaevperiodictable}
\begin{eqnarray}
H_{eff} &=& \sin k ~ \sigma^x + (h-J) \sigma^z \\
 &\simeq& k \sigma^x + \Delta \sigma^z\\ 
 &\equiv& ( i \frac{\partial}{\partial x}) \sigma^x + \Delta  \sigma^z~,
\end{eqnarray}
where $\Delta = h-J$ and we have expanded the spectrum $\sin k \sim k$ for small wavevectors in the neighbourhood of $k=\pm\pi$ in the second line (the continuum approximation). Thus, at the critical point we have $\Delta = 0$ and get the massless Dirac spectrum with Weyl points. It is also important to take note of the ``$\pi$-mode" state at the Brillouin zone edge $k=\pm\pi$: 
\begin{equation}
H_{\pi} = \Delta (1-2c^{\dagger}_{\pi}c_{\pi})~~,~~ \epsilon_{\pi}= \Delta\equiv h-J~.
\label{piHam}
\end{equation}
Note that this Hamiltonian for the $\pi$-mode is fermion number-conserving, i.e., it possesses the global $U(1)$ symmetry of fermionic Hamiltonians that conserve fermion number, even if modes with wavevector $0<k<\pi$ in the fermionised TFIM Hamiltonian do not. As we shall see later in section \ref{sec:lt}, this $U(1)$ symmetry of the $\pi$-mode Hamiltonian has topological consequences for the nature of the superconducting order achieved for $\Delta <0$ (i.e., $h<J$). Indeed, it will be shown to correspond to the $Z_{2}$ topological invariant for the 1D TFIM given earlier, and we will also show that the breaking of this particular global $U(1)$ symmetry is important in reaching a familiar (topologically trivial) superconducting state of matter.
Clearly, the gapless spectrum at $\Delta=0$ reveals a singular Fermi surface (the Dirac point at the Brillouin zone edge), and the sign of $\Delta$ relates to the occupancy of this state. As we will see in a subsequent section, the occupancy of this state is intimately related to the topological properties of the phase of this model for $J>h$. It is also pertinent to recall that the linear dependence of the gap on $\Delta$ corresponds to Onsager's result for the linear dependence of the free energy cost for generating a domain wall excitation in the 2D Ising model near criticality (the interface tension) on the reduced temperature~\cite{onsager-1944}, as well as identifies the Kramers-Wannier condition for the self-duality of the model~\cite{mullerhartmannzittartz-1977}.
\par 
Finally, in order to explore the topological content of the $\pi$-mode in its fullest, it is important to stress that we can think of the Hamiltonian for this state, $H_{\pi}$, as a two-level system or, equivalently, a spin-1/2 in an external generalised B-field $\vec{B}=(\tilde{h}_{x} \hat{x}, 0, \Delta \hat{z})$~where $\tilde{h}_{x}$ and $\Delta\equiv h-J$ are the longitudinal and transverse B-fields respectively of the TFIM. This can be implemented by transforming from a fermionic to a pseudospin-1/2 description using $\sigma_{z}=1-2c^{\dagger}_{\pi}c_{\pi}$,~$\sigma_{x}=(c^{\dagger}_{\pi} + c_{\pi})$ and $\sigma_{y}=-i(c^{\dagger}_{\pi} - c_{\pi})$. Then, the effective ``qubit" Hamiltonian can (upto a constant) be written as 
\begin{equation}
H_{\pi} = \tilde{h}_{x}\sigma_{x} + \Delta\sigma_{z}~,
\label{qubitHam}
\end{equation}
with eigenvalues $\epsilon_{0}^{\pm} = \pm\sqrt{\tilde{h}_{x}^{2} + \Delta^{2}}$~. It is easily seen, following Ref.(\cite{benderorszag}), that there exist level crossings at $\Delta =\pm i \tilde{h}_{x}$ represented by square-root branch-point singularities. The two eigenvalues $\epsilon_{0}^{\pm}$ can be defined on a two-sheeted Reimann surface, such that $\epsilon_{0}=\epsilon_{0}^{+}$ on the upper sheet and $\epsilon_{0}=\epsilon_{0}^{-}$ on the lower sheet. The level-crossing corresponds to an analytic continuation around either of the two square-root branch-point singularities shown above, leading to an exchange of the identities of the two eigenvalues as the sign of the square-root is changed. Such singularities are also referred to as exceptional points in the literature~\cite{rotter-2009}, and have also been shown to correspond to resonance phenomena arising from bound states in atomic systems~\cite{moiseyev-book}. It is even more non-trivial to note that for the special case of $\tilde{h}_{x}=0$ (i.e., the case of the $\pi$-mode of the 1D TFIM), the two square-root branch-point level-crossing singularities coalesce at $\Delta = 0$ (a real number) into a single level-crossing event. This coalescing corresponds to the conversion of a zero of the fermionic quasiparticle propagator for the $\pi$-mode at the Fermi surface ($E_{F}\equiv E=0$) for the case of a gapped spectrum into a pole at criticality. In section \ref{sec:lt}, this will be seen as leading to an important consequence on the nature of the quantum phase transition at $\Delta = 0$. 
\par
Further, the Berry phase accrued under a cyclic adiabatic excursion of the Hilbert space is given by
\begin{equation}
\gamma_{0}  = \pi\left[ 1 - \frac{\Delta}{\sqrt{\tilde{h}_{x}^{2} + \Delta^{2}}}\right] = \pi\left[ 1 - \frac{\Delta}{|\epsilon_{0}^{-}|}\right]~.
\end{equation}
The Berry phase is $\gamma_{0}\to 0$ for $(\tilde{h}_{x}=0, \Delta\geq 0)$,  $\gamma_{0}\to 2\pi$ for $(\tilde{h}_{x}=0, \Delta\leq 0)$ and $\gamma_{0}\to\pi$ for $(\tilde{h}_{x}\neq 0, \Delta =0$). For $(\tilde{h}_{x}=0,\Delta=0)$, i.e.. at the gapless point of the TFIM and with no longitudinal field, this Hamiltonian vanishes. However, the coherent state path-integral for this spin-1/2 state can be written in terms of a Wess-Zumino-Novikov-Witten (WZNW) term~\cite{fradkin-book}. This topological term in the action of the $\pi$-mode theory of the TFIM characterises the integer coverings of the Bloch sphere arising from the non-trivial homotopy group of the non-Abelian $SU(2)$ group, $\pi_{3} (SU(2)) = Z$. Very generally, the action 
for the dynamics of a spin-$S$ in the presence of an external field $\vec{h}$ (as that given by equn.(\ref{qubitHam}) above) is
\begin{equation}
{\cal S}= -4\pi S W_{0} + \int dt H 
\label{wznwaction}
\end{equation}
where we can use equn.(\ref{qubitHam}) for the Hamiltonian $H$, and the topological WZNW term for $\vec{S} = S \vec{n}$ is given by~\cite{fradkin-book}
\begin{equation}
W_{0} = \frac{1}{8\pi}\int_{0}^{1}d\rho \int_{0}^{\beta} \epsilon^{\mu\nu} \vec{n}\cdot \left [\partial_{\mu}\vec{n}\times \partial_{\nu}\vec{n}\right ] \equiv k~,~ k\in Z~,
\label{wznwterm} 
\end{equation}
and where $(\mu,\nu) = (t,\rho)$, periodic boundary condition in time leads to $\vec{n} (\beta) = \vec{n} (0)$ and $\rho$ is an auxiliary coordinate $\rho\in [0,1]$. The present problem has $S=1/2$, and the coefficient of the WZNW term $4\pi S\equiv 2\pi = \gamma_{0} (\Delta <0) - \gamma_{0}(\Delta >0)$. For finite $\vec{h}$, the Landau-Lifshitz equation of motion for the precession of a spin is obtained from the above action
\begin{equation}
\partial_{t} \vec{n} = \vec{n}\times \vec{h}~.
\end{equation}
\par
The above discussion reveals that the contribution of the $\pi$-mode to the partition function at the critical point ($\Delta =0$) is in terms of a phase purely topological in origin, i.e., the free energy for the $\pi$-mode at criticality contains an imaginary piece.
Importantly, it reflects the emergent $SU(2)$ symmetry at $J=h$ (as can also be analogously noted for the case of a massless Dirac fermion spectrum). In sections \ref{sec:lt}-\ref{sec:dual}, we will observe that this emergent $SU(2)$ symmetry of the $\pi$-mode theory has important consequences for the topological transition of the TFIM, as well as the topological properties of its ordered phase. Further, as shown in section \ref{sec:univ}, this $\pi$-mode theory is also identical to the effective quantum system obtained from the classical-quantum correspondence for the 1D Ising model, revealing the holographic nature of the correspondence for the 1D and 2D Ising models.
\par
We can now offer preliminary insight into the nature and stability of the critical point characterised by the emergent $SU(2)$ symmetry of the $\pi$-mode. For this, we treat the longitudinal field $\Delta$ in equation(\ref{qubitHam}) as possessing slow fluctuations with a Gaussian probability distribution, $P(\Delta)\sim e^{-\beta \Delta^{2}/2\sigma}$, where $\langle\Delta\rangle =0$ and $\langle\Delta^{2}\rangle = \sigma/\beta$ and $\sigma$ is size of the typical fluctuations in $\Delta$. The combined Hamiltonian for the $\pi$-mode and field is now~\cite{chandler-book}
\begin{equation}
H_{0} = \tilde{h}_{x}\sigma_{x} + \Delta\sigma_{z} + \frac{\Delta^{2}}{2\sigma} ~,
\label{qubitHamplusfield}
\end{equation}
with the eigenvalue equation $\epsilon^{\pm} = \frac{\Delta^{2}}{2\sigma}\pm\sqrt{\tilde{h}_{x}^{2} + \Delta^{2}}$. It is easily seen from these two eigenvalue equations that there is an avoided level crossing at $\epsilon_{+} (\Delta=0)=\epsilon_{-}(\Delta=0) + 2\tilde{h}_{x}$. For $\sigma < \tilde{h}_{x}$, $\epsilon_{-}(\Delta=0)=0$ represents the minimum eigenvalue, while it becomes a maximum for $\sigma > \tilde{h}_{x}$. Instead, for $\sigma > \tilde{h}_{x}$, $\epsilon_{-}$ has two new minima at $\Delta=\pm\sigma$. A level-crossing is found for $\tilde{h}_{x}=0=\Delta$ (the case of the 1D TFIM), and the two new minima at $\Delta=\pm\sigma$ represent the self-trapping (or noise-induced stabilisation) of the $\pi$-mode by the breaking of the $SU(2)$ symmetry evident for the case of $\Delta=0$. A similar analysis can be carried out for the case of $\Delta=0$ and a Gaussian noise in the field $\tilde{h}_{x}$, and similar results obtained. Indeed, in section \ref{sec:rg-2d-1d-ising}, a more sophisticated RG analysis will be shown to reach the same conclusions. 

\section{\label{sec:lt}Lifshitz transition and topological phase of the TFIM}
We have seen in a previous section that the 1D TFIM is connected to the p-wave superconducting spinless SC model (pWSC) in 1D via the Jordan-Wigner (JW) transformation. This transformation is non-local, reminiscent of the flux attachment useful in understanding the fractional quantum Hall effect. 
We begin here by recalling the Hamiltonian for the anisotropic 1D XY spin-1/2 model in a transverse field with periodic boundary conditions
\begin{eqnarray}
H_{XY} &=& \sum_{i=1}^{N}J\left[ (\frac{1+\delta}{2})\sigma_{i}^{x}\sigma_{i+1}^{x} + (\frac{1-\delta}{2})\sigma_{i}^{y}\sigma_{i+1}^{y}\right ]\nonumber\\ 
&&+ h_{z}\sum_{i=1}^{N}\sigma_{i}^{z}~,
\label{transversefieldxy}
\end{eqnarray}
where the spin-space anisotropy $\delta \geq 0$. The cases of $\delta = 0$ and $\delta =1$ correspond to the one-dimensional XX  and Ising models in a transverse field respectively~\cite{perkcapel-1977}. Note that the $Z_{2}$ invariant shown earlier, $Z = \Pi_{i=1}^{N} \sigma_{i}^{z}$, commutes with the anisotropic 1D XY Hamiltonian,~$\left[H_{XY},Z\right] = 0~$. Indeed, the Hamiltonian in equn.(\ref{transversefieldxy}) is invariant under the combined operations 
$\sigma_{i}^{x}\to -\sigma_{i}^{x}$, $\sigma_{i}^{y}\to -\sigma_{i}^{y}$ and $\sigma_{i}^{z}\to \sigma_{i}^{z}$. 
This is achieved by the operator ${\cal S}=e^{i\frac{\pi}{2}\sum_{i=1}^{N}\sigma_{i}^{z}}$, and corresponds to a global rotation in spin-space of all spins about the z-axis by $\pi$~\cite{dejonghvanleeuwen-1998}. Further
\begin{eqnarray}
{\cal S} &=& e^{i\frac{\pi}{2}\sum_{l=1}^{N}\sigma_{i}^{z}}~=~\Pi_{l=1}^{N} i\sigma_{l}^{z} = W\times Z~, 
\label{twistoperZ}
\end{eqnarray}
where $W=e^{i\frac{\pi}{2}N}$ and $Z=\Pi_{l=1}^{N} \sigma_{l}^{z}$. Clearly, as $W$ and $Z$ both commute with $H_{XY}$, so does ${\cal S}$. $W$ defines a sensitivity to the total no. of spins $N$, e.g., for $N=4M$ and $N=4M+2$, ${\cal S}= \pm Z$ respectively, while for $N=4M+1$ and $N=4M+3$, ${\cal S}= \pm iZ$ respectively. Henceforth, we will assume $N=4M$.
We can see, therefore, that the consequences of this $Z_{2}$ symmetry will be felt for all $0<\delta\leq 1$. By carrying out a Jordan-Wigner transformation, we can write the Hamiltonian for the fermionized model as
\begin{eqnarray}
H &=& - h N +2 h {\sum_{i=1}^{N}} {c_{i}^{\dagger} c_{i}}  + {\sum_{i}}  \big[ J_2 ( c_{i}^{\dagger} c_{i+1}^{\dagger} + c_{i+1} c_{i} ) \nonumber \\
&& + J_1 ( c_{i}^{\dagger} c_{i+1} +  c^{\dagger}_{i+1} c_{i} )\big] ~,
\end{eqnarray}
where $J_{1}=J$ and $J_{2}=\delta J$.
Upon Fourier transforming this Hamiltonian as well as defining the spinor $(c_{k}~c^{\dagger}_{-k})$, we can write this Hamiltonian as~\cite{anderson-1958}
\begin{eqnarray}
H =\sum_{k=-\pi}^{\pi}~(c^{\dagger}_{k}~c_{-k})\left[ d_{x}\sigma_{x} + d_{z}\sigma_{z}\right] {\begin{pmatrix}
c_{k} \\
c^{\dagger}_{-k} \end{pmatrix}}
\end{eqnarray}
where $d_x (k) = -J_{2} \sin k$ and $d_z (k) = h + J_{1} \cos k$~, and where $d_x, ~ d_z$ have been written with $k=\pm \pi$ as the reference point. The energy-momentum dispersion relation is then
\begin{eqnarray}
\epsilon_k &=& \pm \sqrt{(J_{2}\sin k)^2 + (J_{1}\cos {k} + h)^2} \\
&=& \pm \sqrt{d_x^2 + d_z^2} 
\end{eqnarray}
and the wavefunctions for the positive and negative branches of the dispersion are given by 
\begin{eqnarray}
|\psi_{+}\rangle &=& \frac{1}{\sqrt{2}}{\begin{pmatrix}
\sqrt{2 - \frac{\tilde{\gamma}}{\pi}}\\
sgn (k) \sqrt{\frac{\tilde{\gamma}}{\pi}}\end{pmatrix}}~~,\\
|\psi_{-}\rangle &=& \frac{1}{\sqrt{2}}{\begin{pmatrix}
sgn (k) \sqrt{\frac{\tilde{\gamma}}{\pi}}\\
-\sqrt{2 - \frac{\tilde{\gamma}}{\pi}}\end{pmatrix}}
\end{eqnarray} 
respectively, where $\frac{\tilde{\gamma}}{\pi} = 1 - \frac{d_{z}(k)}{\sqrt{d_{x}^{2}(k) + d_{z}^{2}(k)}}$. 
Note that $\tilde{\gamma}$ is adiabatically connected to the Berry phase $\gamma_{0}$ of the $\pi$-mode of the TFIM described earlier for $J_{1}=J_{2}\equiv J$, $\tilde{\gamma}\to\gamma_{0}=\pi(1-sgn(\Delta))$ for $k=\pi$. Further, the critical point of the 1D TFIM for $\delta=1$ extends as a line of critical points for $0\leq \delta \leq 1$ all the way to the 1D XX model at $\delta=0$.

\begin{widetext}

\begin{figure}[!htb]
\centering
\includegraphics[width=0.45\textwidth]{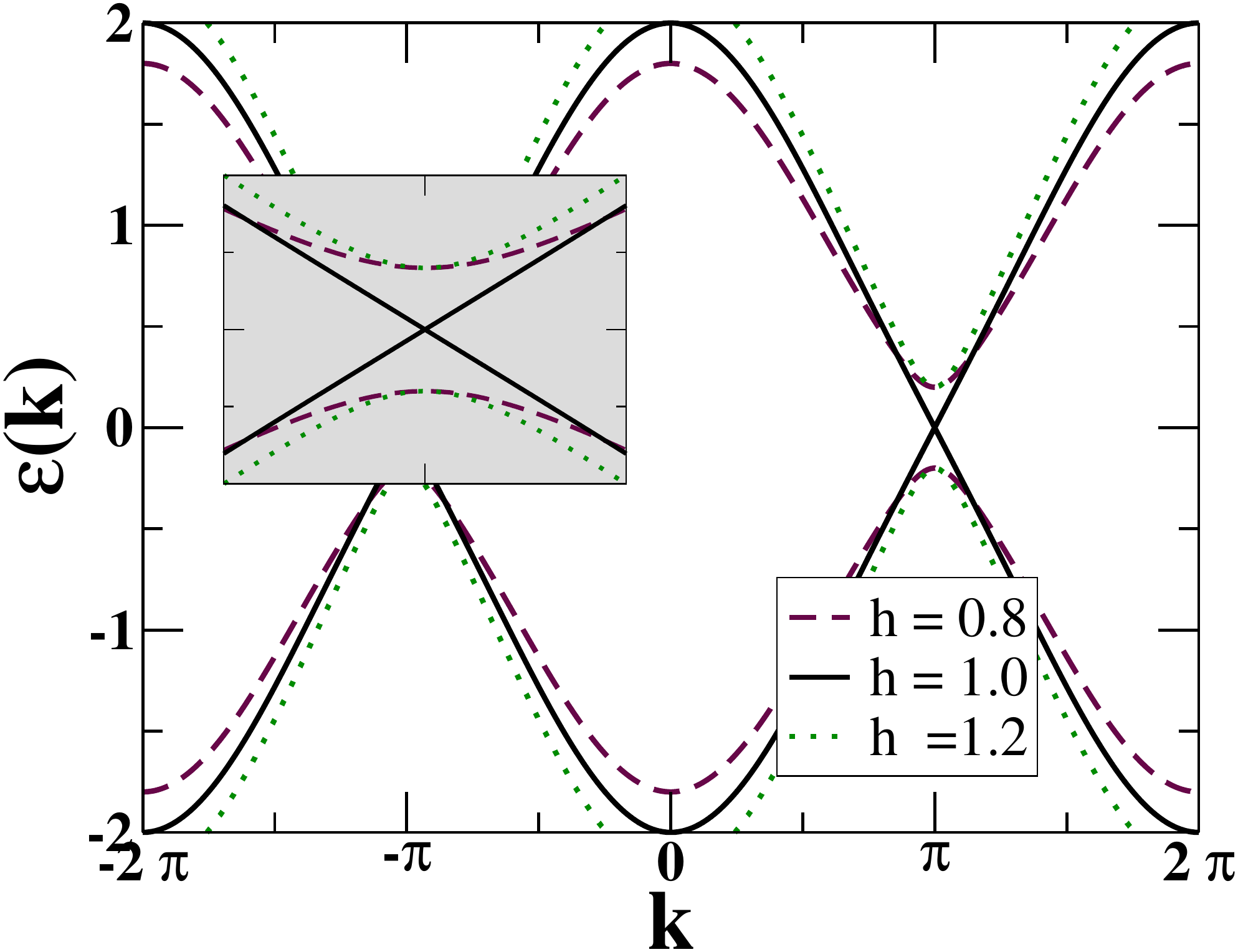}\qquad
\includegraphics[width=0.45\textwidth]{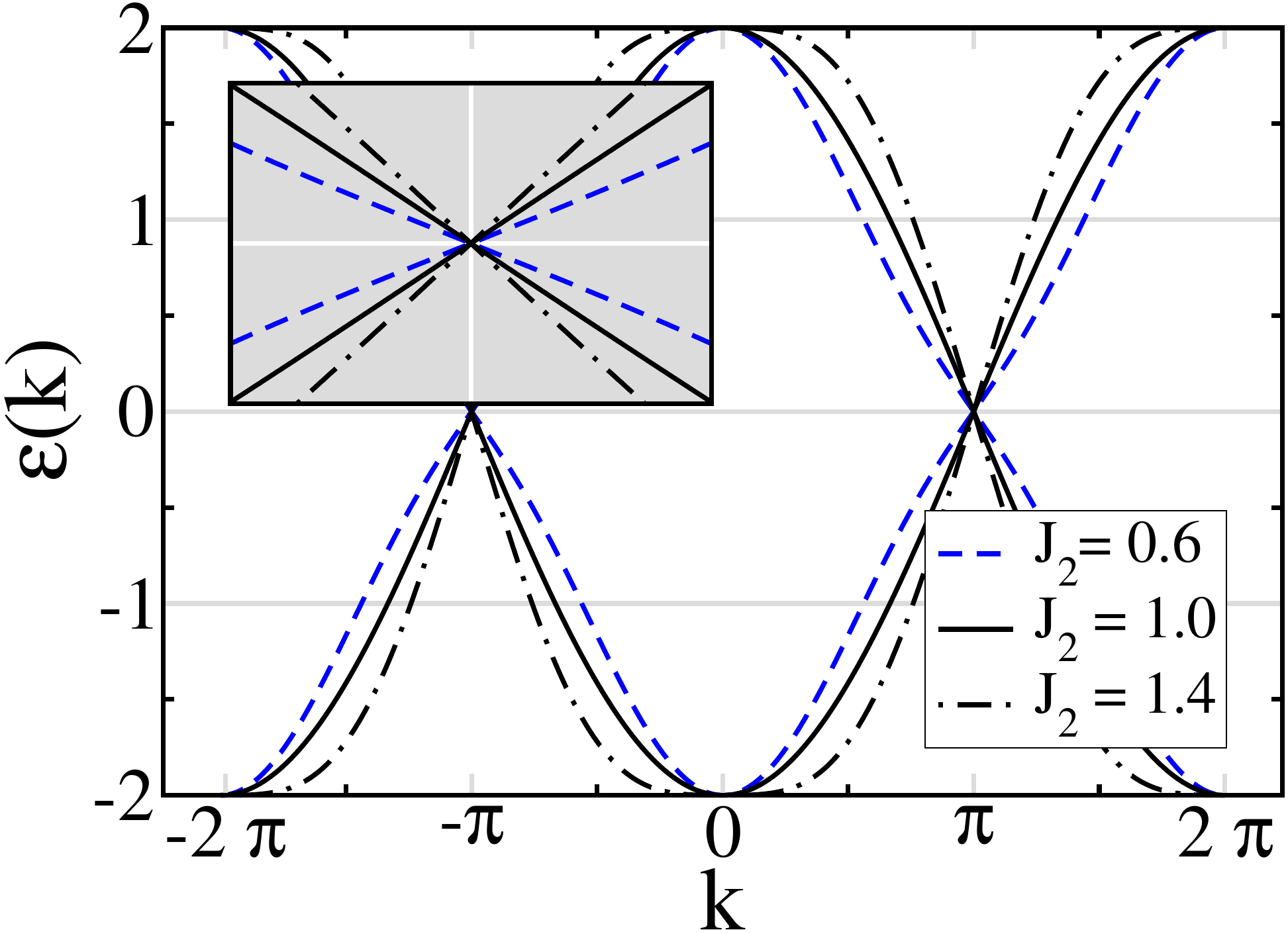}
\caption{(Color online.)~
Left: Plot of dispersion relation for different $h$ values (see left panel) keeping $J_1 =1.0$ and $J_2 = 0.8$. Inset: zoom of the gapless point. Right: Plot of dispersion for different $J_2$ values keeping $J_1 = h  = 1 $. $J_1$, $J_2$ and $h$ are hopping strength, pairing interaction strength and transverse field respectively. When $J_1 = J_2$, we have transverse field Ising model; when $J_2 = 0$, but $J_1 \neq 0$, we have instead the isotropic XY model or XX model with a transverse field. In the anisotropic XY limit ($J_1 \neq J_2$), the spectrum remains gapless for $J_1 = h$ even as $J_{2}$ is varied. Inset: zoom of the gapless point.
}
\label{fig:}
\end{figure}

\end{widetext}

Precisely the same dispersion relation can also be obtained for the Su-Schriffer-Heeger model (SSH) model~\cite{ssh-1979} of fermions hopping on a dimerised 1D lattice: $d_x = - (t-\delta t) \sin k$ and $d_z = 2 \delta t + 2 (t-\delta t ) \sin^2(k/2)$~,~ where $t$ is the nearest neighbour hopping amplitude and $\delta t$ the strength of the dimerisation. This arises from the fact that both models are equivalent to the same generalised two-band lattice fermion Hamiltonian in 1D: the two bands can be represented using a fermionic spinor notation (the Nambu spinor for the pWSC and the two sublattices for the SSH model). The equivalence between the fermionised TFIM/pWSC and SSH models is  easily seen to be: $J \equiv (t-\delta t)$~,~$h \equiv (t+\delta t)$~,~$\delta t \equiv (h-J)/2$ and the gap scale $E_{gap} = 4 \delta t \equiv 2 (h-J)$~. We shall see below that while the fermionised TFIM/pWSC captures a topological transition between topological and non-topological superconducting states of matter, the SSH model does so for a topological and non-topological insulating states of matter. A similar topological transition in a finite-sized Haldane-Fermi-Hubbard model was investigated numerically in Ref.(\cite{varney-2011}).

\subsection{Order parameters, critical exponents and the nature of criticality}
Thus, the 1D anisotropic XY, the 1D TFIM and the SSH model all possess the same Lifshitz topological quantum phase transition associated with a gap closing event. Given the wealth of results known for the two-band model, we can quantify the Lifshitz transition of the the fermionised TFIM/pWSC model via a Berry phase $\gamma$ accrued for an adiabatic parallel transport carried out over the entire Brillouin zone for the case of a gapped dispersion~\cite{zak-1989,sqshen}:
\begin{eqnarray}
\gamma &=& \int_{-\pi}^{\pi}~dk~\langle \psi_{-}|i\partial_{k}|\psi_{-}\rangle\nonumber\\
&=& \frac{1}{2}\int_{-\pi}^{\pi}~dk~(i\partial_{k}\ln(sgn(k)))\frac{\tilde{\gamma}}{\pi}\nonumber\\
&=& \frac{\pi}{2} \left[ sgn (t+\delta t) - sgn (\delta t) \right] ~~\mathrm{for~SSH~model}\nonumber\\
&=& \frac{\pi}{2} \left[ sgn (h) - sgn (h-J) \right]~~\mathrm{for~pWSC~model} \nonumber\\
&=& 0 ~ \text{for} ~  h>0 ~ \& ~  h>J \\
&=& \pi ~ \text{for} ~ h>0 ~\& ~ h< J
\label{BerryThoulessChargePump}
\end{eqnarray}
This identifies $\tilde{\gamma}$ as the Berry connection, which leads to the non-trivial Berry phase $\gamma$ when integrated over the full Brillouin zone. Similiarly, the Lifshitz transition can also be quantified in terms of a topological winding index $\Omega$ as~\cite{sqshen} 
\begin{eqnarray}
(-1)^{\Omega} &=& sgn (\delta t) sgn (t+\delta t)~~\mathrm{for~SSH~model} \\
&=& sgn (h-J) sgn (h) ~~\mathrm{for~pWSC~model}\\
&=& 1 ~ \text{for} ~  h>0 ~ \& ~  h>J : \Omega = 0\\
&=& -1 ~ \text{for} ~ h>0 ~\& ~ h< J : \Omega = 1
\end{eqnarray}
As we are interested in $h\geq 0$, the Berry phase $\gamma$ and the winding no. $\Omega$ can also be directly related to the $\pi$-mode energy $\epsilon_{\pi}=\Delta\equiv h-J$:
\begin{eqnarray}
\gamma &=& \frac{\pi}{2}(1 - sgn (\Delta)) = \frac{\pi}{2}(1 - \frac{\epsilon_{\pi}}{|\Delta|})\\
(-1)^{\Omega} &=& sgn (\Delta) = 1- \frac{2\gamma}{\pi}= \frac{\epsilon_{\pi}}{|\Delta|}~.
\end{eqnarray}
We will now show that the quantity $(-1)^{\Omega}$ is also related to the $Z_{2}$ symmetry operator $Z$ of the TFIM defined earlier. This can be done by rewriting equn.(\ref{twistoperZ}) as follows
\begin{eqnarray}
\Pi_{l=1}^{N}\sigma_{l}^{z}\equiv Z &=& W^{*}\times S\nonumber\\
&=& e^{i\pi(\frac{1}{2}\sum_{l=1}^{N}\sigma_{l}^{z} - \frac{N}{2})}\nonumber\\
&=& e^{i\pi (S_{Tot}^{z} -S)}~,
\end{eqnarray}
where $S=\frac{N}{2}$ and $S_{Tot}^{z} = \frac{1}{2}\sum_{l=1}^{N}\sigma_{l}^{z}$~.
Now, following Ref.(\cite{mcgreevylectures}), it can be shown that spin-flip/fermion parity operator~\cite{fradkin-book} is given by
\begin{eqnarray}
Z 
&=& (-1)^{\sum_{k} c^{\dagger}_{k} c_{k}} = (-1)^{N}~,
\end{eqnarray}
where $N$ is the total no. of fermions that occupy the states in the dispersion spectrum of the fermionised TFIM/pWSC. Given that all states with $k\neq 0,\pi$ are doubly occupied and the state with $k=0$ is never occupied for $h\geq 0$, we can reduce the operator $Z$ to
\begin{eqnarray}
Z = (-1)^{c^{\dagger}_{\pi} c_{\pi}} \equiv (-1)^{\Omega} = e^{-i\pi\Omega}~,
\label{eqZ}
\end{eqnarray} 
where $c^{\dagger}_{\pi} c_{\pi}$ defines the occupancy of the state at $k=\pi$ and $\Omega$ is the topological winding no. defined earlier. In this way, we find 
\begin{eqnarray}
\Omega = S - S_{Tot}^{z}~,
\label{oyacrit}
\end{eqnarray}
such that $\Omega=0$ gives $S_{Tot}^{z}=S$ for $h>J$, and $\Omega=1$ gives $S_{Tot}^{z}=S-1$ for $h<J$~. Both of these are special cases of the Oshikawa-Yamanaka-Affleck (OYA) criterion for obtaining ordered, gapped phases in spin-1/2 chain and ladder systems~\cite{oya-1997}
\begin{equation}
n\times(S - m) = \mathrm{Integer}~,
\end{equation}
where $n$ characterises the no. of degenerate ground states in the gapped phase, $S$ and $m$ are the maximum spin value and magnetisation respectively for a unit cell of the system. For any finite-sized 1D TFIM with periodic boundary conditions, we have $n=1$ for $h>J$ as well as $h<J$. That our result identifies the properties of a unit cell in the original OYA formulation with global quantities ($m\equiv S_{Tot}^{z}$ and $S=N/2$) indicates the topological nature of the problem at hand.
Thus, using the Berry phase $\gamma$ and the winding index $\Omega$, we can see the 1D TFIM undergoes a Lifshitz transition at $h = J$, from a topologically trivial phase for $h>J$ to a topologically non-trivial phase for $h<J$. We note that a winding no. topological invariant has similarly been used to characterise topological phases and Lifshitz transitions in Ising models with extended (three-spin) interactions~\cite{niu-2012,zhangsong-2015} as well as in certain spin-ladder systems.~\cite{degottardi-2011,feng-2007}~We expect, therefore, that the importance of a singular mode in the fermionised spectrum can be realised in these models as well.
\par
We can also relate Ferrell's observation~\cite{ferrell-1973,ferrell-1981} of an equivalence between Onsager's result for the logarithmic divergence of the specific heat~\cite{onsager-1944} with vanishing reduced temperature for the 2D Ising model and a nonlocal dependence of the specific heat for the fermionised TFIM on the wavevector $k$ to the Lifshitz transition and related topological quantities. We can rewrite Ferrell's result for the singular part of the specific heat (i.e., its scaling form) as
\begin{eqnarray}
C (k)&\sim& - \ln (1-\frac{\tilde{\gamma}}{\pi}) \sim - \ln (\frac{d_{z} (k)}{\sqrt{d_{x}^{2}(k) + d_{z}^{2}(k)}})\nonumber\\
&\sim& -\ln \frac{\Delta}{\sqrt{k^{2} + \Delta^{2}}} \sim \ln \frac{k}{\Delta}~~\mathrm{for}~k>>\Delta~,
\end{eqnarray}
where we have used the linearised (Dirac-like) dispersion in the second line. By writing $\Delta \equiv (T-T_{C})/T_{C}$ (the reduced temperature for the 2D Ising model), we obtain Onsager's result for the scaling form of the specific heat~\cite{onsager-1944}. Further, for $h>>J$ (for the TFIM, corresponding to $T>>T_{C}$ for the 2D Ising model), $\Delta >> k$, $\tilde{\gamma}=0$, leading to $C (k<<\Delta)\to 0$. On the other hand, for $h\to J+$ (for the TFIM at criticality, corresponding to the $T=T_{C}$ for the 2D Ising model), $\Delta\to 0$ for $k$ finite, $\tilde{\gamma}\to\pi$, leading to $C \to\infty$. This is easily seen as arising from the vanishing of the $\pi$-mode energy $\epsilon_{\pi}\equiv \Delta =0$  at $h=J$. Finally, for $h<J$, this singular part of the specific heat $C$ becomes a complex valued object, picking up a phase of $\pi$ as it crosses a branch cut at $h\leq J$~\cite{ferrell-1981}. This phase of $\pi$ is precisely the non-trivial Berry phase of $\gamma=\pi$ obtained earlier for the topologically non-trivial ordered phase at $h<J$. As we shall see below, the branch cut arises from the non-trivial homotopy group of the $SU(2)$ topological WZNW theory for the $\pi$-mode, $\pi_{3}(SU(2))=Z$. 
\par
We comment here on the critical exponents associated with the Lifshitz transition. First, the manifest Lorentz invariance of the dispersion spectrum of the 1D TFIM at $h=J$ (i.e., a vanishing Hamiltonian for the ]$\pi$-mode) gives us the dynamical critical exponent $z=1$. Next, a logarithmic dependence of the specific heat on the gap scale ($|h-J|$) suggests that the critical exponent $\alpha=0$. Further, the correlation length $\xi\sim 1/|h-J|^{\nu}$ yields the critical exponent $\nu=1$. Both relations have been shown as related to properties of the $\pi$-mode theory. It is easily seen that these exponents satisfy the Josephson hyperscaling law: $2-\alpha = \nu d$, with $d=2$ being the spatial dimensionality of the classical Ising model (or the space-time dimensionality of the TFIM). The dynamical scaling law $y=z \nu$ is also satisfied, where $y=1$ is the gap exponent ($\Delta\sim |h-J|^{y}$) and also yields the divergence of relaxation timescales $\tau\sim |h-J|^{-y}$ (i.e., critical slowing down). While it is tempting to derive all other critical exponents via other scaling inequalities, we would like to clarify that by working at zero longitudinal field ($h_{x}=0$), we are able to connect only these exponents directly to the Lifshitz transition.
\par
Note that the topological underpinning of the log divergence of the specific heat is manifested by closing the spectral gap first (i.e., taking $\Delta\to 0$, or correlation length $\xi\to\infty$) while holding $k$ finite. This is very different from the usual strategy in which continuous second order transitions are observed: take the thermodynamic limit (constituent no. $N\to\infty$, system volume $V\to\infty$ but density $N/V$ fixed) first and then close the spectral gap (i.e. $\Delta\to 0$, or correlation length $\xi\to\infty$) next. 
Does this violate the Yang-Lee theorem, i. e., the requirement that the thermodynamic limit be taken for the existence of continuous transitions~\cite{yang-lee}, as seen through zeros of the partition function/ non-analytic behaviour of the free energy? Following Ref.(\cite{guralnik:0710.1256}), we learn that this is not necessarily the case. When some of the couplings of a zero-dimensional quantum field theory vanish, the accumulation of Yang-Lee zeros in its partition function can take place in the form of a branch cut  in the complex space associated with the remaining non-trivial couplings. The vanishing couplings are seen to be analogous to the thermodynamic limit in the Yang-Lee theorem~\cite{guralnik:0710.1256}. 
\par
With reference to the problem at hand, the $\pi$-mode of the TFIM is an example of such a zero-dimensional QFT, and the vanishing of the fields $\Delta$ and $\tilde{h}_{x}$ (i.e., $\Delta=0\equiv h=J$ and $\tilde{h}_{x}=0$) give rise to the $SU(2)$ topologically non-trivial WZNW theory for a spin-1/2. As mentioned above, the branch cut corresponds to the non-trivial homotopy group $\pi_{3}(SU(2))=Z$. The branch cut can also be visualised as follows~\cite{fradkin-book}: the WZNW term for a spin-1/2 can be rewritten in terms of the dynamics of a charged particle moving on the surface of a sphere with a (Dirac) magnetic monopole (of strength suitably quantised corresponding to the $S=1/2$ value of the spin) placed at the origin. Such a magnetic monopole is coupled to infinity by a Dirac string (corresponding to a branch cut), resulting in the non-trivial homotopy group $\pi_{3}$. 
This branch discontinuity was also encountered in section \ref{sec:2d-classical-1d-QI}, where we found that the case of  $\Delta=0=\tilde{h}_{x}$ corresponds to the coalescing of two complex square-root branch-point level-crossing singularities into one level-crossing event on the real axis. Indeed, following Refs.(\cite{kortman-1971,fisher-1978}), we see that these coalescing branch points are examples of Yang-Lee edge singularities.
\par
As the physics of a level-crossing is associated with that of a first-order phase transition, the verification of the Josephson and dynamical scaling relations needs understanding. For this, we note that the non-local order parameter valid away from the critical point (the $Z_{2}$ topological invariant $\Omega$) is completely different from that at criticality (the WZNW invariant $W_{0}$ associated with the emergent $SU(2)$ symmetry). This signals the transition as falling outside the Ginzburg-Landau-Wilson paradigm, and shows that there can be critical exponents and scaling relations associated with such phase transitions as well. More insight will be available in section \ref{sec:rg-2d-1d-ising} where the renormalisation group analysis of this transition is laid out.
\par
We can now also point out the important consequence of our results for transitions in topological insulators and superconductors. Following Refs.(\cite{kitaevperiodictable,ryuschnyderfurusakiludwig,leclairbernard}), it is known that for translation invariant systems in $d-$spatial dimensions whose Hamiltonian can be written in terms of free fermions, the topological properties can be determined from consideration of the effective Dirac operators that are emergent at low energies and is the pathway to clarifying their bulk-boundary correspondence. As observed earlier, for the $d=1$ case, the Dirac operator is given by $D = \gamma\partial_{x} + M$, where the $\gamma$ matrix is given by $\gamma=i\sigma_{x}$ and the mass term is $M=\Delta\sigma_{z}$. 
As we have discussed at length, the case of $\Delta\rightarrow 0$ signals the Lifshitz transition. In a $d=2$ topological insulator, a similar transition can be studied for the edge states~\cite{fukaneJJ} using the $4\times 4$ Dirac operator $D=\gamma \partial_{x} + M$ where
\begin{eqnarray}
\gamma = && {\begin{pmatrix}
I & 0 \\
0 & -I \end{pmatrix}} ~~,~~ M = {\begin{pmatrix}
-h_{z}(i\sigma_{y}) & m \\
-m^{T} & h_{z}(i\sigma_{y}) \end{pmatrix}}~,\nonumber\\
&&m=-\tilde{h}_{x}(i\sigma_{y}) + h_{y} I - (\mathrm{Re}\Delta)\sigma_{x} - (\mathrm{Im}\Delta)\sigma_{z}~.
\end{eqnarray}
For the case of $h_{z}=0$, the mass gap is determined entirely by $m$ and can be shown to vanish when $det(m) = \tilde{h}_{x}^{2} + h_{y}^{2} - |\Delta|^{2} \equiv 0$~. Again, the appearance of an emergent $SU(2)$ symmetric massless Dirac spectrum at the transition is observed. Another example of a similar Lifshitz transition in a $d=2$ system can be found in the anomalous Hall insulator with competing superconducting and ferromagnetic orders~\cite{arxiv:1407.6539}, where the emergent Dirac operator has $\gamma_{a}=\sigma_{x}, \gamma_{b}=\sigma_{y}$ and $M=m \sigma_{z}$ and $m$ is the mass function.
\par 
Indeed, Ref.(\cite{kitaevperiodictable}) offers a K-theory based classification for 
translation invariant free fermion Hamiltonians in $d-$spatial dimensions. Kitaev shows that when such Hamiltonians possess a spectral gap, they are topologically equivalent to
the Dirac operator $D = \sum_{a}\gamma_{a}\partial_{a} + M$, where the $\gamma_{a}$ matrices satisfy $\gamma_{a}\gamma_{b} + \gamma_{b}\gamma_{a}=-\delta_{ab}$ and $M$ is a symmetric mass matrix that anticommutes with $\gamma_{a}$ ($\gamma_{a} M = -M\gamma_{a}$  for all $a$) and is nondegenerate (i.e., has no vanishing eigenvalues). Non-trivial symmetry-protected gapless boundary states that arise from the bulk-boundary correspondence are sought from the existence of {\it textures}, i.e., topological constructs arising from the spatial variations of the mass matrix $M$ that vanish at the boundaries~\cite{leclairbernard}. In this way, the classification of gapped phases of free fermion Hamiltonians into 10 classes has been achieved using the ideas of symmetries (time-reversal, parity and particle-hole) and dimensionality alone~\cite{ryuschnyderfurusakiludwig,kitaevperiodictable,leclairbernard}. Much less has been understood about the transitions into these topological phases of matter, and this is something that can now be repaired. In keeping with our detailed analysis of the Lifshitz transition in the 1D TFIM, we can see that the transition for the Dirac operator $D$ in $d$-spatial dimensions given above will again be signalled with at least one vanishing eigenvalue of the mass matrix $M$ (such that either $M\equiv 0$ or $det(M)=0$~\cite{kitaevperiodictable,leclairbernard}), and will lead to the appearance of massless Dirac fermions in the theory with an emergent $SU(2)$ symmetry.

\subsection{Topological properties of the ordered phase \label{subsec:topprop}}
We have already seen that the ordered state is characterised by the Berry phase $\Gamma$ and the topological winding index $\Omega$. From its equivalence to the low-energy theory of the SSH model, we can also obtain several other results for the fermionised TFIM/pWSC model. The topologically non-trivial phase for $J>h$ is doubly degenerate (in the thermodynamic limit) and separated from all other excited states by $E_{gap}$. Further, it possesses fractionally charged excitations whose charge is given by~\cite{stone-1985,dhlee-2007}
\begin{equation}
Q = \frac{\gamma (J>h)-\gamma (J<h)}{2\pi} q= \frac{\pi}{2\pi} q= \frac{q}{2}~,
\label{fraccharge}
\end{equation}
where $q$ is the elementary charge of the Dirac fermions. These fractionally charged excitations appear in the form of domain walls that separate sections of the two ground states in a given mixed ground state.
The non-trivial charge polarisation obtained for the SSH model during one cycle of an applied modulated $E$-field corresponds to the adiabatic Thouless charge pump for the effective low-energy massive Dirac fermions of the pWSC model; this will be demonstrated in detail in a following section. Here, we provide the heuristic picture in terms of domain wall excitations in Fig.(\ref{fig:gap-vs-t}) below.
\begin{figure}[!t] 
   \centering
   \includegraphics[width=0.4\textwidth]{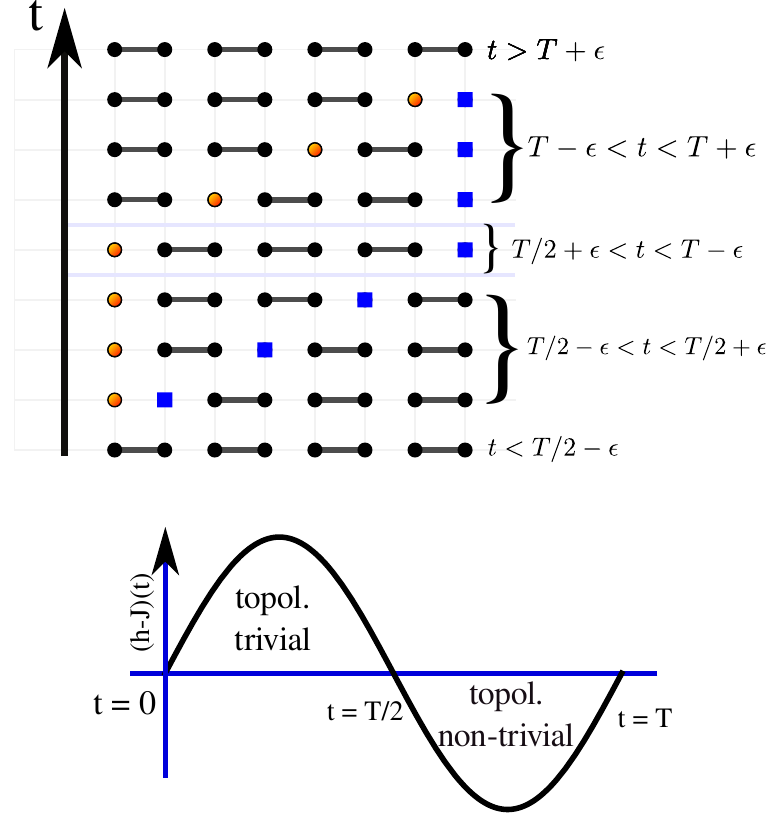}  
   \caption{(Color online.)~(a)The pumping of a single fermion across the ends of a finite-sized system corresponds to the transfer of two domain walls from end to end in the following sequence: (i) we start with a system that has no end states ($J<h$) for time $t<T/2$, (ii) upon crossing the gap-closing at $t=T/2$, we create single ground state with two domain walls ($J>h$); these two deconfine and move out to the two ends of the system within a short time-interval $\epsilon$: $T\leq t < T/2 + \epsilon$, (iii) for $T/2+\epsilon \leq t \leq T -\epsilon$, we have a ground state with domain walls at the two ends (iv) within a short time-interval $\epsilon$ of the gap-closing at $t=T$ (i.e., $T-\epsilon < t \leq T$), the two domain walls come together (are confined), and (v) for $t\geq T$, we are again in a ground state without any end states. (b) Gap as a function of time for the TFIM.}
   \label{fig:gap-vs-t}
\end{figure}
\par
In this way, we obtain a topological Chern no. defined by change in the Berry phase over a one time period (in units of $\pi$ ) 
\begin{eqnarray}
\text{Chern no} ~  {\cal C} = (\gamma(T-\epsilon) - \gamma (0+\epsilon))/\pi = 1~.
\label{chernThoulessChargePump}
\end{eqnarray}
This indicates that the Lifshitz transition into a topologically non-trivial phase involves a change in this Chern no. ${\cal C}$. Indeed, ${\cal C}$ corresponds to the pumping of a single electron across the ends of the system, but it can be written in terms of a product of the Atiyah-Singer topological index~\cite{atiyahsingera,atiyahsingerb} $n_{AS}= 2$ involved in the spectral flow pump and the fractionally charged excitations $Q=1/2$:
\begin{equation}
{\cal C} =n_{AS} \times Q~.
\label{ChernAtiyahSinger}
\end{equation} 
Given that the Atiyah-Singer index is associated with the spectral flow properties of massless Dirac fermions, the above relation between ${\cal C}, n_{AS}$ and $Q$ reveals the importance of the critical point in the cyclic pumping process. As will be discussed in the next section, it is related to a hidden supersymmetry (SUSY)~\cite{akhourycomtet-1986} of the theory at criticality.
\par
We can also show the existence of boundstates at the ends of an open TFIM chain (or the equivalent SSH model in the A-phase) by taking the continuum limit of the lattice Dirac Hamiltonian obtained from the fermionised TFIM/pWSC~\cite{sqshen}. These bound states are found to be localized within a length scale $\xi $ from the ends points
\begin{eqnarray}
\xi &=& \frac{\hbar}{m v} = \frac{J}{|h-J|} \equiv \frac{t-\delta t}{2 |\delta t|}~,
\end{eqnarray}
such that $\xi \to \infty$ as $|h - J| \to 0$ (for TFIM) and $\delta t \to 0$ (for SSH). This shows the edge states merge with the bulk at the Lifshitz transition ($J=h$). 
Further, the existence of edge states for the fermionised TFIM/pWSC can also be demonstrated by a numerical computation of its bandstructure in the presence of a periodically time-varying gap (i.e., Dirac mass) function. This is shown in Fig.(\ref{fig:2.1}) below.
\begin{widetext}

\begin{figure}[hbtp]
\centering
\includegraphics[width=0.3\textwidth]{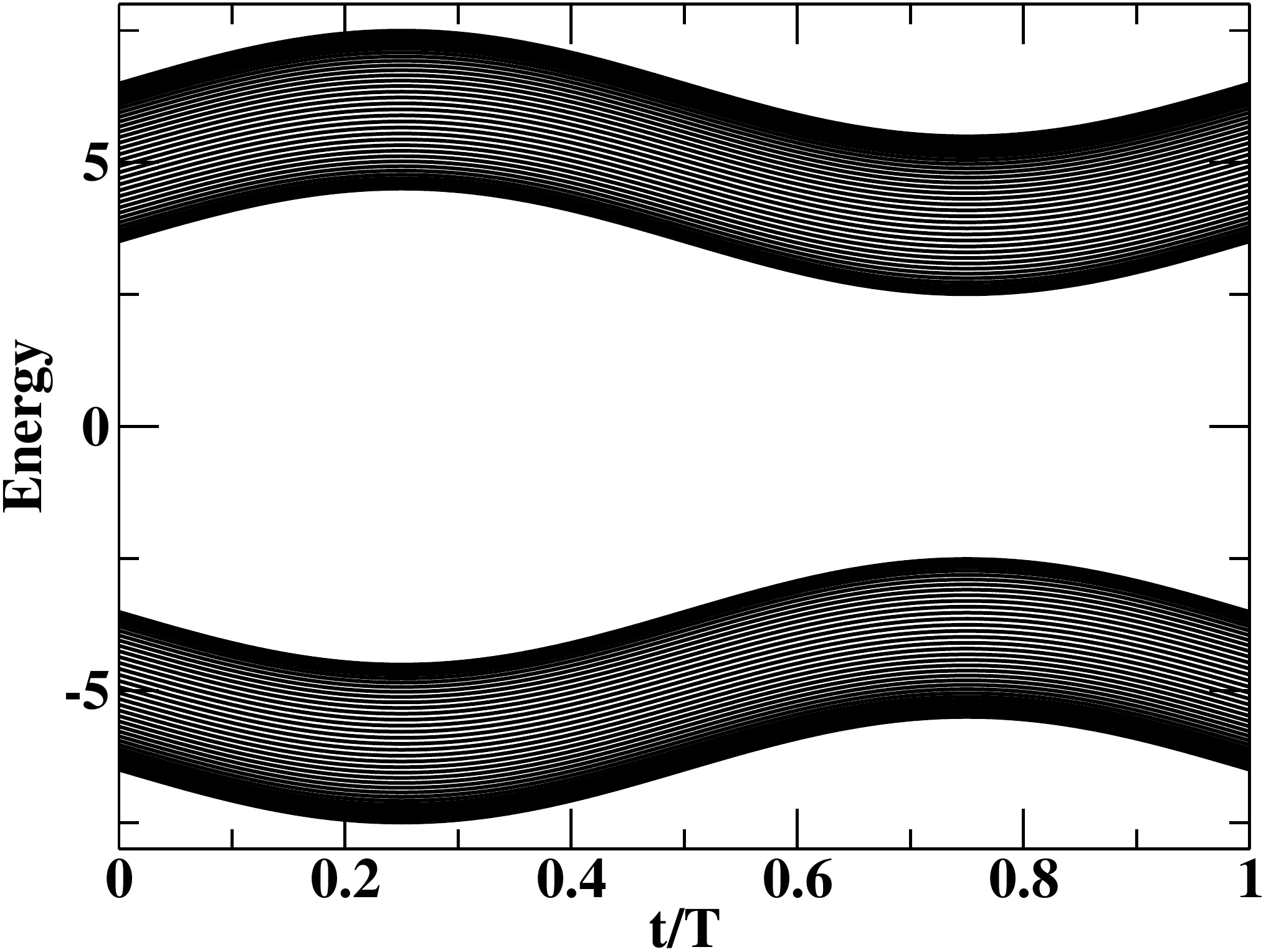}
\includegraphics[width=0.3\textwidth]{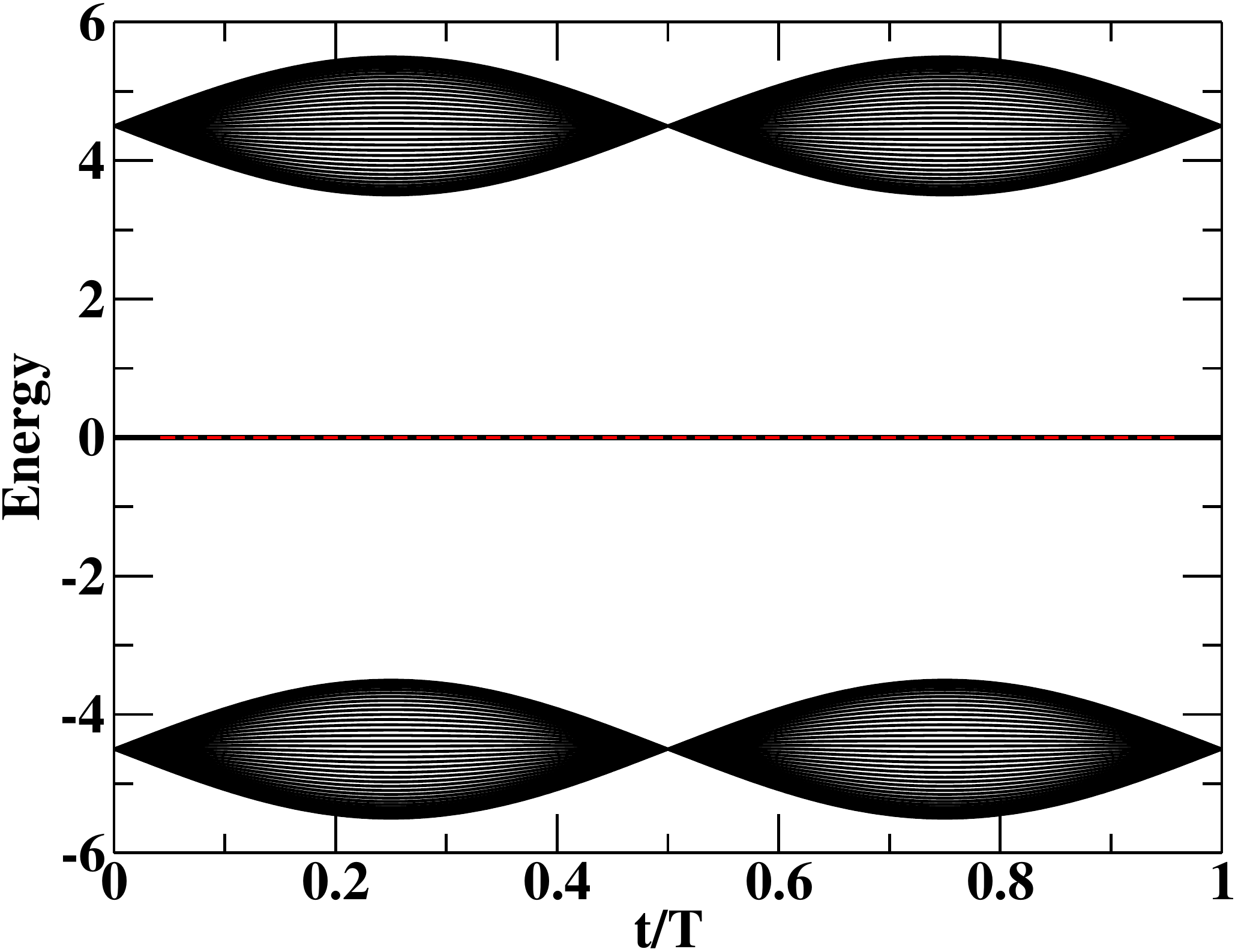}
\includegraphics[width=0.3\textwidth]{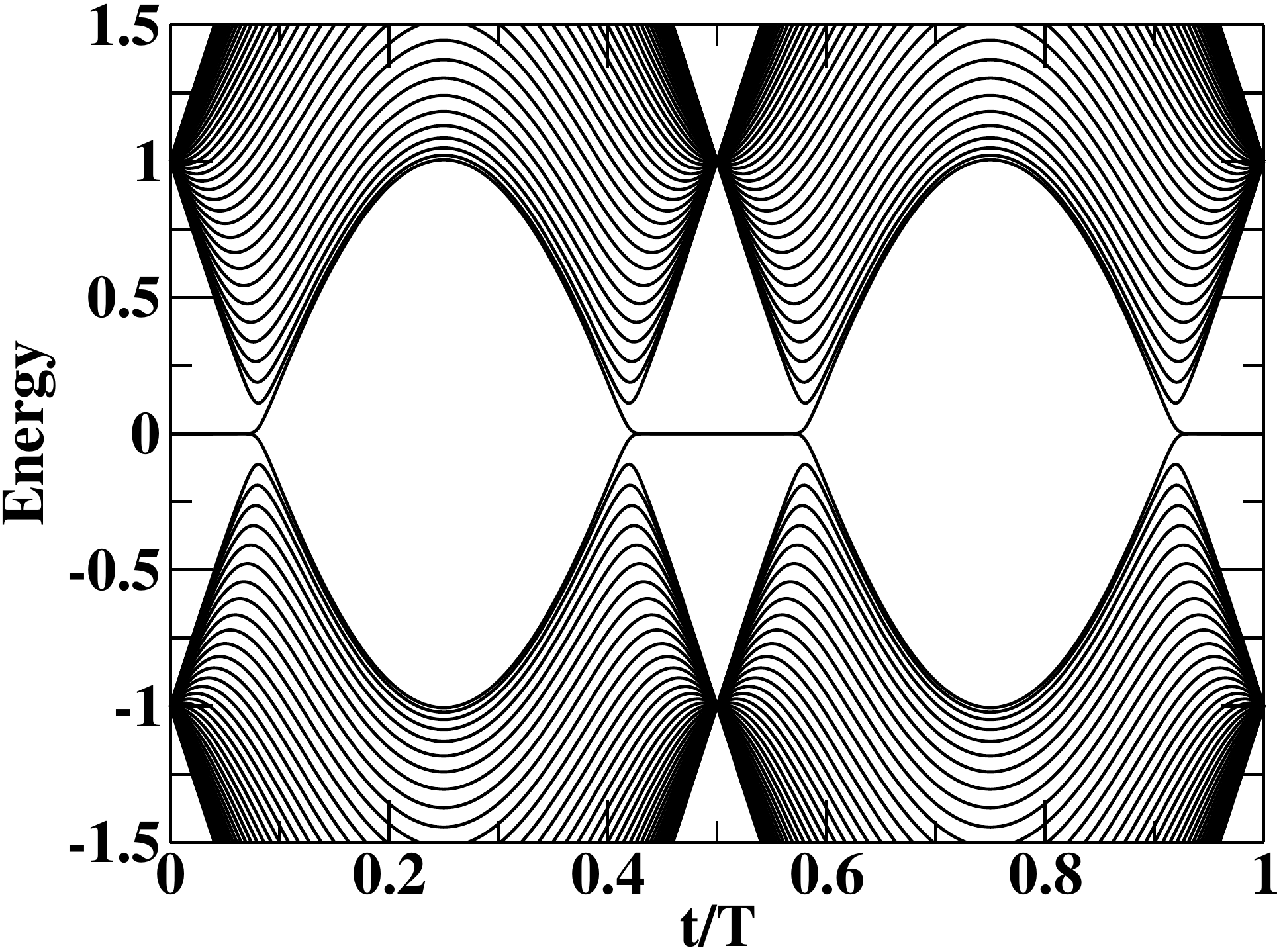}
\caption{(Color online.)~Plot of eigenspectrum as a function of rescaled time for different real space indices. Left: $h = 5 + \sin (\frac{2 \pi t}{T}), J = 1.5$~, the system is always in a topologically trivial phase and no edge states are seen. Middle: $h = \sin (\frac{2 \pi t}{T}), J = 4.5 $, the system is always in the ordered phase (or, topologically non-trivial phase) and we see only zero energy spectrum. Right: $h = 2 \sin (\frac{2 \pi t}{T}), J = 1.0$, the system is periodically cycled between the topologically trivial ($h>J$) and topologically non-trivial ($h < J$) phases. We then observe edge states going from bulk to zero energy periodically with time.}
\label{fig:2.1}
\end{figure}
\end{widetext}

The connection between the $\pi$-mode and the non-trivial end states is found by noting that Pfeuty's end spin correlation for the TFIM~\cite{pfeuty-1970} can be rewritten as
\begin{eqnarray}
\langle \sigma_1^x \sigma_L^x \rangle &=& \frac{J^{2}-h^{2}}{J^{2}} \qquad \text{for} ~~ J>h\nonumber\\
&=& -\epsilon_{\pi}\frac{J+h}{J^{2}}\nonumber\\
&=& (-1)^{\Omega + 1} (1 - \frac{h^{2}}{J^{2}})~,~\Omega=1~,~J>h\\
&=& 0 \qquad \text{for} ~,~\Omega=0~,~J<h~.
\end{eqnarray}
This relation shows that an eigenstate of the bulk (the $\pi$-mode eigenstate) connects the two end spins in the topologically non-trivial phase through a topological winding no. $\Omega$ (the bulk-boundary correspondence). Thus, a $\pi$-mode energy/gap scale $\epsilon_{\pi}\equiv\Delta = h-J$ that varies periodically in time will lead to a spectral flow process involving a periodic change of the end-to-end spin-spin correlation in time. This is the way in which a periodically changing end-to-end correlation leads to the Thouless charge pumping process discussed above. Further, as the Berry phase $\gamma$ is related to Pfeuty's end-spin correlation, 
\begin{eqnarray}
\gamma (t) &=& \frac{\pi}{2}\left[ 1 - \frac{J^{2}}{h^{2}-J^{2}}\langle \sigma_1^x \sigma_L^x \rangle (t)~\right]~, 
\end{eqnarray}
as well as the fact that the Chern no. ${\cal C}$ arises from a periodic excursion in time, we can identify a natural time-scale for the time period $T=\beta\hbar$ (where 
$\beta=1/\textrm{k}_{\textrm{B}}$T is the inverse temperature) from the Kubo-Martin-Schwinger (KMS) periodicity~\cite{kubo-1957,martinschwinger-1959} of the end-spin correlation function $\langle \sigma_1^x \sigma_L^x \rangle (t) = \langle \sigma_1^x \sigma_L^x \rangle (t+T)$. We also comment briefly on the fact that the ``bulk" correlation function $\langle \sigma_{i}^{x}\sigma_{i+n}^{x}\rangle$ has been computed in the limit $n\to\infty$ and found to be~\cite{wu-1966,mccoy-1968,pfeuty-1970,perkyang-2009}
\begin{eqnarray}
\lim_{n\to\infty} \langle \sigma_{i}^{x}\sigma_{i+n}^{x}\rangle &=& (\frac{J^{2}-h^{2}}{J^{2}})^{1/4}~~~\textrm{for}~\Delta<0\\
&=& 0~~~\textrm{for}~\Delta>0~.
\end{eqnarray}
This expression has been interpreted as the expectation value/thermal average of the square of the spontaneous magnetisation $\langle M_{x}^{2}\rangle $ in the ordered phase (even as $\langle M_{x}\rangle=0$ due to the Ising symmetry $\sigma_{i}^{x}\to -\sigma_{i}^{x}$)~\cite{wu-1966,mccoy-1968,pfeuty-1970}. Strikingly, $\lim_{n\to\infty} \langle \sigma_{i}^{x}\sigma_{i+n}^{x}\rangle$ does not coincide with the end-spin correlation expression for the open-chain given above, $\langle \sigma_{1}^{x}\sigma_{L}^{x}\rangle$, in the thermodynamic limit $L\to\infty$; the former is also sufficiently smaller in value than the latter, and vanishes much faster as the critical point is reached. All of this can be ascribed to the topological nature of the end-spin correlation, and we will therefore continue to focus on this aspect in what follows.
\par
While the $\pi$-mode Hamiltonian does not have a $\tilde{h}_{x}$ field (i.e., $\tilde{h}_{x}=0$ in equn.(\ref{qubitHam})) and must be introduced by hand, we can now realise its role. A non-zero, externally imposed, $\tilde{h}_{x}$ will lead to a non-vanishing gap when $\Delta\to 0$ in a finite-sized system, $E_{gap}=2\sqrt{\tilde{h}_{x}^{2}+\Delta^{2}}\to 2|\tilde{h}_{x}|$. 
By writing the theory with open boundary conditions in terms of lattice Majorana fermions~\cite{kitaevmajoranachain}, an intrinsic source of $\tilde{h}_{x}$ can also be realised as the tunneling coefficient between Majorana fermions at the two ends of the chain in the gapped topological phase: here, $\tilde{h}_{x}\sim e^{-L/\xi}$, where the correlation length $\xi=\hbar/E_{gap}$. This tunneling corresponds to fluctuations in the occupancy of the $\pi$-mode (a non-zero $\tilde{h}_{x}\sigma_{x}=\tilde{h}_{x}(c_{\pi}^{\dagger} + c_{\pi})$ term breaks the global $U(1)$ symmetry of the $\pi$-mode Hamiltonian, rendering it number non-conserving), as well as in the end-spin correlation $\langle \sigma_1^x \sigma_L^x \rangle$. Clearly, this intrinsic $\tilde{h}_{x}$ vanishes upon taking the thermodynamic limit in the standard manner (i.e., $L\to\infty$ before $\xi\to\infty$~\cite{bogoliubov-book}). However, the gapless dynamics of the boundary degrees of freedom of the ordered phase (and thus its topological nature) and the end-spin correlation $\langle \sigma_1^x \sigma_L^x \rangle$ has been suppressed in taking the thermodynamic limit in this way. Then, instead of the topological Lifshitz transition discussed above, we expect to find the standard Ginzburg-Landau-Wilson paradigm of quantum phase transitions~\cite{sachdev-book}: a spontaneous magnetisation order parameter $\bar{m}_{x}\sim (T-T_{C})^{-\beta}$, a susceptibility $\chi\sim (T-T_{C})^{-\gamma}$, the importance of the ``bulk" correlation function $\lim_{n\to\infty} \langle \sigma_{i}^{x}\sigma_{i+n}^{x}\rangle$ restored etc.
\par
Further, if $\tilde{h}_{x}$ stays finite in the thermodynamic limit, the gap never closes as $J\to h$ and the topologically nontrivial superconducting state for $J>h$ will be replaced by a topologically trivial superconductor in which no charge pumping is possible. Instead, following Refs.(\cite{kohn-1964,swz-1993}), we expect that this gapped spectrum will respond to twisted boundary conditions (by the application of a flux $\Phi$ to a system with periodic boundary conditions) by giving rise to a finite superfluid stiffness (or superfluid weight) $D_{S}\sim (\partial^{2} E/\partial \Phi^{2})_{\Phi=0}$, where $E$ is the ground state energy in the thermodynamic limit. We also comment briefly on the role of dissipation on such end-to-end tunneling. If the $\Delta \sigma_{z}$ term in Hamiltonian equn.(\ref{qubitHam}) is coupled to a bath of harmonic oscillators with an Ohmic spectral function, the problem becomes precisely that of the well-studied spin-boson problem~\cite{leggettreview-1987}. The RG treatment for the spin-boson problem is standard~\cite{chakravarty-1982,braymoore-1982}: it reveals that if the damping is less than a critical value (corresponding to a Berezinskii-Kosterlitz-Thouless (BKT) transition~\cite{BKT1,BKT2}), the tunneling prevails and leads to the suppression of the topological properties. On the other hand, if the damping is greater than the critical value, the tunneling is quenched. The latter possibility leads to the startling conclusion that the topological properties of this system can be secured via dissipation~\cite{bardyn-2013}. 

\section{\label{sec:electric-field} Lifshitz transition as spectral flow}
In the previous section, we have derived the effective 1D massive Dirac Hamiltonian which describes the passage through the critical point of the TFIM. Following Ref.(\cite{manton-1985,thacker-2014}), we can include a periodically time-varying electric field, thereby adding an extra compactified dimension to the problem. We know that the phase acquired by a charged particle as it goes around the Brillouin zone adiabatically is an example of geometric or Pancharatnam-Berry phase ($\theta$). This phase $\theta$ is actually a result of the electric field which creates a discrete momentum (and thus an effective Brillouin zone) due to periodicity along discrete time direction. This is essentially the physics of the Thouless adiabatic charge pumping phenomenon~\cite{thouless}, where the difference between value of bulk topological parameter $\theta$ before and after a gap-closing transition is equal to the amount of charge that has flowed from the bulk to the boundary. This is also known as the bulk-boundary correspondence for topological insulators. Let us now see how this transpires.
\par
We begin with the effective 1D Dirac Hamiltonian achieved from the critical point of the TFIM. When we put a time-periodic electric field, it acts like a second compactified dimension, as well as generates a time-dependent spectral/Dirac mass gap~\cite{manton-1985,thacker-2014}. The effective Hamiltonian can be written as
\begin{eqnarray}
H=  i \sigma^x\frac{\partial}{\partial x} + (\Delta - E t) \sigma^z~, {\label{eq:thacker}}
\end{eqnarray}
where $E$ is the electric field which will give the quasi time-periodicity $T$ as $E = 1/T$. The Dirac wavefunction is now periodic in x-direction and quasiperiodic in time-direction $t$
\begin{equation}
\psi(x,t+T) = e^{ix}\psi(x,t)~,
\end{equation}
with a zero-eigenvalue eigenfunction that has a Bloch-like form by using a Bloch wavenumber $n\in Z$ and a rescaled time lattice momentum in the compactified time direction as $k = t/T$
\begin{eqnarray}
\psi_{0} (x,k) &=&  \sum_{n = - \infty}^{\infty} |\psi_{\pm}\rangle e^{- \frac{T}{2} (n+k)^2} e^{i n x} \\
&=& \sum_{m = -\infty}^{\infty} |\psi_{\pm}\rangle e^{-\frac{1}{2 T} (x - x_m)^2} e^{-ik (x-x_m)}
\label{thacker:wavefunction}
\end{eqnarray} 
by the Poisson resummation method. The spinor wavefunctions in the helicity basis, $|\psi_{\pm}\rangle$, are given by
$|\psi_{+}\rangle = \frac{1}{\sqrt{2}}{\begin{pmatrix}
1\\
1\end{pmatrix}}~,~
|\psi_{-}\rangle = \frac{1}{\sqrt{2}}{\begin{pmatrix}
1\\
-1\end{pmatrix}}$~.
As we shall see below, the periodicity in $k$ has the information of transport of charge around the effective Brillouin zone, and represents a topological charge~\cite{manton-1985,thacker-2014}.
\par
We observe that this wavefunction looks exactly like the Landau wavefunctions encountered in the physics of the integer quantum Hall effect (IQHE). Below we recall the physics  of the adiabatic charge pumping mechanism and its connections to the physics of the IQHE. As will shortly become apparent, the connection lies in the 2-torus Hilbert space manifold for both problems, and the non-commutativity of momentum operators leading to a spectral flow process on this manifold. This leads to the non-trivial charge polarisation for the Thouless charge pump and the Hall conductivity for the IQHE as topologically quantised measurables. It is also interesting to note that the similarity of this idea to recent proposals for Floquet topological insulators, arising from the application of a time-perioidic electric-field to band insulators of electrons as a driven system (see Ref.(\cite{cayssol-2013}) and references therein).
\subsection{Adiabatic charge pumping on a two-torus: the integer quantum Hall effect}
We begin by recalling the Laughlin-Halperin gedanken~\cite{laughlin-1981,halperin-1982} for the two-dimensional integer quantum Hall effect (IQHE). 
\begin{widetext}

\begin{figure}[hbtp] 
\centering
\includegraphics[width=0.85\textwidth]{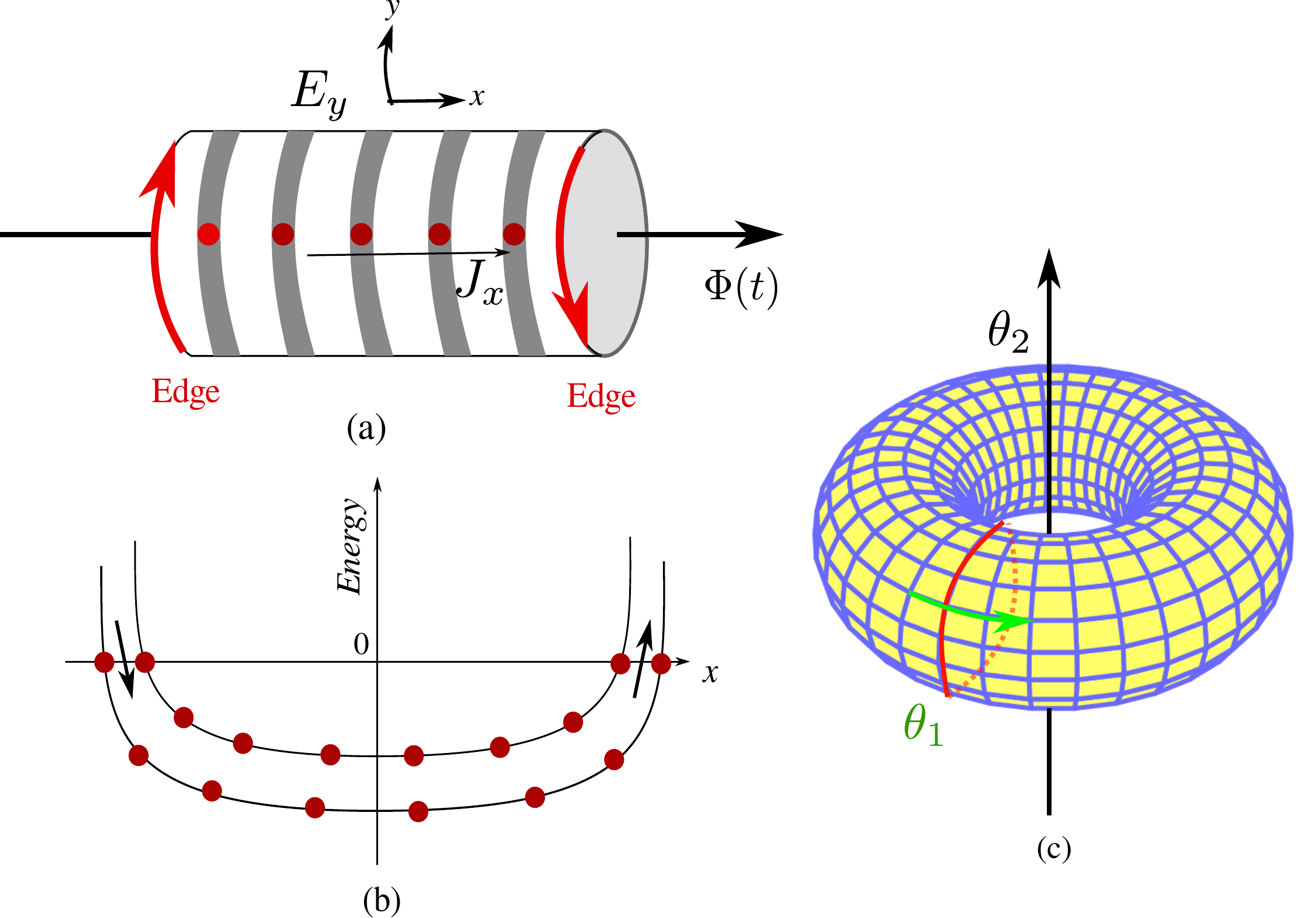}
\caption{(Color online.)~The quantum Hall set up. (a) The is 2D set up wrapped to a cylinder and the flux is going through he cylinder axis. The two edge states flowing charge current in the opposite direction is shown. (b) The spectral flow mechanism as the flux changes by one flux quanta is schematically depicted for the QHE. The energy of the states in each Landau level is shown as a function of the length of the cylinder ($x$), with the edge states at the chemical potential (zero energy). (c) The Berry phase is picked up due to charge transport due to flux change through a torus. The two different fluxes and non-commutative and give rise to non-zero Berry phase and thus topological Chern number.}
\label{fig:hall}
\end{figure}

\end{widetext}
Here, the 2D electron gas is placed on the surface of a cylinder (or the topologically equivalent Corbino disk) with the external B-field placed perpendicular everywhere to the surface (see Fig.({\ref{fig:hall}(a))). A solenoidal flux $\Phi$ pierces the cylinder axially, and is tuned from $\Phi=0$ to $\Phi=\Phi_{0}$ adiabatically in time. This flux corresponds to the application of an $E$-field across the two ends of the cylinder, and mimics the action of the battery in the external circuit to which the system is connected. The electronic wavefunctions of the 2D Landau problem are taken to be stripe-like in nature (with the Landau gauge chosen appropriately) such that they encircle the cylinder (i.e., they are transverse to the cylinder's axis) and form a one-dimensional lattice along the axial direction. Recall that in the Landau problem, we can define the canonical momenta $\Pi_\alpha = p_\alpha - e A_\alpha~,~\alpha = x,y$ such that 
\begin{equation}
\left[ \Pi_x , \Pi_y \right] = \frac{i e B}{\hbar c}~,
\end{equation}
where $B = \vec{\nabla} \times \vec{A} = B \hat{z}$ and in the Landau gauge $\vec{A} = (0,B x , 0) ~ {\text{or}} ~ (-B y , 0, 0)$~. 
This non-trivial commutation is responsible for the Chern number topological invariant (given by filling factor ${\cal C}$) that characterizes the Landau levels. 

The adiabatic change in the time-dependent flux then tunes the motion of this 1D lattice of stripe wave functions in the form of a spectral flow mechanism (see Fig.({\ref{fig:hall}(b))). The transfer of one electron for every Landau level below the chemical potential for flux change by $\Phi_{0}$ corresponds to the development of a Hall voltage drop in the transverse direction and arises from the non-commutativity of $\Pi_{x}$ and $\Pi_{y}$.
By defining the magnetic length $l_{b}$ and the cyclotron frequency $\omega_{c}$ in terms of the magnetic field $B$, $l_B^{-2} = \frac{e B}{\hbar c}~,~
\omega_c = \frac{e B}{m c}$, we can see that $l_B^{-2} = \frac{m}{\hbar^2} E_{gap}$~,~where~$E_{gap} = \hbar \omega_c$. Then, we can rewrite the non-trivial commutation relation given above as  
\begin{equation}
\left[ \Pi_x , \Pi_y \right] = \frac{i}{l_B^2} = \frac{i m}{\hbar^2} E_{gap}~.
\end{equation}
Following Ref.(\cite{tao-haldane}), we can see that $\sigma_{xy}$ is obtained from a Kubo relation involving this non-trivial commutator when defined for the components of the centre of mass momentum $[\Pi_{x}^{c},\Pi_{y}^{c}]$ acting on the ground state wavefunction $|\Psi_{0}\rangle$ and integrating out all excited states $|\Psi_{1}\rangle$ to which the ground state is connected. The Hall conductivity is related to a Berry phase $\bar{\gamma}$ accrued over a closed adiabatic circuit of the Hilbert space
\begin{eqnarray}
\sigma_{xy} &=& \frac{\sigma_{0}}{2\pi}\left(\ln\bar{\gamma}\right)~,~\nonumber\\ 
\bar{\gamma} &=& e^{iN_{e}/\left(L_{x}L_{y}\times\left[ \Pi_x , \Pi_y \right]\right)}~,
\end{eqnarray}
where $N_{e}$ denotes the total no. of electrons, $L_{x}, L_{y}$ are the spatial extents of the 2D system and $\sigma_{0}= e^{2}/h$ is the conductance quantum.

Alternatively, this can be thought of as carrying out a parallel transport gedanken in flux space through $\Pi_{c}^{\mu}=\frac{\partial H}{\partial \alpha_{\mu}}$ ($\theta_{\mu}$ indicates the two fluxes of the torus) over a full circuit. This is shown in Fig.(\ref{fig:hall}(c)). In this manner, the Hall conductivity $\sigma_{xy}$ is revealed to be a Berry phase obtained from a dynamical correlation function of $\left[\frac{\partial H}{\partial \theta_{\mu}},\frac{\partial H}{\partial \theta_{\nu}}\right]$ in flux ($\theta_{\mu}$) space. When quantised, this Berry phase becomes a Chern no~\cite{thouless,tao-haldane}. 
\subsection{Thouless adiabatic charge pump in 1D}
As seen above, the quantised Hall conductivity of the IQHE is basically the adiabatic pumping of one charge for every Landau level below the chemical potential over a torus composed of two compactified spatial dimensions. As we will now see, the Thouless charge pump in a one-dimensional system is the same over a torus of a compactified spatial dimension and a compactified time dimension (see Fig.~(\ref{fig:hall})). 
The flux is changing in a periodic manner (i.e., generates an electric field) such that the charge is transferred in the x-direction. As the t-direction is compactified, this can be done by using a flux tagged within the  cylinder, $\Phi_x (t)$. Then any wavefunction that are extended is the t-direction will feel an E-field, and undergo spectral flow in the x-direction. As observed for the Landau problem earlier, this can be related to the non-trivial commutator 
\begin{equation}
\left[ \Pi_x , \Pi_t \right] = i \hbar e \frac{ET}{L_{y}}~,
\end{equation}
defined on the compactified $(x,t)$ 2-torus with $\Pi_x = p_x + e A_x$~,~$\Pi_t = i \frac{T}{L_{y}}\frac{\partial}{\partial t} + e A_t$~,~$\vec{A} = (-Et ,0,0)$ and  $\vec{\nabla} = \hat{x}\frac{\partial}{\partial x}  +  \hat{y}\frac{T}{L_{y}}\frac{\partial}{\partial t}  + 0 \hat{z}$~. By defining a lengthscale $l_E^{-2} = \frac{e E T}{\hbar c L_{y}}$ and a frequency scale $\omega_E = \frac{e E T}{m c L_{y}}$~,~ we can rewrite the commutator in a familiar fashion 
\begin{equation}
\left[ \Pi_x , \Pi_t \right] ~= ~\frac{i m}{\hbar^2} E_{gap} ~,
\end{equation}
where $E_{gap}=\hbar\omega_{E}$. Following our calculations for the Landau problem, we can see that this commutation relation leads to a non-trivial charge polarisation between the ends of the system in the spatial dimension~\cite{thouless,manton-1985,thacker-2014} as the periodic $E$-field in modulated over one time period $T$. This polarisation is proportional to the change in the Berry phase $\gamma (t)$ (\ref{BerryThoulessChargePump}) computed over a time interval $0\leq t\leq T$, and yields the quantised Chern no.${\cal C}$ (equn.(\ref{chernThoulessChargePump})).
\par
Note that any such extended wavefunctions along the compactified $t$ dimension are equivalent to the stripe wavefunction of the Laughlin cylinder in the y-direction (refer to Fig.~(\ref{fig:hall})). Similarly, 
the $i \frac{\partial}{\partial t}$ operator in the quasiperiodic t-direction are equivalent to the operator $i \frac{\partial}{\partial y}$ along the compactified y-direction of the Laughlin cylinder. Thus, the discrete wavevector in the t-direction (equivalent to a set of integer winding nos. labelled by $n\in Z$) is the analog of the wavevector in the y-direction $k_y$ in the Laughlin-Halperin gedanken. For this argument to work, it is important to note that we are probing the Chern no. via an adiabatic tuning of the flux. Then, the time period $T$ over which the $E$ field is modulated must be much larger than all intrinsic energy scales of the system, e.g., the gap scale $E_{gap}$ or the Dirac mass scale. As the mass scale for the effective Dirac fermions obtained from the 1D TFIM is given by $|J-h|$, we must have $T \gg \hbar/|J-h| $ for adiabaticity, i.e., the E-field must be $\ll E_{gap}~\equiv~(|J-h|)$. For $J \rightarrow h$, $E_{gap}\rightarrow 0$ and $T \rightarrow \infty $~\cite{thacker-2014}. 
\subsection{Anomaly cancellation and the Chern number}
Having recognised above that the space-time 2-torus as being equivalent to two compactified spatial dimensions, we will now demonstrate the importance of the non-trivial dynamics of massive 2D Dirac electrons on this topologically non-trivial manifold in determining the Chern no. related to the quantised Hall conductivity of the Landau problem as a realisation of the exact cancellation of anomalies between the edge and bulk of the system~\cite{stone-1991}. 
\begin{widetext}

\begin{figure}[!htb]
\centering
\includegraphics[width=0.9\textwidth]{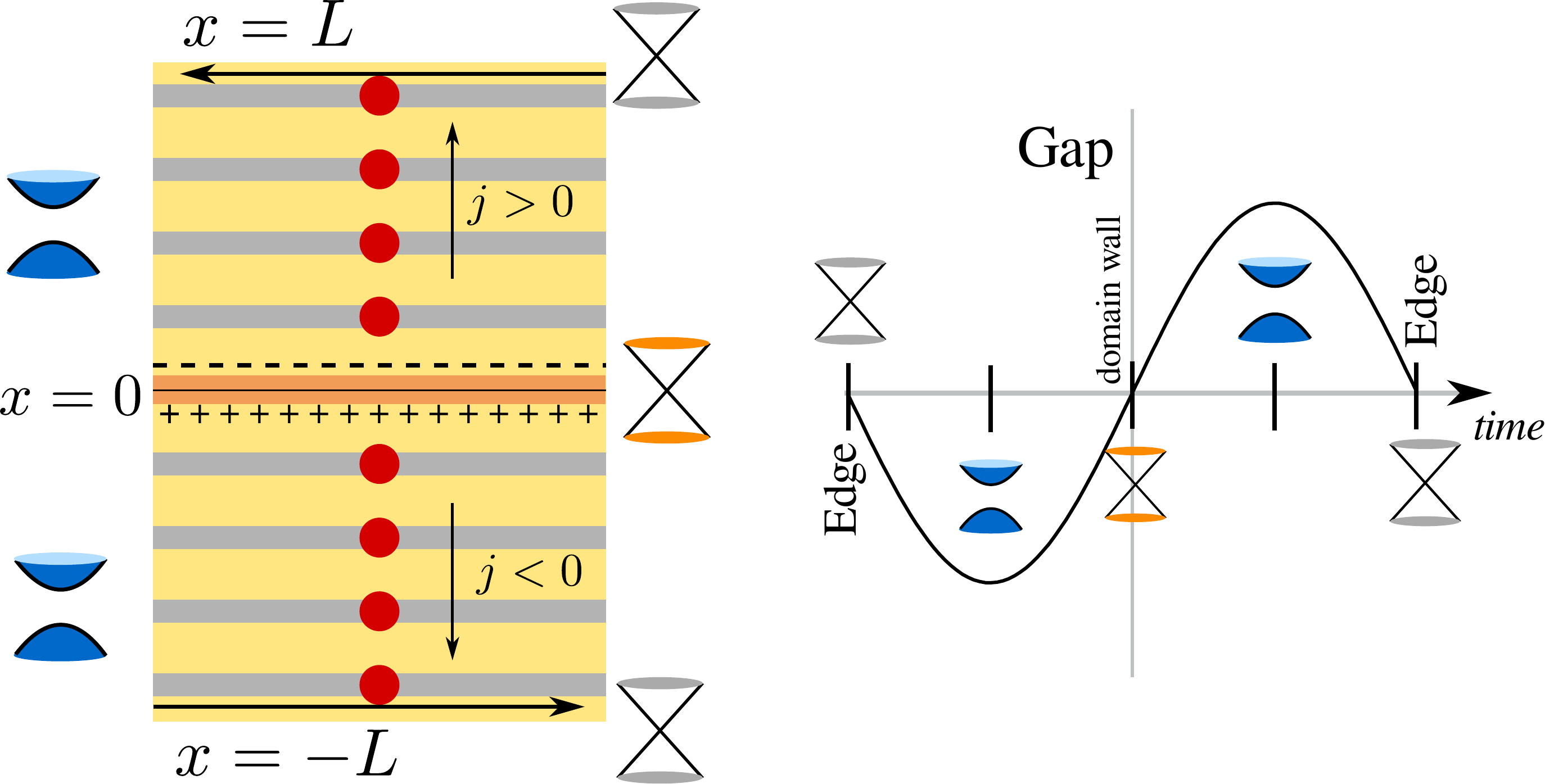}
\caption{(Color online.)~Equivalence between the quantum Hall setup and the adiabatic cycling of the gap in the 1D TFIM. (left panel) Real-space representation of the QHE. The three gapless states (massless Dirac-cones)  are to be identified with one domain wall in the bulk (the ``capacitor") and two oppositely-directed edge states at the boundaries. Spectral (current) flow through the gapped bulk is shown, leading to the cancellation of anomalies between the bulk domain wall and the edge states. (right panel) Variation of gap function $\pm (J-h)$ in the 1D TFIM. Below the domain wall is a topological phase (gap is negative) whereas above it is a topologically trivial phase (gap is positive). Gapless critical points reached in time here can be mapped onto the gapless states of the QHE in real-space. Adiabatic cycling of the gap in time recreates the spectral flow of the QHE.}
\label{fig:domain}
\end{figure}

\end{widetext}
 We start with the 2D Dirac Hamiltonian where we have a mass term that changes sign across a boundary
\begin{equation} 
H = -i \sigma_3 \partial_x -i \sigma_2 \partial_y + m(x) \sigma_1~.
\label{eq: Q12}
\end{equation}
This particular form of the equation emerges from the fact that we have chosen Landau gauge with $ A = (0, B x, 0) .$ The wavefunction in this gauge is the expected Gaussian in the x direction and a plane wave in the y-direction
\begin{equation}
\psi(x,y)_{Landau} = e^{i k_y y} e^{-\frac{(x-x_0)^2}{2 l^2}}
\end{equation}
The Quantum Hall(QH) plane has a left and a right going edge current as shown in figure \ref{fig:domain}. Concentrating on the Dirac Hamiltonian, the eigenstates are in $ \sigma_2 $ basis such that $ \sigma_2 u_0 = \epsilon u_0 $ with 
\begin{equation}
\psi(x,y) = e^{i k_y y} \; u_0 \; e^{\epsilon \int_{0}^{x} m(x') dx'}~.
\end{equation}
Here,~$\epsilon = -\frac{1}{2}\left[m(\infty) - m(-\infty)\right] = m(\infty) sgn(x)$, $m(\infty) = E_{F} - \hbar\omega_{C}$ is the cyclotron frequency and $E_{F}$ is the Fermi level. The state $ u_0 $ corresponds to one that is bound to the domain wall where the mass $m$ goes through a zero~\cite{stone-1991}. The states on the edge have a definite chirality $ \epsilon $ and if an electric field $ E_y $ is applied, the extended state in the y-direction responds. The momenta in the y-direction $ k_y $ are thereby changed to $ k_y + e E_y t. $ The flow of states can be regarded as an apparent anomaly where charges are being created out of the vacuum. The continuity equation for the edge can be written as
\begin{equation}
\partial_\mu j^\mu = \epsilon \frac{e^2}{4 \pi} \epsilon_{\sigma \tau}~, F^{\sigma \tau}
\end{equation}
where $ \epsilon = -1 $ as $ m=1 $ for $ x>0 $ and vice-versa. $ F^{\sigma \tau} $ is the general covariant field tensor. The present problem has only the a non-zero $ E_y $ component. The anomaly at the edge is a result of the charge fed from the bulk; this is the essence of the bulk-boundary correspondence~\cite{callanharvey-1985}. Thus, there should be an equal and opposite anomaly contribution from the bulk. Each $ k_y $ is like one-dimensional fermion theory. To calculate that, we observe that the current contribution in the system is the rate of change of phase. The phase change is given by the $ \sigma_2 $ and $ \sigma_3 $ terms in the Hamiltonian, with the net phase given by $ \tan^{-1} \frac{\phi_y}{\phi_x} .$ Then, the current is found to be~\cite{stone-1991}
\begin{eqnarray} 
\frac{e}{2\pi} \partial_t \tan^{-1}(\phi_2/\phi_1) &=& \frac{e}{2\pi} \partial_t \tan^{-1}\left(\frac{k_y + e E_y t}{m}\right)|_{t=0}\nonumber\\
&=& \frac{e^2 E_y}{2\pi m} \left(\frac{1}{1 + k_y^2/m^2}\right)~.
\label{eq: Q11} 
\end{eqnarray}  
This current leads to a flux which can be calculated by integrating over all the momenta
\begin{eqnarray}
j(x) &=& \frac{e^2 E_y}{2\pi m} \int \left(\frac{1}{1 + k_y^2/m^2}\right) dk_y\nonumber\\
&=& \frac{e^2 E_y}{4\pi}\; sgn(m)~.
\end{eqnarray}
\par 
The flux changes sign across $ x=0 $ along with the mass $m$. For, $ x<0 $, $ j(x) = -\frac{e^2 E_y}{4\pi} $ whereas for $ x>0 $, $ j(x) = \frac{e^2 E_y}{4\pi} $. The value of $j$ goes to $0$ at $x=0$. There is no flow of charges across $ x=0 $ and this tells us there is a domain-wall like structure at $ x=0$ (see Fig.(\ref{fig:domain}) (left panel)). The anomaly cancellation is easily observed when we calculate the current change across $ x=0 $ i.e. 
\begin{eqnarray}
j(x>0) - j(x<0) &=& \frac{e^2 E_y}{4\pi} - (- \frac{e^2 E_y}{4\pi}) \nonumber \\
&=& \frac{e^2 E_y}{2\pi}~.
\end{eqnarray}  
In a more standard notation, we can write this as
\begin{equation}
\partial_\mu j^\mu = \frac{e^2}{4 \pi} \epsilon_{\sigma \tau} F^{\sigma \tau}~,
\end{equation}
which is identical to that found at the edge. This calculation gives the same result irrespective of whether it is computed between the edges or across the two sides of the domain wall
\begin{equation}
\left[j(x=+\infty) - j(x=-\infty)\right] = \left[j(x=+\epsilon) - j(x=-\epsilon)\right]~.
\end{equation}
The Hall conductance can then be calculated  
\begin{equation}
\sigma_{xy} = \frac{e^2}{2\pi} = \frac{e^2}{h}\;\;\; \text{with}\;\;\; \hbar =1~.
\end{equation}
Note that the fact that the Hall conductance $\sigma_{xy}$ is quantized in units of $\sigma_{0}=e^{2}/h$ is actually of topological origin. Indeed, $\sigma_{xy}/\sigma_{0}={\cal C}$, where ${\cal C} =1$ is being assumed for the lowest Landau level in this calculation. We know that $\nu$ is a Chern index and now, we have identified this topological index as a property of the domain wall. This calculation shown here also reveals the bulk-boundary connection of this domain wall to the edge states. In this way, we can see that the Atiyah-Singer index topological quantum number for the edge states~\cite{hatsugai} is twice the Chern index ${\cal C}$ of the domain wall. This is basically the relation shown earlier in equation (\ref{ChernAtiyahSinger}). Further, following Ref.(\cite{ishikawamatsuyama}), the Chern index ${\cal C}$ can be shown to be related to a Ward-Takahashi identity of Dirac fermions on the space-time 2-torus, and is robust against scattering from disorder as well as electronic interactions. 
\par
While Ref.(\cite{stone-1991}) focuses on the anomaly cancellation between the bulk and a given edge of the system, the Laughlin-Halperin thought experiment\cite{laughlin-1981,halperin-1982} shows that the current flow ultimately happens between the two opposite edges of the system, leading to the Hall conductivity. The other edge, therefore, has to have a contribution with opposite chirality such that
\begin{equation}
\partial_\mu j^\mu = -\epsilon \frac{e^2}{4 \pi} \epsilon_{\sigma \tau} F^{\sigma \tau}~.
\end{equation}
\par
This anomaly is carried forward to the bulk and then to the other edge. The domain wall at $ x=0 $ acts like a capacitor with no net charge flowing through it. As soon as the charge from the lower part($ x<0 $) hits the domain wall, the upper part($ x>0 $) receives the charge from the domain wall (as shown in Fig.(\ref{fig:domain}) (left panel)). This domain wall is a requirement for chirality flip and hence cannot be affected by any local disorder. Indeed, from this calculation, it is clear that only non-local scattering events between the two edge-states can affect the topologically protected stability of the domain wall and the IQHE. The opposite directionality of the currents is a hallmark of the Quantum Hall effect and this requires a domain wall structure. Our finding of the domain wall gives evidence for the conjecture in Ref.(\cite{halperin-1982}) on the existence of the extended state that interpolates between the two edge states as being the key to Laughlin's demonstration of the topological nature of the QHE~\cite{laughlin-1981}. It also clarifies, as shown in Fig.(\ref{fig:domain} (right panel)), the connection between the Thouless charge pumping (via the adiabatic modulation of the spectral gap) in time for the fermionised 1D TFIM problem discussed earlier and the spectral flow of the IQHE system in real space. The critical points of the former at time $t=0$ and $t=T$ correspond to the gapless edge states of the latter, the critical point at time $t=T/2$ of the former to the domain wall within the bulk of the latter (which is holographically connected to the edge states via the anomaly cancellation mechanism) and the gapped phases of the former to the bulk of the latter.
\subsection{Thouless Charge Pump as CUT-RG\label{CUTRGsection}}
In order to complete our discussion of the holography at the heart of the spectral flow process on the 2-torus, we will now show that the Chern no. ${\cal C}$ for the Thouless charge pumping in the 1D TFIM can be obtained purely from the $\pi$-mode Hamiltonian, Eq.(\ref{piHam}), which we have shown tracks the Lifshitz transition. We begin by using the $U(1)$ symmetry of eq.(\ref{piHam}) to rewrite it as a problem of a quantum particle on a circle (POC) coupled to an Aharanov-Bohm flux $\Phi$
\begin{eqnarray}
H_{\pi} &=& \Delta \sigma_{z} = \frac{1}{2}(\sigma_{z} + \Delta)^{2} - \frac{\sigma_{z}^{2}}{2} - \frac{\Delta^{2}}{2}\nonumber\\
&=& \frac{1}{2I}\left(p_{\phi} + \frac{\Phi}{\Phi_{0}}\right)^{2} + {\textrm const.}~,
\label{pocHam}
\end{eqnarray}
where $p_{\phi}$ is the angular momentum of the POC, $\sigma_{z}$ (with eigenvalues $\pm 1$) is mapped onto $p_{\phi}$ ($\sigma_{z}\equiv p_{\phi}$), the gap $\Delta \equiv\Phi/\Phi_{0}$, ${\textrm const.}=-\frac{1}{2}(\sigma_{z}^{2} + \Delta^{2}) = -\frac{1}{2}(1 + \Delta^{2})$ and the moment of inertia of the particle $I=1$ for the fermionised 1D TFIM.   
\par
The Hilbert space here corresponds to that belonging to the $p_{\phi}$ operator, $|n\rangle = (1/\sqrt{L})~e^{in\phi}$, where the winding quantum no. $n$ is quantised ($n\in Z$) due to single-valuedness of $|n\rangle$ and $L$ is the perimeter of the circle. This Hilbert space is a sub-space of the full Hilbert space of the 1D TFIM. In this way, we can see that the tracking of the Lifshitz transition about $\Delta = 0$ (i.e., the degeneracy of the two levels) corresponds to the level-crossing of the eigenvalues $\pm 1$ of $p_{\phi}$ in the POC for $\Phi=0$.
We can now define rotation by azimuthal angle $\delta\phi$ on the circle via the operator $T_{\delta\phi}=e^{\frac{i}{\hbar}p_{\phi}\delta\phi}$ and ``flux-insertion" operator $U=e^{\frac{i}{\hbar}\frac{\Phi}{\Phi_{0}}\phi}$~. Then, we can show that 
\begin{eqnarray}
UT_{\delta\phi} &=& T_{\delta\phi}U~e^{\frac{\delta\phi}{\hbar}\frac{\Phi}{\Phi_{0}}\left[p_{\phi},\phi\right]}\nonumber\\
&=& T_{\delta\phi}U~e^{-i\delta\phi\frac{\Phi}{\Phi_{0}}}~.
\end{eqnarray} 
As before, this leads to a Berry phase $\gamma$ defined by taking a closed circuit on the circle $\delta\phi=2\pi$ which encloses the flux $\Phi$
\begin{eqnarray}
\gamma &=& -\frac{1}{2\pi}\textrm{Im}\left[\ln(U^{-1}T_{\delta\phi=2\pi}^{-1}UT_{\delta\phi=2\pi})\right]\nonumber\\
&=& -\frac{1}{2\pi}\textrm{Im}\left[\ln(e^{-i\delta\phi\frac{\Phi}{\Phi_{0}}})|_{\delta\phi=2\pi}\right]~=~\frac{\Phi}{\Phi_{0}}~.
\end{eqnarray}
$\gamma$ can be visualised as arising from the non-commutation of the $T_{\delta\phi}$ and $U$ operations on the Hilbert space $|n\rangle$. As the operation $U$ twists the boundary condition on $|n\rangle$ (from periodic boundary condition for $\Phi=0$ to anti-periodic for $\Phi=\Phi_{0}$), $\gamma$ measures the geometric phase collected by an adiabatic close-circuit excursion of the Hilbert space by the $U$ and $T$ (real-space rotation) operations. This is made evident by rewriting $\gamma$ as a boundary term obtained by integrating over a total derivative with respect to $\delta\phi$
\begin{eqnarray}
\gamma &=& \int_{0}^{2\pi}\frac{d(\delta\phi)}{2\pi}\frac{\partial}{\partial(\delta\phi)}\textrm{Tr}_{\phi}(\ln U)\nonumber\\
&=& \frac{1}{2\pi}\int_{0}^{2\pi~}d(\delta\phi)~\textrm{Tr}_{\phi}(U^{-1}\partial_{\delta\phi}U)~,
\end{eqnarray}
where the trace $\textrm{Tr}_{\phi}$ is taken over the complete Hilbert space $\sum_{|n\rangle}T^{-1}|n\rangle\langle n| T$~. In this way, we can see that the Berry connection is given by 
\begin{eqnarray}
A_{\phi}&=&{\textrm Tr}_{\phi}(U^{-1}\partial_{\delta\phi}U)\nonumber\\
&=& \langle n (\Phi)| \partial_{\delta\phi} |n (\Phi)\rangle~,
\end{eqnarray}
$|n (\Phi)\rangle \equiv U|n\rangle$~, and the Berry phase is a line integral taken over the Berry connection.
For a quantised flux $\Phi=\Phi_{0}$, the Berry phase $\gamma$ becomes the quantised Chern number ${\cal C}=1$. Indeed, by carrying out a Hopf map for the Berry flux equivalent to $A_{\phi}$, this Chern number for $\Phi=\Phi_{0}$ can be shown to be equivalent to the Wess-Zumino-Novikov-Witten term for a single spin-1/2 given earlier in eq.(\ref{wznwterm})~\cite{fradkin-book}, ${\cal C} \equiv 2S =1$ for $S=1/2$.
\par
We are now in a position to see that the Thouless charge pumping spectral flow process can be regarded as a continuous set of unitary transformations on the $\pi$-mode/POC Hamiltonian. Denoting the time variation of the flux by $\Phi(\tau)$, we can write
\begin{eqnarray}
H(\Phi(\tau)) &=& \frac{1}{2I}(p_{\phi}+\frac{\Phi(\tau)}{\Phi_{0}})^{2}\nonumber\\
&=& T_{\delta\phi}^{-1}U(\tau)^{-1}~H(\Phi=0)~U(\tau)T_{\delta\phi}~.
\label{tdependentPhi}
\end{eqnarray}
Then, taking a derivative of this Hamiltonian with respect to $\tau$ gives
\begin{eqnarray}
\frac{dH}{d\tau} &=& \frac{1}{\Phi_{0}}\frac{d\Phi}{d\tau}T_{\delta\phi}^{-1}(p_{\phi} + \frac{\Phi}{\Phi_{0}})T_{\delta\phi}\nonumber\\
&=& \frac{\dot{\Phi}}{\Phi_{0}}T_{\delta\phi}^{-1}U^{-1}~p_{\phi}~UT_{\delta\phi}\nonumber\\
&=&-i\frac{\dot{\Phi}}{\Phi_{0}}\left[ \phi,H(\Phi(\tau))\right]~,
\end{eqnarray}
where we have used the relations $p_{\phi} + \Phi/\Phi_{0}=U^{-1}p_{\phi}U=-i\left[ \phi,H(\Phi(\tau))\right]$~. Using the fact that $\frac{\Phi_{0}}{\dot{\Phi}}\frac{d}{d\tau} \equiv \frac{d}{d(\Phi/\Phi_{0})}$, we can rewrite the above as
\begin{equation}
\frac{dH}{d(\Phi/\Phi_{0})}= -i\left[ \phi,H(\Phi)\right]\equiv J_{\Phi}~,
\end{equation}
where $J_{\Phi}$ is the persistent current obtained by flux-insertion/applying twisted boundary conditions to the Hilbert space. Thus, by tuning the flux in time adiabatically, we are modifying the Hamiltonian $H(\Phi(\tau))$ through a set of continuous unitary transformations (CUT) $U(\Phi(\tau))$. Indeed, this is a special case of the CUT formalism~\cite{wegner-1994,kehrein-book}, where unitary transformations $U=e^{\eta (B)}$ are used towards diagonalising a Hamiltonian  $H(B)$ with inter-particle interactions in a step-by-step fashion, $\eta(B)=-\eta^{\dagger}(B)$ and $B$ is the tuning parameter for the CUT. This is regarded as a set of renormalisation group (RG) transformations, as it helps in obtaining the effective Hamiltonian that governs low-energy/long-wavelength dynamics through the following ``RG"-like relation
\begin{equation}
\frac{dH(B)}{dB} = \left[\eta(B), H(B)\right]~.
\end{equation}
For $H(B) = H_{0}(B) + H_{int}(B)$ where $H_{0}(B)$ is already diagonal and non-interacting and $H_{int}(B)$ contains inter-particle interactions, Wegner suggested~\cite{wegner-1994}~$\eta(B) = \left[ H_{0}(B), H_{int}(B)\right]$~. For the problem at hand, we can see that $B=\Phi/\Phi_{0}$, $\eta(B)=-i\phi(\Phi/\Phi_{0})$. While $H$ does not contain any inter-particle interactions in the present problem, the unitary evolution under CUT results simply from $\left[\phi,H\right]\neq 0$. Therefore, spectral flow of the Hilbert space due to cyclic passage through the critical point of the 1D TFIM (see Fig.(\ref{fig:gap-vs-t})) is equivalent to CUT-RG flow, i.e., the evolution of $H(\Phi=0)$ through a one-parameter family of Hamiltonians $H(\Phi)$ that are unitarily equivalent to it. 
\par
The Chern number topological invariant can now be obtained as an integral over the full $(\phi,\Phi/\Phi_{0})$ 2-torus for the change in the persistent current $J_{\Phi}$ with respect to a flux $\Phi$ that varies between $-\Phi_{0}/2$ and $\Phi_{0}/2$
\begin{eqnarray}
{\cal C} &=& \frac{1}{2\pi}\int_{0}^{2\pi}d\phi~\int_{-1/2}^{1/2}d(\frac{\Phi}{\Phi_{0}})~\frac{dJ_{\Phi}}{d(\Phi/\Phi_{0})}\nonumber\\
&=& \frac{1}{2\pi}\int_{0}^{2\pi}d\phi~\int_{-1/2}^{1/2}d(\frac{\Phi}{\Phi_{0}})~\frac{d}{d(\frac{\Phi}{\Phi_{0}})}(p_{\phi} + \frac{\Phi}{\Phi_{0}}) = 1~.
\label{chernhall}
\end{eqnarray}
Thus, the Chern number ${\cal C}$ of the Thouless charge pump can be considered as the quantity that is conserved even as the topological invariant $Z\equiv sgn(\Delta)= sgn(\Phi/\Phi_{0})$ is changed in value between $1$ and $-1$ during the adiabatic CUT-RG evolution. As mentioned in the last section, the mapping of the $\pi$-mode Hamiltonian onto that of a POC reveals a supersymmetry (SUSY) of the theory at criticality $\Delta\equiv \Phi/\Phi_{0} =0$~\cite{correaplyushchay-2007,rau-2004}, with a topological quantum no. called the Witten index $W=1$~\cite{witten-1982}. On the other hand, at flux $\Phi=\pm\Phi_{0}/2$ in the POC, the SUSY is broken and the Witten index $W=0$. Thus, from our discussion above, we can see that the Chern no. ${\cal C}=1$ of the Thouless charge pump corresponds to the Witten index $W=1$ of the SUSY critical state at $\Phi=0$ as the flux $\Phi$ is tuned from one broken SUSY state at $\Phi=-\Phi_{0}/2$ to another at $\Phi=\Phi_{0}/2$~\cite{akhourycomtet-1986,bollesimon-1987,lewississon-1975}. As the POC is the simplest quantum mechanical system which shares the $\theta$-vacua structure of a gauge theory~\cite{manton-1985,asorey-1983,rajaraman}, the CUT-RG shown here can be interpreted as equivalent to a passage from one $\theta$-vacuum to another (i.e., $\theta$ renormalisation) as the flux $\Phi\equiv\Delta$ is tuned through a large gauge transformation from negative values to positive values across the topological transition at $\Phi=0$~\cite{apenko-2008,bulgadaev-2006,bogachek-1990}.   
\par 
Finally, this Chern no. is also connected to the no. of kinks/edge states that are generated during the cyclic pumping process $n=1-(-1)^{\Omega}$ (where $\Omega$ is the winding no. describing the topological phase of the TFIM):~${\cal C} = \frac{1}{T}\int dt~ n(t)$~,~where $T$ is the time period of the periodic pumping process (see Fig.(\ref{fig:gap-vs-t})). Clearly, as $n(t)$ cycles between $0$ (trivial phase) and $2$ (topological phase), ${\cal C} = \frac{1}{T}\times \frac{T}{2}\times 2=1$~. As noted in Ref.(\cite{martindelgado-2010}) using the Majorana fermion formalism, this process is basically the adiabatic equivalent of the Kibble-Zurek mechanism~\cite{kibble-1976,zurek-1985}, where a finite density of topological excitations are generated during quench dynamics of a system, i.e., as it is taken through a continuous transition at a finite (non-adiabatic) rate. It should be noted, however, that the Kibble-Zurek mechanism is a feature of a second-order phase transition with a broken symmetry and a local order parameter, while our result applies to the case of a topological transition with a non-local order parameter. 
\section{\label{sec:dual}Duality, SPT order and entanglement in the TFIM}
It is well-known that the 1D TFIM can be written in terms of dual order and disorder spin/pseudospin operators (see, for instance, Refs.(\cite{kadanoffceva-1971,fradkin-susskind-1978})). The order operators are the original spins defined on the lattice sites while the disorder pseudospins are defined on the links that lie between the sites. Importantly, this duality of the TFIM is also known to be equivalent to the famous Kramers-Wannier duality of the 2D Ising model. The transformation between these two sets of spin/pseudospin operators is non-local and keeps the form of the TFIM Hamiltonian invariant. This is also reflected in the invariance of the BdG quasiparticle excitation spectrum of the fermionised TFIM/pWSC (upon taking the continuum limit in the neighbourhood of $k=\pi$) for $J < h$ and $J>h$: both are the same massive Dirac spectrum. The important question, then, is: what does this duality does tell us about the topological properties of the various phases of the TFIM? The answer is found by noting the relation between the ``$\pi$-mode" Hamiltonian given earlier at $k=\pi$, $H_{k=\pi}$, and the $Z_{2}$ symmetry operator of the TFIM defined earlier in equn.(\ref{eqZ}), $Z = (-1)^{c^{\dagger}_{\pi} c_{\pi}} \equiv (-1)^{\Omega} = sgn(\Delta)$. $Z$ is also a special case of the string operator used in the duality transformation between the order and disorder operators\cite{kadanoffceva-1971,fradkin-susskind-1978}: this string operator spans the entire system. 
The eigenvectors of the symmetry operator $Z$ are also the eigenstates of $H_{\pi}$ with different boundary conditions~\cite{mcgreevylectures}: periodic boundary conditions (PBC) for when $\Delta <0$ (the state at $k=\pi$ is occupied, $\Omega=1$, $N$ is odd) and anti-periodic boundary conditions (APBC) for when $\Delta >0$ (the state at $k=\pi$ is unoccupied, $\Omega=0$, $N$ is even). As will be brought to use later in this section, the duality reveals that this transition can equivalently be captured in terms of the disorder operators, whose ``ordered" phase (for $h>J$) possesses a non-trivial value of a dual $Z_{2}$ symmetry operator.
\par
The Lifshitz transition, however, takes place when $\Delta=0$, such that $sgn(\Delta)\equiv Z$ is ill-defined. Instead, the emergent $SU(2)$ symmetry of the $\pi$-mode characterises the TFIM critical point (known to be self-dual in terms of the order and disorder spin/pseudospin operators). The observation that the topological order parameter $Z$ changes abruptly across the transition, together with the finding that an altogether different topological order parameter characterises the critical point itself, suggest that the transition cannot belong to the Ginzburg-Landau-Wilson paradigm. We will confirm this from a RG analysis in the next section. As shown in another section, the same transition is also observed for the 1D Ising model at $T=0$. This is seen to arise simply from the fact that the $\pi$-mode of the fermionised TFIM is precisely the same theory as that obtained from the classical-quantum correspondence for the 1D Ising model, and reveals the holographic nature of the correspondence for the 1D and 2D Ising models. Further, it is interesting to note that a similar quantum phase transition was found in Ref.(\cite{arxiv:1407.6539}) between two anomalous Hall insulator phases of a 2D electron gas with Rashba spin-orbit coupling and competing orders arising from ferromagnetism and s-wave superconductivity. 

Now, the eigenstates of $Z$ correspond to the ground states of the finite-sized TFIM in its two phases, and are given by a $h\neq 0$ adiabatic continuation of the following $h=0$ states~\cite{zengwen,mcgreevylectures}
\begin{eqnarray}
|+\rangle &=& \frac{1}{\sqrt{2}}(|\rightarrow\rangle^{\otimes N} + |\leftarrow\rangle^{\otimes N})~,~ \Omega=0,\Delta >0,\textrm{APBC},\nonumber\\
|-\rangle &=& \frac{1}{\sqrt{2}}(|\rightarrow\rangle^{\otimes N} - |\leftarrow\rangle^{\otimes N})~,~ \Omega=1,\Delta <0,\textrm{PBC},
\label{ghzsptwavefunc}
\end{eqnarray}
where $|\rightarrow\rangle^{\otimes N}$ denotes all $N$ spins pointing along the $+$x direction in spin space, and so on. The ground state wavefunction at a finite $h<J$ includes a finite number of domain wall excitations that involve pairs of spin-flip excitations on the GHZ states given above such that they do not change the topological number $Z$. These are the lattice fermions $(c_{i},c^{\dagger}_{i})$ we studied via the JW transformation in section \ref{sec:lt}, and which obey boundary conditions given by~\cite{mcgreevylectures}
\begin{equation}
c_{N+1} = (-1)^{\Omega+1}c_{1} = -Z c_{1}~.
\end{equation}
Note that these excitations are very different in nature from the bosonic spin-wave excitations above a broken symmetry ferromagnetic ground state with all spins aligned.
\par
In this way, we see that the robustness of the topologically non-trivial state for $h<J$ to local quantum fluctuations (arising from the transverse field $h$) is lost only when a change in boundary conditions is felt by the ground state across the Lifshitz transition at $J=h$. These two ground states are, however, degenerate in the thermodynamic limit for $J>h$ (as also observed in subsection \ref{subsec:topprop}). There exists no term in the Hamiltonian, however, to take the system from one ground state to the other; instead, an external push is needed for this. Thus, 
for $J>h$, the system is naturally in the PBC ground state, even though the APBC ground state is degenerate with it. Both these ground states are separated from all others by a finite spectral gap (as obtained from the BdG quasiparticle dispersion). These ground states can also be distinguished in terms of “edge states” as follows. The PBC and APBC ground states can be constructed in terms of the bonding and antibonding combinations of the two completely aligned ferromagnetic states in the $z$ direction in spin space. These are Greenberger-Horne-Zeilinger (GHZ) states~\cite{zengwen}, where the anti-bonding state possesses unpaired Majorana end states (which together make up the one extra fermion corresponding to the $\pi$-mode occupancy), while the bonding state does not~\cite{mcgreevylectures}. Thus, the occupation of the fermionised $\pi$-mode of the energy-momentum dispersion leads to the creation of an extra fermion which can be split into two (unpaired) Majorana fermions at the ends of the system in real space. This is precisely the same in one of the two phases of the Su-Schrieffer-Heeger (SSH) model, which, as mentioned in section \ref{sec:lt}, is not surprising given that both the fermionised TFIM/pWSC and SSH models can be mapped onto the same effective two-band lattice Hamiltonian (and thence its continuum Dirac equivalent). 
\par
Further, it is shown in Ref.(\cite{zengwen}) that the two degenerate GHZ states, $|\pm\rangle$ (shown earlier for $J>h$ in the thermodynamic limit $N\to\infty$), correspond to symmetry protected topological (SPT) order with short-ranged entanglement in real-space. As the two degenerate GHZ states are eigenstates of the $Z_{2}$ symmetry operator $Z$, it becomes clear that the short-ranged entanglement properties of these GHZ states is related to this non-local symmetry of the TFIM model~\cite{zengwen}. Having shown that $Z$ is a topological invariant for the anisotropic 1D XY model for anisotropy $0<\delta\leq 1$, the above conclusions extend to the $U(1)$ SPT states of those models as well. In keeping with their $U(1)$ nature, such states are actually Bloch states obtained from an angular variable $0\leq\phi\leq 2\pi$ (which lives on the XY equatorial plane of the spin-space Bloch sphere) and labelled by a (winding) quantum no. $n$
\begin{eqnarray}
|\psi\rangle_{n} &=& \frac{1}{{\cal N}}\int_{0}^{2\pi}d\phi~e^{in\phi}~|\phi\rangle~,~n\in \textrm{integer}\\
&=& \frac{1}{{\cal N}}\int_{0}^{\pi -}\hspace*{-0.2cm}d\phi\left[ |\phi\rangle + |\pi+\phi\rangle\right ]~e^{i2m\phi}~,~n\in\textrm{even},\nonumber\\
&=& \frac{1}{{\cal N}}\int_{0}^{\pi -}\hspace*{-0.2cm}d\phi\left[ |\phi\rangle - |\pi+\phi\rangle\right ]~e^{i(2m+1)\phi}~,~n\in\textrm{odd},\nonumber
\label{u1sptwavefunc}
\end{eqnarray}
where the direct-product state $|\phi\rangle \equiv |\nearrow_{\phi}\rangle^{\otimes N}$ corresponds to all $N$ spins pointing along a direction ($\nearrow$) in the XY plane with angle $\phi$ with respect to the y-axis, and ${\cal N}$ is a normalisation constant. Thus, the $Z$ topological invariant partitions the Bloch wavefunctions into two sets with winding no. $n=$ odd and $n=$ even respectively. The ground state in the topologically non-trivial phase is given by $|\psi\rangle_{n=1}$, i.e. with a non-trivial winding no. $n=1$, with the GHZ ground state $|-\rangle$ for the 1D TFIM obtained by fixing $\phi=\pi/2$ (the $+$ x-axis). 

The authors of Ref.(\cite{zengwen}) also show that one can always convert such short-ranged entangled states into product states via the application of local, invertible transformations that are not necessarily unitary (dubbed gSL transformations in Ref.(\cite{zengwen})). Specifically, it can be shown that the gSL transformations for the TFIM which transform a GHZ state for $N$ spins into a product state involves the breaking of its $Z_{2}$ symmetry:
\begin{eqnarray}
W_{N} &=& \Pi_{j=1}^{N} \hat{O}_{j}~,~\mathrm{where}\\
\hat{O}_{j} &=&  {\begin{pmatrix}
0 & a \\
1 & 0 \end{pmatrix}}_{j} 
\end{eqnarray}
acts on the $j$-th spin, with $0<a<1$ (such that the local transformation is invertible). Clearly, $\hat{O}_{j}^{\dagger}\hat{O}_{j} < \mathrm{I}$, where $\mathrm{I}$ is the $2\times 2$ unit matrix. By acting with this gSL operation on the GHZ states $|\pm\rangle$, we obtain a product state as $N\to\infty$
\begin{eqnarray}
W_{N} |\pm\rangle &=& \frac{1}{\sqrt{2}}(|\rightarrow\rangle^{\otimes N} \pm a^{N}|\leftarrow\rangle^{\otimes N} )\nonumber\\ 
&&\lim_{N\to\infty} \frac{1}{\sqrt{2}}|\rightarrow\rangle^{\otimes N}~.
\end{eqnarray}
Similarly, the action of the gSL transformation 
\begin{eqnarray}
\tilde{W}_{N} &=& \Pi_{j=1}^{N} \hat{P}_{j}~,~\mathrm{where}\\
\hat{P}_{j} &=&  {\begin{pmatrix}
0 & 1 \\
b & 0 \end{pmatrix}}_{j} 
\end{eqnarray}
acts on the $j$-th spin, with $0<b<1$ such that 
\begin{eqnarray}
\tilde{W}_{N} |\pm\rangle &=& \frac{1}{\sqrt{2}}(b^{N} |\rightarrow\rangle^{\otimes N} \pm |\leftarrow\rangle^{\otimes N} )\nonumber\\ 
&&\lim_{N\to\infty} \pm\frac{1}{\sqrt{2}}|\leftarrow\rangle^{\otimes N}~.
\end{eqnarray} 
In this way, we can see that the entanglement lowering action of a gSL transformation is equivalent to putting in a field that breaks this symmetry. Further, this can only be achieved by overcoming the spectral gap associated with the BdG quasiparticle spectrum of the fermionised TFIM/pWSC and, as discussed earlier in section \ref{sec:lt}, replacing it with a gap that does not possess the topological invariant Z. This is precisely the role played by the $\tilde{h}_{x}$ field in the $\pi$-mode Hamiltonian (which suppresses the topological degrees of freedom in the system), and is the pathway to the Ginzburg-Landau-Wilson paradigm. Finally, as the critical point of the TFIM is $SU(2)$ symmetric, the ground state must involve the superposition of GHZ states defined in terms of every possible quantisation axes on the Bloch sphere. This is a $SU(2)$-symmetric SPT state. 
\par
While the entanglement content of the ground state of the 1D TFIM and its change across the phase transition has been probed using various measures of entanglement (see Ref.(\cite{amicovedral-2008}) and references therein), we will instead focus on identifying the non-local entanglement that arises simply from the topological features of the two phases and the Lifshitz critical point studied here. For this, the approach taken in Ref.(\cite{ryuhatsugai-2006}) will prove useful; there, a relation between the lower-bound of the many-body entanglement entropy ($S_{0}$) and Berry phase ($\gamma$) computed with respect to the ordered phase ground state of a two-band fermionic model was obtained and then applied to the SSH model. Indeed, this lower bound was found to arise from the non-local contribution of the edge states that arise in the ordered ground state. As observed in section \ref{sec:lt}, we have already identified the universality between the SSH and 1D TFIM systems via the two-band fermionic model. Thus, we employ the relation~\cite{ryuhatsugai-2006}
\begin{eqnarray}
\frac{S_{0}}{2} &=& -\frac{\gamma}{2\pi}\ln(\frac{\gamma}{2\pi}) - \frac{(2\pi - \gamma)}{2\pi}\ln(\frac{2\pi - \gamma}{2\pi})~,
\label{entangentropy}
\end{eqnarray}
where $\gamma$ is given by equn.(\ref{BerryThoulessChargePump}). Then, $S_{0}=2\ln 2$ for $\gamma (\Delta<0)=\pi$ (the ordered phase of the 1D TFIM) and $S_{0}=0$ for $\gamma (\Delta >0)=0$ (the disordered state). Importantly, the same results can also be found from a dual Berry phase $\gamma_{Dual}$ (which is related to the dual $Z_{2}$ topological order parameter defined with respect the disordered ground state)~\cite{ryuhatsugai-2006}:
\begin{equation}
\gamma_{Dual} (\Delta) = -\pi - \gamma (\Delta)~,
\end{equation}
such that $\gamma_{Dual} (\Delta>0)=-\pi$ and $\gamma_{Dual} (\Delta<0)=-2\pi$. This allows us to define the Chern number topological index coefficient $N_{c}$ of the WZNW term in the action for the $\pi$-mode at criticality (equn.(\ref{wznwaction})) as
\begin{equation}
4\pi S \equiv 2\pi N_{c} = \gamma (\Delta <0) - \gamma_{Dual} (\Delta >0) = 2\pi~
\label{volovikrelation}
\end{equation}
i.e., $N_{c} \equiv 2S =1$ for spin $S=1/2$. This is yet another display of the fact that the emergent $SU(2)$-symmetric critical point is described by a non-local topological order parameter different from the non-local (and duality-related) Berry phase order parameters which describe the two phases on either side~\citep{arxiv:1407.6539}. Indeed, from subsection (\ref{CUTRGsection}), we know that the Chern no. ${\cal C}$ computed earlier for the Thouless charge pump, ${\cal C}\equiv 2S=N_{c}$~. Thus, equn.(\ref{volovikrelation}) is, much like equn.(\ref{ChernAtiyahSinger}), a statement about the holographic relationship between the Lifshitz transition and its associated ordered and disordered phases.  
\par
We are now in a position to compute the non-local entanglement content of the state at criticality. This can be seen simply from a consideration of the occupancy of the $\pi$-mode. When viewed from a particle-hole viewpoint, the $\pi$-mode is occupied by a hole for $\Delta >0$, while it is occupied by a particle for $\Delta <0$. A time-dependent unitary transformation that changes $\Delta$ (see equn.(\ref{tdependentPhi})) commutes with the topological quantity Z, i.e., it does not affect the occupancy of the $\pi$-mode as long as either $\Delta>0$ or $\Delta<0$. As these correspond to pure states in the particle-hole basis, carrying out a partial trace of all states other than the $\pi$-mode will yield a reduced density-matrix, $\hat{\rho}$, that reflects the occupancy of the $\pi$-mode. For $\Delta\neq 0$, this gives (in the particle-hole basis)
\begin{equation}
\hat{\rho}~(\Delta <0) = {\begin{pmatrix}
1 & 0 \\
0 & 0 \end{pmatrix}}~~,~~ \hat{\rho}~(\Delta >0) = {\begin{pmatrix}
0 & 0 \\
0 & 1 \end{pmatrix}}~.
\end{equation}
Hence, $\hat{\rho}~(\Delta \neq 0)$ yields no additional entanglement entropy over and above the lower-bound $S_{0}$ computed above in equn.(\ref{entangentropy}). However, for the particle-hole symmetric critical point $\Delta=0$, $Z\equiv sgn(\Delta)$ is ill-defined and the $\pi$-mode occupancy has equal contributions from particle and hole, i.e., the many-body state possesses equal-amplitude linear combinations of particle and hole for the $\pi$-mode. Clearly, this will lead to a reduced density-matrix for the $\pi$-mode with equal weightage for particle and hole sectors (i.e., a mixed state) 
\begin{eqnarray}
\hat{\rho}_{c} &=& {\begin{pmatrix}
\frac{N_{c}}{2} & 0 \\
0 & \frac{N_{c}}{2} \end{pmatrix}} = {\begin{pmatrix}
\frac{1}{2} & 0 \\
0 & \frac{1}{2} \end{pmatrix}}~~~\textrm{for}~N_{c}=1~,
\end{eqnarray}
yielding an entanglement entropy at criticality
\begin{equation}
S_{c} = -2\frac{N_{c}}{2}\ln(\frac{N_{c}}{2})=\ln 2~~~\textrm{for}~N_{c}=1~. 
\end{equation}
Thus, we find a sudden jump in the entanglement entropy lower-bound $\Delta S_{0}$ across the transition, and the lower bound at criticality $(S_{c}(\Delta=0))$ being the average of the lower bounds on either side
\begin{eqnarray}
\Delta S_{0}\equiv S_{0}(\Delta\to 0-) &-& S_{0}(\Delta\to 0+)=2\ln 2~,\\
S_{c} = \frac{1}{2}(S_{0}(\Delta\to 0-) &+& S_{0}(\Delta\to 0+))=\ln 2~.
\label{entholo}
\end{eqnarray} 
\par
The state at criticality is maximally mixed, as seen from the purity ($\textrm{Tr}(\hat{\rho}_{c}^{2})$)
\begin{eqnarray}
\textrm{Tr}(\hat{\rho}_{c}^{2}) &=& 2\frac{N_{c}^{2}}{4} = \frac{1}{2}\equiv \frac{1}{d}~~~(\textrm{for}~N_{c}=1)~,
\end{eqnarray}
where $d=2$ corresponds to the dimensionality of the $\pi$-mode Hilbert space.
This leads immediately to an additional measure of the entanglement at criticality, the concurrence $C_{Crit.}$~\cite{wootters-2001} 
\begin{equation}
C_{Crit.}= \sqrt{2\left[1-\textrm{Tr}(\hat{\rho}_{c}^{2})\right]} = 1~.
\end{equation}
The gapless spectrum about the singular Fermi surface at criticality is linked to the emergent particle-hole symmetry/$SU(2)$ symmetry of $\hat{\rho}_{c}$, i.e., $N_{c}$ is related to the central charge $c$ of the associated conformal field theory (CFT) of the 1D TFIM~\cite{fradkin-book,swingle-2010} 
\begin{equation}
c = \frac{1}{2} = \frac{N_{c}}{2} \equiv S~.
\end{equation}
Thus, following Ref.(\cite{calabresecardy-2004}), $N_{c}$ also appears in the entanglement entropy generated by a real-space partitioning of the system of non-interacting spinless fermions in 1D into sections of length $l$ and $L-l$ in the thermodynamic limit (i.e. $L\to\infty$)
\begin{equation}
S_{l} = \frac{c}{3}\ln(\frac{l}{a})=\frac{N_{c}}{6}\ln(\frac{l}{a})~,
\end{equation}
where $a$ is a microscopic cut-off lengthscale (e.g., the lattice spacing). This relation indicates that $S_{l} (N_{c}=1) = (1/6)\ln(l/a)$~\cite{swingle-2010} arises from the contribution of the $0$-dimensional singular Fermi surface (the point-singularity of the Dirac spectrum) in momentum space.
\par  
Importantly, the singular nature associated with a vanishing Hamiltonian for the $\pi$-mode at $\Delta=0$ (criticality) manifests itself in the fact that unitary operations like equn.(\ref{tdependentPhi}) that were valid for $\Delta\neq 0$ are unimportant at $\Delta=0$. Instead, the reduced density-matrix $\hat{\rho}_{c}$ for the $\pi$-mode is invariant under rotations given by $S=e^{i\frac{\theta}{2}\vec{\sigma}\cdot\hat{n}}$~, where the components of $\vec{\sigma}$ are related to $(c_{\pi}, c^{\dagger}_{\pi})$ as shown earlier in section \ref{sec:2d-classical-1d-QI} and $\hat{n}$ is the unit normal on the Bloch sphere. This reflects the emergent $SU(2)$ symmetry at criticality. Thus, any field that breaks this symmetry will immediately lead to a non-degenerate product state being chosen, unless it is protected by a symmetry (i.e., a SPT). For $h>J$, the TFIM ground state is a unique product state $|\uparrow\rangle^{\otimes N}$. On the other hand, in the presence of a longitudinal field $h_{x}$, simple product states $|\rightarrow\rangle^{\otimes N}$ and $|\leftarrow\rangle^{\otimes N}$ are chosen for $h_{x}>0$ and $h_{x}<0$ respectively. 
\par
The duality transformation for the 1D TFIM has a consequence which extends to higher dimensions: the theory written in terms of the disorder operators corresponds to the Ising lattice gauge field theory~\cite{mcgreevylectures}. Here, the quantum paramagnet ($h_{z} > h_{z}^{*}$) corresponds to the the non-local Wilson line operator (equivalent to the $Z_{2}$ symmetry operator $Z =-1$ of the TFIM) leading to antiperiodic boundary conditions on the disorder operators, and the quantum ferromagnet ($h_{z} < h_{z}^{*}$) to periodic boundary conditions on the disorder operators ($Z=1$). From our earlier discussions, we see that the non-local Wilson line ``order parameter" of the Ising lattice gauge field theory in 1D is also related to the occupancy of the $\pi$-mode ($sgn(\Delta) \equiv Z$) of the 2 band fermionised TFIM/pWSC theory written in terms of domain wall fermions/spinons. It will be interesting to see whether the non-local Wilson loop operator for the TFIM in higher dimensions can be similarly related to the physics of the $\pi$-mode of the domain wall/spinon dispersion. Further, does the emergent $SU(2)$ symmetry of the 1D TFIM at criticality appear for higher dimensional TFIM theories as well? We will explore these questions in section \ref{sec:univ}.

\section{\label{sec:rg-2d-1d-ising} Symmetry breaking and the RG Phase diagram}

Consider a generalised 1D TFIM Hamiltonian:
\begin{eqnarray}
H/J = -\sum_{i=1}^{N-1}\sigma_{i}^{x}\sigma_{i+1}^{x} + \frac{h_{z}}{J}\sum_{i=1}^{N}\sigma_{i}^{z} + \frac{h_{x}}{J}\sum_{i=1}^{N}\sigma_{i}^{x}
\label{eq.ham-J-hz-hx}
\end{eqnarray}
While there exists an exact solution for the case of $h_{z}=J$ and $h_{x}\neq 0$~\cite{zamolodchikov}, with an $E_{8}$ symmetry of the bound states in the spectrum, we will develop the $T=0$ phase diagram from a scaling or renormalisation group (RG) analysis. Note that for $h_{x}>>h_{z}=J$, we have unique symmetry-broken product states which are field-aligned. In order to understand the Lifshitz transition in the TFIM better, we now carry out a renormalisation group (RG) analysis as follows. It is important to note right at the outset that our RG differs from the Abelian bosonisation route taken in Refs.(\cite{banderitzykson,zuberitzykson}). The authors of those works employed the duality which maps two disconnected 1D TFIM chains onto the 1D XY chain~\cite{fradkin-book}. The latter corresponds to the free Luttinger-Thirring model~\cite{fradkin-book} (i.e., non-interacting massive Dirac fermions in 1D), which in turn can be bosonised to give the sine-Gordon model at the $\beta^{2}=4\pi$ point. This yields the bulk correlation functions via bosonisation~\cite{banderitzykson,zuberitzykson}. We will, instead, focus on the effects of interactions between the massless Dirac fermions emergent at criticality.
\par
We begin by noting that at criticality ($J=h$) and $h_{x}=0$, the effective low-energy theory for the TFIM is equivalent to that of Kramers doublet of 1D massless spinless Dirac fermions located near $k=\pm\pi$. The emergent $SU(2)$ symmetry of this theory (equivalent to the emergent time-reversal symmetry of the massless Dirac fermion doublet) is critical to its further analysis. Following the classic works of Witten~\cite{witten}, as well as Knizhnik and 
Zamolodchikov~\cite{knizhnik}, we identify the universal low-energy theory for such a Kramers doublet of interacting 1D massless Dirac system of spinless fermions
as the $SU(2)_{k=1}$ (i.e., the level-1) WZNW theory of 1D Tomonaga-Luttinger liquid of spinless fermions/the Heisenberg spin-1/2 chain with nearest neighbour antiferromagnetic interactions. This identification is crucial in keeping track of the fact that all backscattering processes that cross the Brillouin zone (i.e., have a wavevector $\Delta k\sim 2k_{F}$) must enclose a Dirac point. The latter acts as a singularity in momentum space and, in keeping with its emergent $SU(2)$ symmetry, imparts a $\pi$ Berry phase to all such backscattering processes. Witten's non-Abelian bosonisation method, which manifestly respects $SU(2)$ symmetry, is therefore ideal in dealing with the problem at hand.
\par
Thus, following Witten, the action for the $SU(2)_{k}$ (i.e., the level-$k$) WZNW theory is given by
\begin{equation}
S = \int d\tau dx\frac{1}{2g}\mathrm{tr}[\partial_{\mu}U^{\dagger}\partial_{\mu}U] + k\Gamma [U]~.
\end{equation}
where the field $U$ is an element of the $SU(2)$ group defining a map from $S^{2}$ 
to $S^{3}$. The coupling $g$ of the WZNW theory accommodates $SU(2)$ symmetric interactions 
between the underlying Dirac fermions which can, in principle, lead to a mass gap for this WNZW theory. That such a mass gap is not generated was, indeed, the startling conclusion reached by Witten. He found that this was due to the presence of the non-trivial topological Wess-Zumino-Novikov-Witten (WZNW) term of this theory given by:
\begin{equation}
\Gamma [U]= \frac{i}{12\pi}\epsilon_{\mu\nu\rho}
\mathrm{tr}[(U^{\dagger}\partial_{\mu}U)(U^{\dagger}\partial_{\nu}U)(U^{\dagger}\partial_{\rho}U)]~.
\end{equation}  
This topological WZNW term $\Gamma$ is defined in by the area traced out by the field 
$U$ which encloses the volume $S^{3}$~\cite{fradkin-book}. 
\par
Witten showed that the topological coupling $k$ affects the RG flow of the coupling $g$ in 
this WZNW theory~\cite{witten}
\begin{equation}
\frac{dg}{dl} = [1-(\frac{k~g}{4\pi})^{2}]~(\frac{g}{4\pi})~.
\end{equation}
This RG equation shows the existence of a non-trivial stable fixed point at $g^{*}=4\pi/k$. 
Put together with the fact that at this value of $g^{*}$, the theory can be written using the 
non-Abelian bosonisation formalism in terms of free bosons which satisfy a $SU(2)_{k}$ 
Kac-Moody current algebra, Witten conjectured that the WZNW theory must have an exact 
fixed point at $g^{*}$. Using conformal field theoretic methods, this was shown to be correct 
by Knizhnik and Zamolodchikov~\cite{knizhnik}. In this way, we identify the $SU(2)_{k=1}$ WZNW 
theory with coupling $g^{*}=4\pi$ as the correct universal low-energy theory appropriate for 1D spinless Dirac fermions with zero mass.
\par
Remarkably, this identification links the fermionic theory at the critical point of the 1D TFIM with that for the critical Heisenberg spin-1/2 chain under RG. As mentioned earlier, this is the result of the underlying Lorentz invariance and $SU(2)$ symmetry in both theories, and reflected in the fact that the critical exponents of bulk spin-spin correlation functions in both theories is determined completely by these properties~\cite{zuberitzykson,giamarchi}. Further evidence for this is provided by asking for the effects of perturbations to the Witten fixed point theory. As shown in Ref.[\cite{fradkin-book}], the $SU(2)_{k=1}$ WZNW theory is stable against 
perturbations involving the backscattering coupling between chiral spin currents, 
$g_{1}~\vec{J}_{R}\cdot\vec{J}_{L}$. Here, $\vec{J}_{R/L}$ correspond to the right- and 
left-moving chiral spin currents respectively of the 1+1D theory: 
$J^{a}_{R}(x)=\frac{1}{2}\psi^{\dagger}_{R,\sigma}(x)\tau^{a}_{\sigma,\sigma^{'}}\psi_{R,\sigma^{'}}(x)$~,~
$J^{a}_{L}(x)=\frac{1}{2}\psi^{\dagger}_{L,\sigma}(x)\tau^{a}_{\sigma,\sigma^{'}}\psi_{L,\sigma^{'}}(x)$~, 
where $a=(1,2,3)$, $\sigma=(1,2)$ corresponds to the two Dirac flavours with wavevectors near $k=\pm\pi$ and $\tau^{a}$ are the three Pauli matrices. The RG equation for $g_{1}$ is
\begin{equation}
\frac{dg_{1}}{dl} = -\frac{2}{\pi}g_{1}^{2}~.
\end{equation}
Thus, we can see that the case of a symmetry-preserving perturbation corresponds to 
the coupling $g_{1}>0$: here, $g_{1}$ is marginally irrelevant and can neither break the $SU(2)$ 
symmetry dynamically, nor open a gap in the spectrum. In this way, we find that the stable Witten fixed point $g^{*}=4\pi$ as the 1D Heisenberg spin-1/2 chain theory and the critical fixed point with $g=0$ as the unstable 1D TFIM model at criticality ($J=h$) possessing emergent $SU(2)$ symmetry. The entire 
RG flow from $g=0$ to $g=g^{*}$ is $SU(2)$ symmetric. 
\par
We now assess the dynamical role played in the scaling theory by external magnetic fields in the TFIM along the transverse and longitudinal directions $(h_{z},h_{x})$. 
Given the $SU(2)$-symmetric nature of the Witten fixed point ($g=g^{*}$) and TFIM ($g=0$) theories, and that the $(h_{z},h_{x})$ perturbations break this $SU(2)$ symmetry explicitly, the analysis for 
both can be carried out in precisely the same way and similar conclusions reached. 
Thus, we analyse below only the longitudinal field $h_{x}$. Following Affleck~\cite{affleck}, 
it is convenient to consider here the role played by topological excitations through the sine-Gordon 
version of the WZNW theory reached via Abelian bosonisation (this is equivalent to the 1+1D $\mathrm{O}(3)$ NLSM Lagrangian):
\begin{equation}
L = \frac{1}{2}(\partial_{\mu}\phi)^{2} + \sum_{n, q_{n}}\gamma_{n}\exp^{i(n\sqrt{g}\phi + 2\pi n q_{n}\Theta)} - \frac{m_{x}}{\pi}\partial_{x}\phi~.
\end{equation}
Here, $\phi$ is the scalar field encoding the Ne$\mathrm{\grave{e}}$l order, $g$ is the phase 
stiffness parameter/NLSM coupling, $n$ is the vorticity of the topological excitations in the field $\phi$, $q_{n}$ 
is the charge of the topological excitation and $\gamma_{n}$ the fugacity for an instanton excitation with vorticity $n$, and and $m_{x}$ is 
the magnetisation conjugate to the transverse field $h_{x}$. The topological angle is $\Theta = S - m_{x}$~\cite{tanakatotsukahu}, where $S=1/2$ corresponds to the case of gapless 1D Dirac electrons. We note that we have earlier identified a similar relation between a topological quantity ($\Omega$), $S$ and a magnetisation in equn.(\ref{oyacrit}). There too, it signalled a Lifshitz transition. The difference between these two relations stems from the fact that the magnetisation in equn.(\ref{oyacrit}) changed by an integer value, making $\Omega$ integer-valued as well. On the other hand, the magnetisation $m_{x}$ is continuous in nature, and thus so is $\Theta$. This is the principal difference between a Berry phase-like quantity ($\Theta$) and a integer-valued topological invariant ($\Omega$): quantised values of the former correspond to the latter. 
\par
The first term in the action corresponds to the cost of generating collective excitations, 
the second and third the cost of topological excitations and the fourth the effect of an explicit 
symmetry-breaking $h_{x}$-field (through the magnetisation $m_{x}$).
The bare value of the instanton fugacity can be computed using standard instanton techniques~\cite{rajaraman}, 
and $\gamma_{1}\sim\exp(-S_{0}/\hbar)$, where $S_{0}$ is the classical Euclidean action for the 
instanton of the sine-Gordon problem. As shown by 
Affleck~\cite{affleck}, this simplest instanton excitation has $n=\pm 1$ and each value of $n$ 
has two charges $q_{n}=\pm 1/2$ (which corresponds to the two values of the 
charge $Q$ discussed earlier in section \ref{sec:lt} for the electronic problem, see equn.(\ref{fraccharge})). In the absence of an external $h_{x}$-field, $m_{x}=0$, and with $S=1/2$, 
we have $\Theta=1/2$. Then, in this case, we have
\begin{eqnarray}
&&\sum_{n=\pm 1, q_{n}=\pm 1/2}\gamma_{1}\exp^{i(n\sqrt{g}\phi + 2\pi n q_{n}\Theta)} \nonumber\\
&=&\gamma_{1}(\exp^{i\pi/2}+\exp^{-i\pi/2})(\exp^{i\sqrt{g}\phi} +\exp^{-i\sqrt{g}\phi})\nonumber\\
&=&4\gamma_{1}\cos(\pi/2)\cos(\sqrt{g}\phi)~=~0~.
\end{eqnarray}
In this way, we can see that while these instanton excitations are RG relevant (the coupling $\gamma_{1}$ 
has a scaling dimension $g<1/2$), it is suppressed via a destructive interference mechanism for $2\pi\Theta=\pi$ 
and thus unable to open a gap in the spectrum~\cite{affleck}. 
\par
From the identification of the topological angle $\Theta=S-m_{x}$~\cite{tanakatotsukahu}, we can  perform a RG analysis of the sine-Gordon model for $m_{x}\neq 0$. First, we can see immediately that the 
destructive interference mechanism no longer suppresses the instanton tunnel coupling 
$\gamma_{1}$ for $m_{x}\neq 0$:
\begin{eqnarray}
&&\sum_{n=\pm 1, q_{n}=\pm 1/2}\gamma_{1}\exp^{i(n\sqrt{g}\phi + 2\pi n q_{n}\Theta)} \nonumber\\
&=&\gamma_{1}(\exp^{i\pi(1/2-m_{x})}+\mathrm{h.c.})(\exp^{i\sqrt{g}\phi} +\mathrm{h.c.})\nonumber\\
&=&4\gamma_{1}\cos(\pi(1/2-m_{x}))\cos(\sqrt{g}\phi)~.
\end{eqnarray}
The RG equation for $g$, $\gamma_{1}$ and the magnetisation $m_{x}$ is given as~\cite{horovitz,giamarchi}
\begin{eqnarray}
\frac{dg}{dl} &=& -\gamma_{1}^{2} J_{0}(m\alpha)~~,~~\frac{d\gamma_{1}}{dl} = (2-g)\gamma_{1}\nonumber\\
\frac{dm_{x}}{dl} &=& m_{x} - \frac{\gamma_{1}^{2}}{2\pi\alpha} J_{1}(m_{x}\alpha)~,
\end{eqnarray}
where $J_{0}$ and $J_{1}$ are Bessel functions that arise from the use of sharp cut-off functions while implementing 
the RG transformations~\cite{giamarchi}. The first two RG equations are of the Berezinskii-Kosterlitz-Thouless (BKT) type~\cite{BKT1,BKT2}, and relate to the deconfinement of vortex-antivortex pairs and their subsequent proliferation. Importantly, the RG equation for $m_{x}$ is symmetric under the interchange 
of $m_{x}\rightarrow -m_{x}$. These RG equations show that the presence of a non-zero $m_{x}$ leads to two gapped ground states in which the $SU(2)$ symmetry is explicitly broken and the magnetisation is saturated at $m_{x}=-1/2$ and $m_{x}=1/2$ respectively. This happens even as the deconfined $\gamma_{1}$ vortices proliferate. A large magnetisation will, in turn, lead to a suppression the $\gamma_{1}$. A non-zero longitudinal magnetisation $m_{x}$ corresponds to an imbalance between the two chiralities of fermions in the 1D massless Dirac theory, while a transverse magnetisation $m_{z}$ away from its critical value $m_{z}^{*}=2/\pi$~\cite{pfeuty-1970} corresponds to a mass gap for the 1D Dirac fermion spectrum of the TFIM for $J\neq h$. 
\par
In this way, we can put together the RG phase diagram as shown in Fig.(\ref{fig:rg-tfim-hx}) below. The phase diagram has three axes in the WZNW coupling $g$, the transverse magnetisation $m_{z}$ and the longitudinal magnetisation $m_{x}$. The $SU(2)$-symmetric tricritical TFIM theory (with a $SU(2)$ SPT ground state) lies at the origin $(m_{x}=0,m_{z}=2/\pi,g=0)$, while the $SU(2)$-symmetric Witten saddle fixed point (corresponding to an algebraic spin liquid) lies at $(m_{x}=0,m_{z}=2/\pi,g=g^{*})$ (i.e., on the coupling $g$ axis at the critical value $g=g^{*}$). All non $SU(2)$-symmetric RG flows lead away from the coupling $g$ axis and towards stable fixed points on the $m_{z}$ and $m_{x}$ axis at $(m_{x}, m_{z}, g) = (\pm 1, 0, 0)$, $(0, 0, 0)$  and $(0, 1, 0)$. 
We note in passing that the topology of this RG phase diagram is qualitatively similar to that obtained for the 2D Ising model at finite-$T$ from an RG analysis~\cite{goldenfeld-book}. An important difference, however, is that while the RG flow along the temperature/coupling axis (vertical) axis for the 2D Ising model shows the critical point to be unstable in all directions and involves a second-order transition (with critical exponents belonging to the Ising univerality class), the Witten fixed point is stable on the $SU(2)$-symmetric coupling $g$ (vertical) axis for the 1D TFIM and involves an infinite-order transition belonging to the WZNW universality class. All flows along the field axes in both cases are qualitatively similar (i.e., with respect to topology of the phase diagram) as they are always first-order transitions. In this way, we can see that the Lifshitz transition observed in the RG phase diagram of the TFIM lies on the saddle surface between two first-order transitions, and cannot be described as a Ginzburg-Landau-Wilson-type transition. Indeed, the RG phase diagram obtained is the characteristic to those obtained very generally in gauge-field theories with a $\theta$-vacuum structure~\cite{bulgadaev-2006,apenko-2008}.
\par
We have discussed in section \ref{sec:2d-classical-1d-QI} the equivalence between the criticality observed in a $T=0$ quantum Hamiltonian and the (Euclidean) finite-$T$ transfer matrix of its 2D statistical mechanical counterpart. Based on this, we are led to conjecture that our findings for the $T=0$ RG phase diagram of the 1D TFIM offer new insight on the nature of the finite-$T$ criticality of the 2D Ising model. Our results suggest that the transition at zero-field (i.e., $h_{x}=0=h_{z}$) belongs to the WZNW universality class and involves the change in a topological Chern number (i.e., a real-space non-local order parameter) across a critical point with an enhanced $SU(2)$ symmetry. On the other hand, the transitions for non-zero fields are first-order in nature as they involve the explicit breaking of the $SU(2)$ symmetry, and involve the deconfinement of vortices that change the magnetisations $m_{x}$ and $m_{z}$ away from the critical values. The nature of the vortices (and the phases their condensation results in) are different for transverse field $h_{z} > h_{z}^{*}\equiv J$ and $h_{z}<h_{z}^{*}$: the former case correspond to the disorder operators (quantum paramagnet) and the latter to the order operators (quantum ferromagnet) of the duality transformation for the TFIM~\cite{fradkin-susskind-1978}. As observed earlier, the TFIM theory in terms of the disorder operators corresponds to the Ising lattice gauge field theory in 1D~\cite{mcgreevylectures}. The topological quantum phase transition between the quantum paramagnet ($h_{z} > h_{z}^{*}$) and ferromagnet ($h_{z} < h_{z}^{*}$) corresponds to the the non-local Wilson line operator leading to a change in boundary conditions (from antiperiodic ($Z=-1$) to periodic ($Z=1$)) on the disorder operators.
\par
The emergent $SU(2)$ symmetry of the $\pi$-mode theory at criticality (essentially the physics of a single spin-1/2 with vanishing external fields) is responsible for the same emergent non-Abelian symmetry in the RG phase diagram of the full problem. Does the same emergent $SU(2)$ symmetry continue to appear at criticality for higher dimensional TFIM theories as well? Is it related to the $\pi$-mode degree of freedom? We present some answers in Section VII. Finally, it remains an open challenge to find the 2D $SU(2)$-symmetric finite-$T$ theory whose dynamics and criticality are equivalent to that of the theory of the $SU(2)$-symmetric Heisenberg spin-1/2 1D chain at the Witten fixed point. It is known that the correspondence leads to a theory with non-positive Boltzmann weights for some configurations (i.e., with a complex classical Hamiltonian). Such a complex classical (Euclidean) Hamiltonian likely contains an imaginary term which is of topological origin. A likely candidate is the WZNW term identified for the $\pi$-mode of the TFIM at criticality in section \ref{sec:2d-classical-1d-QI}. It would be interesting to see whether our approach can inspire methods that would allow for the identification and analysis of such a finite-temperature statistical mechanical model.

\begin{figure}[!hbt] 
   \centering
   \includegraphics[width=0.45\textwidth]{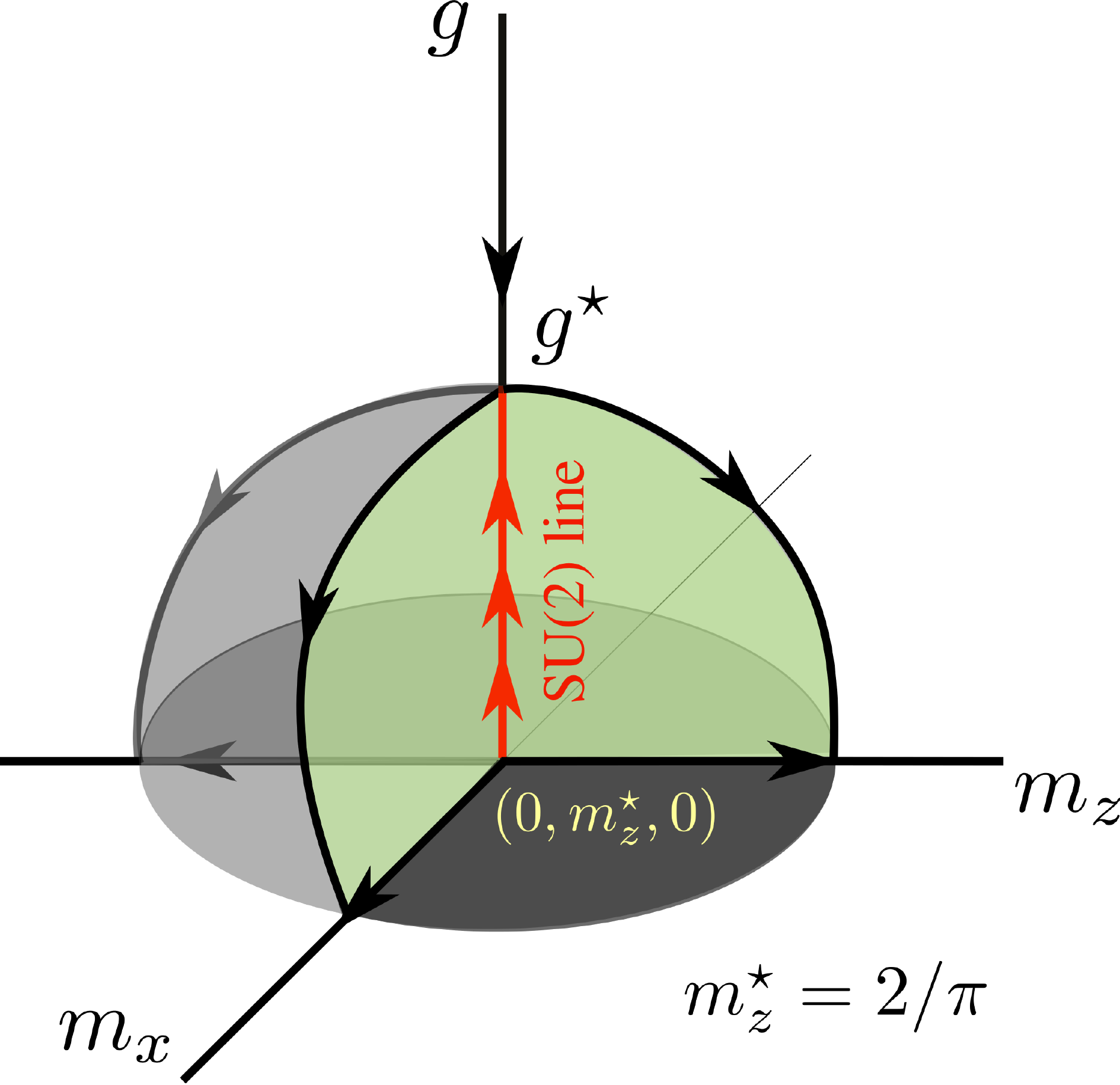} 
   \caption{(Color online.)~Renormalization Group (RG) phase diagram for the generalised 1D TFIM. The vertical axis represent the coupling $g$ for scattering of Dirac quasiparticles across the singular Fermi surface at criticality. $m_{x}$ and $m_{z}$ represent the longitudinal and transverse magnetisation respectively. The critical point of the 1D TFIM $(0,m_{z}^{*},0)$ is seen to be connected to the WZNW theory for the 1D spin-1/2 Heisenberg chain $(0,m_{z}^{*},g^{*})$ through a line of $SU(2)$ symmetric theories. The RG flows (arrows) are driven by topological excitations: hedgehogs for the $SU(2)$ vertical line and merons away from it.}
   \label{fig:rg-tfim-hx}
\end{figure}

\section{\label{sec:univ}Universality and Holography of topological transitions in Ising models}
The Ginzburg-Landau-Wilson approach to Ising models in various dimensions helped in establishing the idea of universality: continuous transitions can be grouped by critical exponents that depend only on the (spontaneously broken) symmetry of the model and the spatial dimensionality. In this section, we will investigate whether the concept of universality continues to hold for topological transitions in Ising models. We will also question whether there exist relations that link topological transitions of Ising models in different dimensions.
\subsection{The $T=0$ phase transition of the 1D Ising model}\label{HolographyinIsing}
The periodic one-dimensional ferromagnetic nearest neighbor Ising model{\cite{fradkin-susskind-1978}} (for convenience and without loss of generality) can be written as
\begin{equation}
H = - \sum_i \left[ J\sigma_i \sigma_{i+1} + h{\sigma_i}\right]
\end{equation}
where $i$ is the site index, and $\sigma_i$ is classical spin on site $i$ which can take values $\pm 1$. $J$ is Ising interaction strength and $h$ is external field acting uniformly on each site. One can write the action ($S$) and path integral ($Z$) respectively as $S = \beta H $ and $Z = {\sum_{\text{conf}}}~e^{-S}$, where $\beta$ is the inverse temperature. The partition function or the path integral can be written as trace over $N^{th}$ power ($N$ being the total number of sites in the classical spin chain) of the transfer matrix $T$. The row and columns of transfer matrix are spin configurations for the neighbouring two spins. Here one most critical concept is coming. We consider that axis of the lattice is the time axis of the quantum mechanics problem. This is where the correspondence between classical to quantum mechanics lies. Thus transfer matrix carries information from one time to next time interval, and hence this is time evolution operator for the quantum mechanics problem .

So the partition function is defined as $Z = {\text{Tr}} ~ T^N $, where the transfer matrix is 
\begin{eqnarray}
T(i,i+1) &=& \sum_{\sigma_i, \sigma_{i+1}} e^{\beta (J\sigma_i \sigma_{i+1} + h/2({\sigma_i} + \sigma_{i+1}))}
\end{eqnarray}
In matrix form 
\begin{eqnarray}
T = 
\left[ \begin{array}{cc}
e^{(K+\bar{h})} & e^{-K} \\ 
e^{-K} & e^{(K-\bar{h})}
\end{array} \right]~~,
\label{tmatrix1DIsing}
\end{eqnarray}
where $K=\beta J$ and $\bar{h}=\beta h$. The eigenvalues of the transfer matrix are given by
\begin{eqnarray}
\lambda_{1,2} = \exp^{K}\left[\cosh(\bar{h}) \pm \sqrt{\sinh^{2}(\bar{h}) + \exp^{-4K}} \right]~,
\end{eqnarray}
and the free energy density is
\begin{eqnarray}
f=\frac{F(K,\bar{h})}{N} &=& -\frac{1}{\beta}\ln(\lambda_{1})\nonumber\\
&=&\lim_{T\to 0} - \frac{1}{\beta}\ln\left[e^{K+|\bar{h}|}\right]\nonumber\\ 
&=& - (J + |h|)~.
\label{1DIsingfreeenergy}
\end{eqnarray}
Thus, while there is no non-analyticity in $f$ for any $T>0$, a non-analyticity is found in the $\lim T\to 0$. It is important to note that this non-analytic behaviour found as $T\to 0$ or $K\to\infty$ does not need the 
thermodynamic limit $N\to\infty$ to be taken~\cite{goldenfeld-book}. This non-analyticity in $f$ for $h=0$ coincides with a degeneracy of the eigenvalues of the $T$-matrix 
\begin{eqnarray}
\lim_{K\to\infty} \lambda_{1} = 2\cosh(K) \rightarrow \lambda_{2} = 2\sinh(K)~,
\end{eqnarray}
and a divergence of the correlation length
\begin{equation}
\xi = \frac{1}{\ln(\lambda_{1}/\lambda_{2})} = \frac{1}{\coth(K)} \to\infty~~\mathrm{as}~~K\to\infty~.
\end{equation}
The order parameter for the transition at $T=0$ is given by the magnetisation
\begin{equation}
m= -\frac{\partial f}{\partial h} = sgn (h)~.
\end{equation}

In order to understand the topological nature of this $T=0$ phase transition, we will compute the $T=0$ effective quantum problem onto which the thermal partition function of the 1D Ising model is mapped. Starting with the transfer matrix given above in equn.(\ref{tmatrix1DIsing}), we take the limit when the classical lattice points are closely spaced or the time evolution of the quantum system is smooth/continuous or time grid is very small. Thus $\beta J$ is large compared to $\beta h$. The classical Ising chain is on a periodic lattice, and thus we have $N$ number of motif which give $N$ number of transfer matrices to make the full partition function. Thus a single transfer matrix can be written as
\begin{eqnarray}
T &=& e^{\beta J} 
\left[ \begin{array}{cc}
e^{\beta h} & e^{-2 \beta J} \\ 
e^{-2 \beta J} & e^{\beta -h}
\end{array} \right] \nonumber
\end{eqnarray}
Dropping the constant factor $e^{\beta J}$ and expanding the $e^{\beta h}$ term
\begin{eqnarray}
T &\sim& \left[ \begin{array}{cc}
1 + \beta h & e^{-2 \beta J} \\ 
e^{-2 \beta J} & 1 - \beta h
\end{array} \right] \nonumber \\
&=& I + \beta h \sigma^z + e^{-2 \beta J}\sigma^x = I - \tau H_{\text{spin}}~,
\end{eqnarray}
where we have equated the $T$-matrix to a propagator  of an effective two-level system/quantum mechanical spin in the presence of external fields over a Euclidean time $\tau$. The Hamiltonian for this spin system is given by
\begin{eqnarray}
H_{\text{spin}} = - \left(\lambda \sigma^x + h_{z} \sigma^z \right)~,~
\end{eqnarray}
where the external fields are $h_{z} = \bar{h}/\tau\equiv \beta h/\tau, \lambda = e^{-2 K}/\tau$~. 
The eigenvalues of this Hamiltonian are $\epsilon_{0}^{\pm} = \pm\sqrt{h_{z}^{2} + \lambda^{2}}$~.
For $h\neq 0$, $K\to\infty$ (i.e., $T\to 0$), we can relate the magnetisation $m$ of the 1D Ising model to the Berry phase of this spin-1/2 system under a cyclic adiabatic excursion of the Hilbert space 
\begin{eqnarray}
\gamma_{0}  &=& \pi\left[ 1 - \frac{h_{z}}{\sqrt{h_{z}^{2} + \lambda^{2}}}\right]\nonumber\\ 
&=&\lim_{K\to\infty, h\neq 0} \pi\left[ 1 - sgn (H)\right]\nonumber\\
&=&\pi\left[1-m\right]~.
\end{eqnarray}
The Berry phase is $\gamma_{0}\to 0$ for $h_{z}>0, m=1$ and $\gamma_{0}\to 2\pi$ for $h_{z}<0, m=-1$. The Hamiltonian vanishes at the critical point given by $h_{z}=0,\lambda=0$. However, the coherent state path-integral for this spin-1/2 state can be written in terms of a Wess-Zumino-Novikov-Witten (WZNW) term~\cite{fradkin-book}. This topological term in the action of the $\pi$-mode theory of the TFIM characterises the integer coverings of the Bloch sphere arising from the non-trivial homotopy group of the non-Abelian $SU(2)$ group, $\pi_{3} (SU(2)) = Z$. As observed for the 1D TFIM, this analysis reveals once again that the partition function at the critical point of the 1D Ising model can be written in terms of phases (of topological origin), as the free energy at criticality contains an imaginary piece.
Importantly, it reflects the emergent $SU(2)$ symmetry at $h_{z}=0, \lambda=0$. 

Thus, we can see that the $T=0$ transition of the 1D Ising model is of the same kind as that observed for the $T=0$ 1D TFIM: both are characterised by an emergent $SU(2)$ symmetry observed precisely at criticality, and a change in a Berry phase across the transition. Indeed, the effective quantum problem attained via the classical-quantum correspondence for the 1D Ising model is identical to the $\pi$-mode of the fermionised TFIM/pWSC. Given that the transition of the 1D TFIM characterises that of the finite-$T$ transition in the 2D Ising model (via the classical-quantum correspondence), we can see that the topological $T=0$ transition of the 1D Ising model is precisely that of the 2D Ising model at $T=T_{C}$. This tells us that the classical-quantum correspondence for the 1D and 2D Ising models (as obtained from their respective transfer matrices) is holographic in nature. The non-trivial thermal dynamics of the 2D Ising model leading to the topological phase transition at $T_{C}$ can be studied via the quantum critical dynamics of an equivalent 1D Ising model at $T=0$. This is yet another manifestation of a \textit{bulk-boundary} (holographic) correspondence: the $T=0$ critical lower-d system can be regarded as the boundary of the finite-$T$ critical bulk higher-d system. Remarkably, this holographic relates one critical state of matter to another. This is unlike the well-known holography in topological insulators and superconductors, where an ordered state of matter (i.e., with a gapped spectrum) in the bulk shares its topological properties with a critical state of matter (i.e., with a gapless spectrum) at its boundaries.
\par
This finding has some important consequences. First, following the classic review by Kogut on spin systems and lattice gauge theories~\cite{kogut-1979}, we know that the partition function of the 2D classical Ising lattice gauge theory is equivalent to that of the 1D classical Ising chain. The gauge theory is denoted by a Hamiltonian consisting of products of Ising spins defined on the links of elementary plaquettes of the square lattice
\begin{eqnarray}
H_{gauge}&=&-J\sum_{n,\mu,\nu} \sigma_{z} (n,\mu) \sigma_{z}(n+\mu,\nu)\nonumber\\
&&\times\sigma_{z}(n+\mu+\nu,-\mu)\sigma_{z}(n+\nu,-\nu)~.
\end{eqnarray}
The equivalence stated above leads to the consequence that this Ising gauge theory cannot have an ordered state at any $T>0$.  Second, from Ref.(\cite{nussinovortiz}), we know that the partition functions of both the Kitaev toric code~\cite{kitaevtoriccode} and Wen's plaquette model~\cite{wenplaquettemodel} are equivalent to two independent copies of the 2D Ising lattice gauge theory. Finally, the free energy density of the nearest-neighbour Ising model on the infinite dimensional Bethe lattice~\cite{eggarter-1974,vonheimburgthomas-1974,mullerhartmannzittartz-1974} has the same form as that for the 1D nearest-neighbour Ising model (see equn.(\ref{1DIsingfreeenergy})). From our findings, we can now also clarify the topological nature (and emergent $SU(2)$ symmetry) of the $T=0$ transition in all these models.
\subsection{Phase transition in the infinite-range Ising model}
\def\Jt{\tilde{J}}
Ising models involving interactions between each spin and every other spin in the system can be considered as increasing the range of the Ising interaction from its usual nearest-neighbour to infinity. In order to maintain the extensivity of the free energy, it is important to scale the Ising exchange coupling ($J$) by the total no. of spins ($N$), i.e., $J\to \Jt=J/N$. Given that the coordination number of the interaction scales with $N$, these models have been usually regarded as exemplifying the exactness of the mean-field approach to phase transitions in the limit of infinite spatial dimensionality $d\to\infty$~\cite{brout-1960,kac-1959}. In keeping with the spirit of the present work, we will investigate the transition in this model in order to probe the existence of Lifshitz transitions that correspond to level-crossing events.
\par
We begin by considering a system of non-interacting Ising spins placed on the vertices of a $d$-dimensional hypercubic lattice, but which is placed in a spatially uniform, time-independent field $h$ as well as a spatially uniform field but slowly fluctuating field ${\cal X}$
\begin{equation}
H = (h + {\cal X})\sum_{i} s_{i}^{z}~,
\end{equation} 
and where the field ${\cal X}$ has a Gaussian probability distribution $P({\cal X})\propto e^{-\beta {\cal X}^{2}/2\Jt}$ with a width $\Jt$ and $\beta=1/k_{B}T$. Integrating out this slowly fluctuating Gaussian random field ${\cal X}$ from the partition function
\begin{equation}
Z = \sum_{s_{i}=\pm s}\int_{-\infty}^{\infty} d{\cal X} P({\cal X}) e^{-\beta H}
\end{equation}
leads to an effective all-neighbour interaction $\Jt$ being generated among the Ising spins~\cite{chandler-book}
\begin{eqnarray}
Z&=&\sqrt{\frac{2\pi \Jt}{\beta}} \sum_{s_{i}=\pm s} e^{\frac{\beta \Jt}{2}(\bar{S_{z}})^{2} - \beta h \bar{S_{z}}}~
\end{eqnarray}
where $\bar{S_{z}}=\sum_{i} s_{i}^{z}$ is the effective large spin governing the dynamics of the system. It is easily seen that $(\bar{S_{z}})^{2}=\sum_{i,j}s_{i}^{z}s_{j}^{z} - Ns^{2}$~, $N$ being the total no. of spins. Thus, the effective Hamiltonian we have at hand is
\begin{eqnarray}
H_{eff} &=& - \frac{\Jt}{2}(\bar{S_{z}})^{2} + h \bar{S_{z}}\nonumber\\
&=& - \frac{\Jt}{2} (\bar{S_{z}} - \frac{h}{\Jt})^{2} + \frac{h^{2}}{2\Jt}~.
\label{IsingLiebMattis}
\end{eqnarray}
This is a ferromagnetic infinte-range Ising Hamiltonian in a field. It can be easily shown that this model is equivalent to the Lipkin-Meshkov-Glick (LMG) model with an antiferromagnetic exchange~\cite{lmg-1965}
\begin{eqnarray}
H_{LMG} &=& \Jt(\bar{S_{x}}^{2} + \gamma \bar{S_{y}}^{2}) +h\bar{S_{z}} - \frac{N\Jt}{4}(1+\gamma)~,
\label{afmlmg}
\end{eqnarray}
where $\bar{S_{\alpha}}=\frac{1}{2}\sum_{l=1}^{N}\sigma_{l}^{\alpha}$, $\Jt >0$ and $\gamma$ is the spin-space anisotropy. The Hamiltonian $H_{eff}$ in equn.(\ref{IsingLiebMattis}) is reached by taking $\gamma=1$ (the isotropic XY point), upto the constant $N\Jt/2=J/2$. At this point, it is important to note that the $Z_{2}$ topological invariant defined in earlier sections, $Z=\Pi_{l=1}^{N}\sigma_{l}^{z}$, commutes with $H_{eff}$ for all values of the spin anisotropy $\gamma$, i.e., $\left[H_{eff},Z\right] =0$~. Therefore, $Z$ (and thereby the topological winding no. $\Omega$, $Z=(-1)^{\Omega}$) continues to label the ordered phase of the LMG model in the same way as observed earlier for various other Ising models. Further, changes in $Z$ and $\Omega$ will signal the Lifshitz transitions here as well.
\par
For the case of an even no. of spins, $H_{eff}$ is easily seen to be minimised by $\bar{S_{z}}=(N/2, -N/2+2m)$ for the critical field $h^{*}=m\Jt$, where $m=p/2$ and $p\in Z$, $0\leq |p| \leq N$. Using equn.(\ref{oyacrit}), we can see that the topological winding no. $\Omega = S - \bar{S_{z}}$ changes in value from $\Omega=0$ (for $h\to h^{*}-$) to $\Omega = N-2m$ (for $h\to h^{*}+$). In this way, we can see that the Lifshitz phase transition at $h=0=m$ involves a change in the ground state from a topologically trivial ground state with $\Omega=0$ for $h\to 0-$ to a topologically non-trivial ground state with $\Omega=N$ for $h\to 0+$. As $h$ is further increased towards positive values, there exists a sequence of excited state Lifshitz transitions corresponding to level-crossing events at various values of $h^{*}$ where the $\Omega=0$ state becomes degenerate with the $\Omega=N-2m$ state.
\par
Any departure from $h^{*}=0$ can be described  by an effective Hamiltonian written in terms of the degenerate two-level subspace (characterised by a spin-1/2 variable $\vec{\Sigma}$) similar to that encountered earlier in equn.(\ref{qubitHam}) : 
\begin{equation}
H_0 = -\frac{\Jt}{2}(\vec{\Sigma})^{2} + h\Sigma^{z}~.
\end{equation}
where $h$ here describes the departure from $h^{*}=0$. Interestingly, this Hamiltonian can also be mapped onto the particle on a circle model (equn.(\ref{pocHam})). Clearly, at $h=0$, the critical theory is again observed to possess an $SU(2)$ symmetry with topological consequences. Indeed, by following the analysis of subsection \ref{CUTRGsection}, we can again obtain a non-zero Chern no. by a cyclic adiabatic excursion around $h^{*}=0$. The change in the entanglement entropy across such a topological transition has already been discussed in section \ref{sec:dual}.
\par
We can now also analyse the LMG model with $2N$ spins and with ferromagnetic exchange~\cite{vidal-2004,vidal-2007} by the transformations $\Jt (=J/2N) \to -\Jt$ and $h\to -h$ (without loss of generality) in equn.(\ref{afmlmg}) and (\ref{IsingLiebMattis}): this effectively changes the sign of equn.(\ref{IsingLiebMattis}). Here, the ground state at $h=0$ is topologically non-trivial with $\bar{S_{z}}=0,~\Omega=N$. As $h$ is tuned towards increasingly positive values, there are a sequence of level-crossing Lifshitz transitions at $h^{*}=m\Jt/2$, where $m=1,3,\ldots,2N-1$ where the ground state $\Omega$ is reduced in value from $N$ ($h\to\Jt/2-)$ to $0$ ($h\to (2N-1)\Jt/2+$) as $\bar{S_{z}}$ increases in value from $0$ to $N$. Importantly, the final transition at $h^{*}=(2N-1)\Jt/2=(J/2)(1-1/2N)$ involves a Lifshitz transition from a topological state ($\Omega=1$) to a topologically trivial state ($\Omega=0$). Note that as $N\to\infty$, this last critical field is at $h^{*}=J/2$~\cite{botet-1982}. We can now also comment on the role of a non-zero positive spin-anisotropy $\gamma$. As noted above, we can rewrite the problem in the neighbourhood of a Lifshitz transition as a particle on a circle (POC) with the field $h$ taking the role of an Aharanov-Bohm (AB) flux. The anisotropy now gives an addition term proportional to $\Jt (1-\gamma) (\bar{S_{x}}^{2} - \bar{S_{y}}^{2})$; this term can be shown to become a $\cos(2\phi)$ potential in the POC, where $\phi$ is the angular position of the particle. Following Refs.(\cite{weigert-1994,rajaraman}), we can use the instanton formalism to compute the ground state energy as a function of the field $h$. Importantly, we know from Ref.(\cite{weigert-1994}) that the level crossing Lifshitz transition events (which correspond to vanishing tunnel splitting in the POC) happen at $h=m\Jt/2$, i.e., half-integer values of the AB-flux in units of the flux quantum. Remarkably, we can see that the inclusion of the spin anisotropy $\gamma$ does not affect the values of the critical field $h^{*}$.
\par
As observed in the previous subsection, the degeneracy in the low-energy subspace at criticality manifests itself here too in the form of a non-analytic behaviour of the free energy density $f(h,{\cal X})$ obtained upon integrating out the fluctuating spins via the transfer-matrix method~\cite{chandler-book}
\begin{eqnarray}
f=\frac{F({\cal X},h)}{N} &=& -\frac{1}{\beta}\ln(\lambda_{1})\nonumber\\
&=&\frac{{\cal X}^{2}}{2\Jt}-\frac{1}{\beta}\ln\left[e^{\beta |{\cal X} + h|}\right]\nonumber\\ 
&=& \frac{{\cal X}^{2}}{2\Jt} - |{\cal X} + h|~.
\end{eqnarray}
In this way, we can conclude that the nature of the transition in the infinite-range Ising models is identical to that observed for the nearest-neighbour Ising model in $d=1,2$: a level-crossing with an emergent $SU(2)$ symmetry and a topological (WZNW) order parameter at criticality. A final word on infinite-range Ising models. While such models may appear to be somewhat artificial in nature, they have been shown to appear quite ubiquitously in several models of quantum magnetism and superconductivity~\cite{vanwezel1,vanwezel2}. In these examples, Hamiltonians such as equn.(\ref{IsingLiebMattis}) are seen to govern the quantum dynamics of collective degrees of freedom describing the system as a whole and are important in assessing the susceptibility of the system to spontaneous symmetry breaking. It is also interesting to ponder whether similar transitions may be encountered in the Sherrington-Kirkpatrick model~\cite{sherkirk} for spin-glasses, where the bond-dependent exchange coupling of an infinite-range Ising model is chosen randomly from a Gaussian ensemble.

\subsection{Topological transitions in the 2D TFIM}
As shown in Ref.(\cite{elliot-1970,fradkin-susskind-1978}), the classical Ising model in $d+1$ spatial dimensions at finite-$T$ maps onto the transverse-field Ising model in $d$ spatial dimensions at $T=0$ via the transfer matrix. Having used this mapping in sections \ref{sec:2d-classical-1d-QI}-\ref{sec:rg-2d-1d-ising} to learn about the topological transition at $T=T_{C}$ in the 2D Ising model via the corresponding $T=0$ transition in the 1D TFIM, we will now do the same for the 3D Ising model through its quantum equivalent (the 2D quantum/transverse-field Ising model). As before, we will begin by considering the two-dimensional nearest-neighbour anisotropic XY model with transverse field along $z$ direction for the sake of generality
\begin{eqnarray}
H_{XY} &=& J_1 \sum_{<i,j>} \left[ \frac{(1+\delta)}{2} \sigma^x_{i,j} \sigma^x_{i+1,j} + \frac{(1-\delta)}{2}  \sigma^y_{i,j} \sigma^y_{i+1,j} \right] \nonumber \\
&& + J_2 \left[ \sum_{<i,j>}  \frac{(1+\delta)}{2}  \sigma^x_{i,j} \sigma^x_{i,j+1}  \frac{(1-\delta)}{2} \sigma^y_{i,j} \sigma^y_{i,j+1} \right] \nonumber \\
&&  + h \sum_{<i,j>} \sigma^z_{i,j} \label{eq:2dxyh}
\end{eqnarray}
where $J_1$ and $J_2$ are the real-space anisotropy along $x$ and $y$ direction respectively, $\delta$ corresponds to the spin-space anisotropy, $h$ is the transverse field and $(i,j)$ denotes sites on the 2D square lattice.
This model simplifies to various other models upon tuning the parameters $J_{1}$, $J_{2}$ and $\delta$. For instance, $\delta = 1$ and $J_1 = J_2$ corresponds to the 2D TFIM. For $J_2 = 0$, it becomes anisotropic 1D $XY$ model with transverse field, while for $J_2 = 0$ and $\delta = 1$, it is the 1D TFIM. 
\par
Does this 2D model possess a symmetry-related topological invariant as found in the equivalent 1D model (see equn.(\ref{z2topinv}))? For this, we define a two dimensional version of the $Z_{2}$ symmetry operator 
\begin{eqnarray}
Z_{2D} = \prod_{i,j} \sigma^z_{i,j} = \left( \prod_{i=1,2,\cdot L_x} \prod_{j=1,2,\cdot L_y} \sigma^z_{i,j} \right)
\end{eqnarray}
where, the $\prod_{j=1,2,\cdot L_x} \sigma^z_{i,j}$ and $\prod_{i=1,2,\cdot L_y} \sigma^z_{i,j}$  non-local operators correspond to Wilson loop operators of the 2D Ising quantum lattice gauge theory that span the system in the x and y directions respectively~\cite{fradkin-book}. Further, $Z_{2D}$ operator commutes with the Hamiltonian (eq.(\ref{eq:2dxyh})), $\left[Z_{2D}, H_{XY}\right ]=0$. For this, one can use the following identity $[A,BC] = [A,B] C + B[A,C]$, the identities for the Pauli spin matrices as $\sigma^\alpha \sigma^{\beta} = i \epsilon^{\alpha \beta \gamma} \sigma^{\gamma}$ where $(\alpha, ~ \beta, ~ \gamma)$ are $(x,~y,~z)$ and the commutation relation as $[ \sigma^\alpha ~ , ~ \sigma^{\beta}] = 2 ~ i ~ \epsilon^{\alpha \beta \gamma} \sigma^{\gamma}$, such that
\def\r{\vec{r}}
\def\del{\vec{\delta}}
\begin{eqnarray}
&& \left[ \sigma^x_{\r} \sigma^x_{\r +\del} ~ , \sigma^z_{\r} \sigma^z_{\r + \del } \right]  \nonumber \\
&& = \left[ \sigma^x_{\r} \sigma^x_{\r +\del} ~ , \sigma^z_{\r} \right] \sigma^z_{\r + \del }  + \sigma^z_{\r} \left[ \sigma^x_{\r} \sigma^x_{\r +\del} ~ , \sigma^z_{\r + \del } \right] \nonumber \\
&& = - \left[ \sigma^z_{\r} ~ , \sigma^x_{\r} \sigma^x_{\r +\del} \right] \sigma^z_{\r + \del } - \sigma^z_{\r} \left[ \sigma^z_{\r + \del } ~ , \sigma^x_{\r} \sigma^x_{\r +\del} \right] \nonumber \\
&& = - \left[ \sigma^z_{\r} ~ , \sigma^x_{\r} \right] \sigma^x_{\r +\del}  \sigma^z_{\r + \del } - \sigma^z_{\r} \sigma^x_{\r}  \left[ \sigma^z_{\r + \del } ~ , \sigma^x_{\r +\del} \right] \nonumber \\
&& = - ~ 2 ~ i ~ \sigma^y_{\r} \times (-i) \sigma^y_{\r +\del } - i ~ \sigma^y_{\r} ~ 2 ~ i ~ \sigma^y_{\r + \del } = 0~,
\end{eqnarray}
and, similarly, $\left[ \sigma^y_{\r} \sigma^y_{\r +\del} ~ , \sigma^z_{\r} \sigma^z_{\r + \del } \right] =0$~. Indeed, it appears that such a $Z_{2}$ topological invariant can be similarly defined for an anisotropic XY model in $d$-spatial dimensions, and that it will commute with the Hamiltonian. As shown explicitly for the 1D case, this has the important consequence that the ordered ground state will be a $U(1)$ SPT Bloch state $|\psi\rangle_{n=1}$ for $0<\delta<1$ (equn.(\ref{u1sptwavefunc})) and a GHZ Ising SPT state $|-\rangle$ for $\delta =1$ (equn.(\ref{ghzsptwavefunc})). Having identified $Z_{2D}$, we can now attempt at finding a topological transition involving a change in $Z_{2D}$ as the transverse-field $h$ is varied. 
\par
While the Jordan-Wigner transformation had helped obtain an exact solution of the 1D TFIM, transforming Eq.(\ref{eq:2dxyh}) into a bilinear non-interacting fermionic form in 2D is subtle due to the presence of non-trivial string operators~\cite{fradkin-1989,fradkin-book}. Indeed, the fermions obtained from this transformation of XY spins on the square lattice are {\it not} free: the highly non-local string operator generates long-range gauge interaction between the fermions. The gauge field is itself a function of the local fermionic density in real-space, and when taken together with the fermionic field operators, they help define new objects which satisfy a hard-core condition and anyonic commutation relations (i.e., anyonic inter-particle exchange statistics). It has been shown that the theory for the gauge field is topologically non-trivial, and corresponds to the Chern-Simons (C-S) gauge-field theory in 2D~\cite{fradkin-1989,lopezrojofradkin-1994}. The Chern-Simons coupling constant $\theta$ is related to the inter-particle exchange statistics angle $\delta$, $\delta = 1/2\theta$. Thus, for $\theta=(2\pi (2k+1))^{-1}$ (i.e., such that $\delta$ is an odd multiple of $\pi$), bosonic statistics are obtained. In this way, Ref.(\cite{fradkin-1989}) shows that the 2D spin-1/2 XY model can be mapped onto a problem of hard-core bosons. This mapping is an example of a gauge transformation leading to the attachment of fluxes to fermions in order to obtain bosons (as seen, for instance, in the fractional quantum Hall effect~\cite{zhanghannsonkivelson-1989}); here, the exponential gauge-field terms (i.e., the Jordan-Wigner phase factors) are the disorder operators of the underlying spin model~\cite{kadanoffceva-1971,fradkin-susskind-1978}. However, it is clear that the 2D fermions experience the effects of a local flux (i.e., the dynamical C-S field) as well as a global flux arising from boundary conditions. The latter is similar to the case in 1D studied earlier; however, the former makes the 2D TFIM intractable to an exact analytical solution. 
\par
Some analytic progress can be made by making a mean-field/saddle-point approximation where a homogeneous fermion density is assumed, such that an average value of the C-S statistical gauge-field is obtained. For the hard-core boson case, $\theta=(2\pi (2k+1))^{-1}$, the average-field approximation (AFA) corresponds to taking an average half-flux quantum of the C-S field within each plaquette of the 2D square lattice (this is also referred to as a staggered-flux phase~\cite{fradkin-book}), and has allowed for studies of some 2D spin models on square lattices~\cite{wang-feb-1991, wang-june-1991, wang-1992,lopezrojofradkin-1994, derzhko-2003} as well as models of quantum antiferromagnets on frustrated lattices~\cite{yang-1993,misguich-2001}. The consensus from these studies is that while the results of the AFA are qualitatively correct for an incompressible system with a spectral gap, a description of transitions out of such a phase (i.e., when the spectral gap closes, and we obtain a compressible state of matter) needs a careful field-theoretic analysis of the dynamical fluctuations of the C-S gauge-field around the AFA ground state.  
\par
With this in mind, we undertake the AFA for the model given in Eq.(\ref{eq:2dxyh}) and try to see if the linear gap-closing mechanism happens here as observed earlier in section \ref{sec:lt} for the 1D TFIM. Without attempting a field-theoretic analysis of the transition (as carried out for the 1D TFIM in section \ref{sec:rg-2d-1d-ising}), we direct the reader towards efforts made in similar directions (see Refs.(\cite{lopezrojofradkin-1994,arxiv:1407.6539,senthilfisher-2006})). Here, instead, we will attempt to identify the topological invariant of the 2D TFIM, $Z_{2D}$, with a topological index obtained from the time-reversal symmetric points of the equivalent fermionised model~\cite{fukane-2007,sqshen}. In this way, even as an exact solution of the partition function remains elusive via the fermionisation route, we will show that the topological nature of the transition can be captured exactly within the AFA. The 2D Jordan Wigner transformation is a mapping from spin to spinless fermions written as
\begin{eqnarray}
\sigma^+_{l,m} = e^{- i \alpha_{l,m} } \ddag_{l,m} \\
\sigma^-_{l,m} = e^{ i \alpha_{l,m} } \d_{l,m}
\end{eqnarray}
where 
\begin{eqnarray}
\alpha_{l,m} = \sum_{f \neq l} \sum_{g \neq m} \textrm{Im}\left[\ln (f-l + i (g-m))\right] ~\ddag_{f,g}\d_{f,g}~,
\end{eqnarray}
is the string operator~\cite{wang-feb-1991}. We can write the spin Hamiltonian in terns of fermionic variable as follows
\begin{eqnarray}
H &=& J_1 \sum_{l,m}\big[ ~ ( \ddag_{l,m} \d_{l+1,m} e^{-i (\alpha_{l,m} - \alpha_{l+1,m})} \nonumber \\
&& + \d_{l,m} \ddag_{l+1,m} e^{i (\alpha_{l,m} - \alpha_{l+1,m})}) ~ \nonumber \\
&& + \delta ( \ddag_{l,m} \ddag_{l+1,m} e^{-i (\alpha_{l,m} + \alpha_{l+1,m})} \nonumber \\
&& + \d_{l,m} \d_{l+1,m} e^{i (\alpha_{l,m} + \alpha_{l+1,m})})\big] + \nonumber \\
&& J_2 \sum_{l,m}\big[ ~ ( \ddag_{l,m} \d_{l,m+1} e^{-i (\alpha_{l,m} - \alpha_{l,m+1})} \nonumber \\
&& + \d_{l,m} \ddag_{l,m+1} e^{i (\alpha_{l,m} - \alpha_{l,m+1})}) ~ \nonumber \\
&& + \delta ( \ddag_{l,m} \ddag_{l,m+1} e^{-i (\alpha_{l,m} + \alpha_{l,m+1})} \nonumber \\
&& + \d_{l,m} \d_{l,m+1} e^{i (\alpha_{l,m} + \alpha_{l,m+1})})\big] + h\sum_{l,m}\ddag_{l,m}\d_{l,m}~.
\end{eqnarray}
The staggered site-dependent flux phase (corresponding to half-flux quantum in every plaquette) has the form
\begin{eqnarray}
\alpha_{l+1,m} - \alpha_{l,m} = \pi\\
\alpha_{l+1,m+1} - \alpha_{l+1,m} = 0\\
\alpha_{l,m+1} - \alpha_{l+1,m+1} = 0\\
\alpha_{l,m+1} - \alpha_{l,m} = 0~.
\end{eqnarray}
The choice of a staggered flux phase {\cite{derzhko-2003, wang-feb-1991, wang-june-1991, wang-1992}} gives the above Hamiltonian as
\begin{eqnarray}
H &=& J_1 ~ \sum_{l,m}(-1)^{(l+m)} \big[ (\ddag_{l,m} \d_{l+1,m} - \d_{l,m} \ddag_{l+1,m} ) \nonumber \\
&& + \delta ( \ddag_{l,m} \ddag_{l+1,m} - \d_{l,m} \d_{l+1,m} ) \big] + \nonumber \\
&& J_2 \sum_{l,m}\big[ ~ ( \ddag_{l,m} \d_{l,m+1} - \d_{l,m} \ddag_{l,m+1} ) ~ \nonumber \\
&& + \delta ( \ddag_{l,m} \ddag_{l,m+1} - \d_{l,m} \d_{l,m+1} )\big] + h\sum_{l,m}\ddag_{l,m}\d_{l,m}~.
\end{eqnarray}
Upon Fourier transforming the fermionic operators 
\begin{eqnarray}
d_{l,m} = \frac{1}{\sqrt{L_x ~ L_y}} \sum_{k_x , k_y} e^{i (k_x l + k_y m)} d_{k_x , k_y}
\end{eqnarray}
and by working in the particle-hole (Nambu) basis, we can write the bilinear Hamiltonian for the 2D TFIM problem. Finally, we can once again diagonalize the Fourier-transformed Hamiltonian via a Bogoliubov transformation to obtain four distinct quasiparticle dispersion relations
\begin{eqnarray}
\epsilon_1(\vec{k}) &=& \big[ (J_2 \cos {k_y} + \delta J_1 \cos{k_x} + h)^2 \nonumber \\
&+& (J_1 \sin {k_x} + \delta J_2 \sin{k_y})^2 \big]^{1/2}  {\label{eq.e1}} \\
\epsilon_2(\vec{k}) &=& \big[ (J_2 \cos {k_y} - \delta J_1 \cos{k_x} + h)^2 \nonumber \\
&+& (J_1 \sin {k_x} - \delta J_2 \sin{k_y})^2 \big]^{1/2} {\label{eq.e2}} \\
\epsilon_3(\vec{k}) &=&  -\epsilon_1(\vec{k})~~,~~
\epsilon_4(\vec{k}) =  -\epsilon_2(\vec{k}) 
\end{eqnarray}
The dispersion relations are plotted in Fig.({\ref{fig:dis-TRS}}). These dispersion relations are found to be independent of the precise value of the staggered flux, i.e., a C-S flux different from the half flux-quantum per plaquette assumed earlier. This shows that the AFA yields dispersion relations that are sensitive to the effects of the global boundary-condition changing flux, but not to the effects of the local C-S flux. However, as we will see below, this is sufficient in providing insight into the topological nature of the transition. The effective Dirac Hamiltonian which will give us same quasiparticle spectrum as Eq.(\ref{eq.e1} - \ref{eq.e2}), can be written as 
\begin{eqnarray}
H &=& \left(J_1 \sin {k_x} \pm \delta J_2 \sin {k_y}  \right) \alpha_x \nonumber \\
&+& \left( h \pm \delta J_1 \cos {k_x} + J_2  \cos {k_y}  \right) \beta \label{eq.hdirac2d} \\
&=& A \alpha_x + B \beta
\end{eqnarray}
where $A = \left(J_1 \sin {k_x} \pm \delta J_2 \sin {k_y}  \right) $, $B = \left( h \pm \delta J_1 \cos {k_x} + J_2  \cos {k_y} \right)$ and the matrices $\alpha_{x,y} = \left[ \begin{array}{cc}
0 & \sigma_{x,y} \\ 
\sigma_{x,y} & 0
\end{array} \right] $ and $\beta = \left[ \begin{array}{cc}
I & 0 \\ 
0 & -I
\end{array}  \right]$ are $4 \times 4$ matrices~\cite{sqshen}, and $\sigma_{x,y}$ are the usual Pauli matrices. (The reader should not mix up this $\beta$ matrix with the inverse temperature $\beta=1/\textrm{k}_{\textrm{B}}$T which we have used elsewhere in this manuscript.) As observed earlier in section \ref{sec:lt} for the fermionised 1D TFIM, it is not difficult to see here too that the effective Dirac Hamiltonian describes excitations about $k$-modes that are singular with respect to the Bogoliubov transformation: these are the high-symmetry points of the first Brillouin zone (B.Z.) at $(k_{x},k_{y}) = (0,0)$, $(0,\pm\pi)$, $(\pm\pi,0)$, $(\pm\pi,\pm\pi)$. The dispersion relations are seen to possess linear gap-closing Lifshitz transitions at these time-reversal symmetric (TRS) points for special values of the transverse field $h^{*} = |\delta J_1 - J_2|$ and $h^{*} = | \delta J_1 +  J_2 | $ values (see Figs.(\ref{fig:dis-TRS},\ref{fig:contour-ising})). 

\begin{widetext}

\begin{figure}[htp] 
\centering
\includegraphics[width=0.984\textwidth]{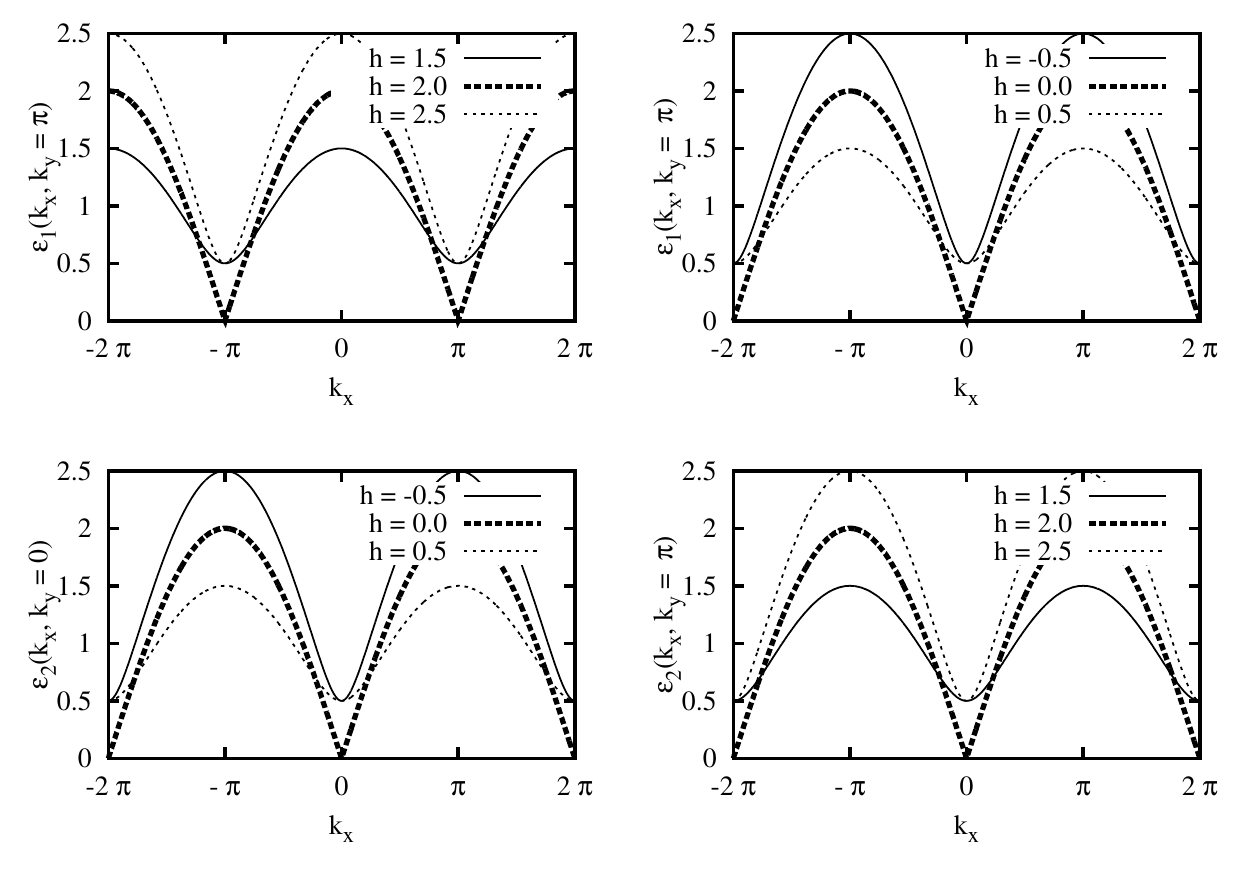}
\caption{(Color online.)~
The spectrum of Eq.(\ref{eq.e1}) is plotted within Average Field Approximation (AFA) as a function of $k_x$ for different $h$ values keeping $J_1 = J_2 = \delta = 1$. Thus shows for 2D-TFIM the gap closes near the TRS symmetric point. They are $(\pi,0)$, $(-\pi,0)$ for the left panel and $(0,\pi)$ and $(0,-\pi)$ for the right panel. The spectrum for Eq.(\ref{eq.e2}) look the same and gap closes at $(0,0)$ and $(\pi,\pi)$.}
\label{fig:dis-TRS}
\end{figure}

\begin{figure}[htp] 
\centering
\includegraphics[width=0.45\textwidth]{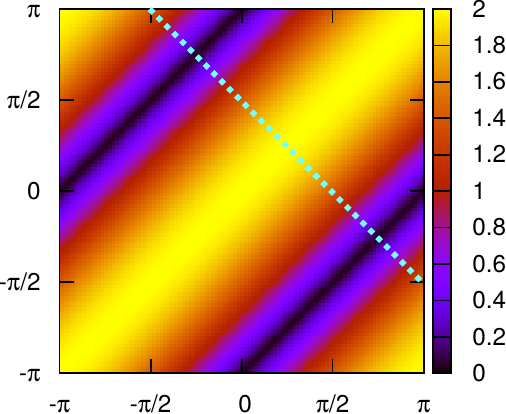} \qquad  \includegraphics[width=0.45\textwidth]{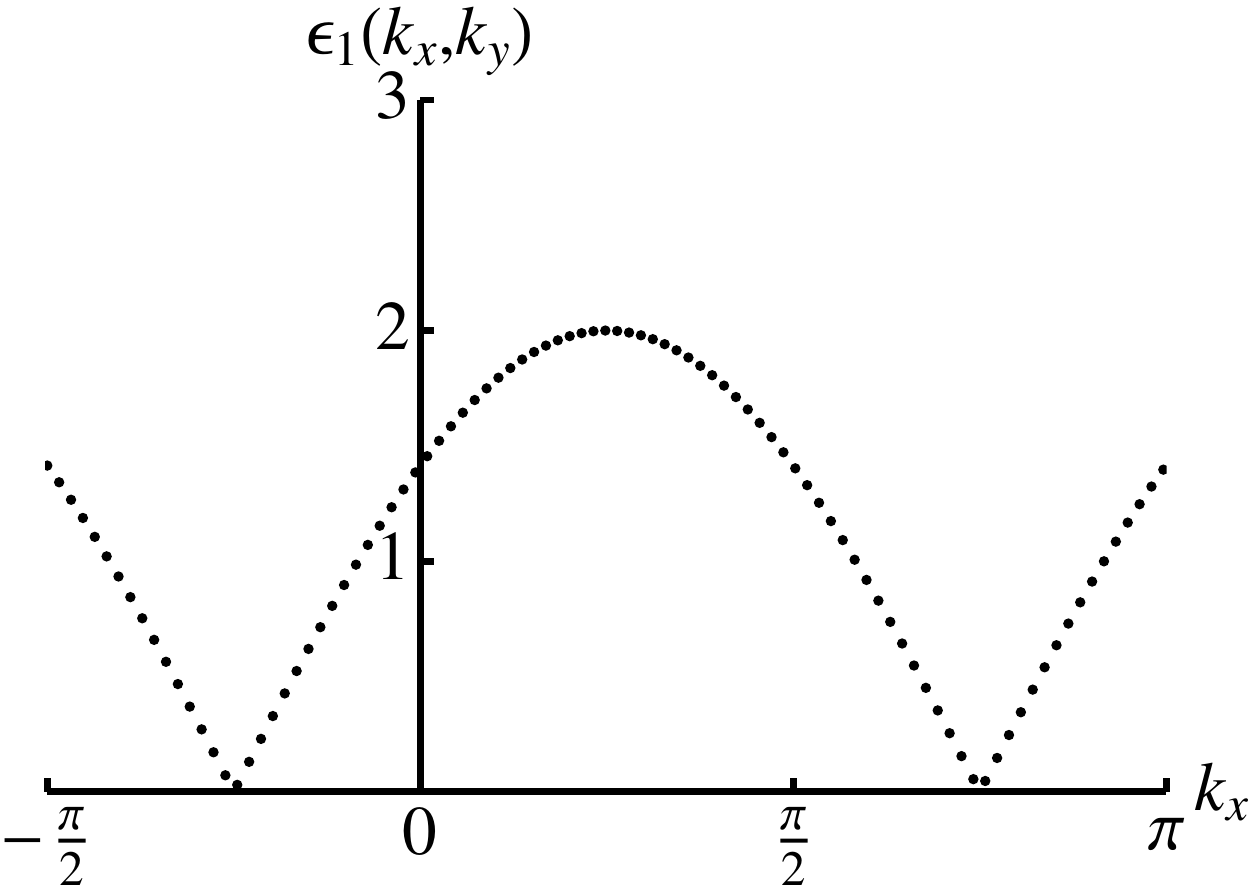}

\includegraphics[width=0.45\textwidth]{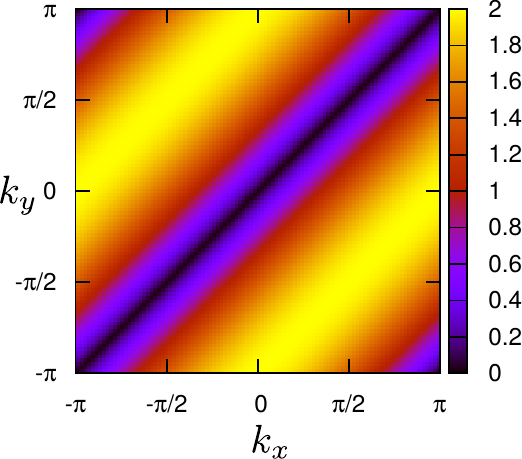} \qquad \includegraphics[width=0.45\textwidth]{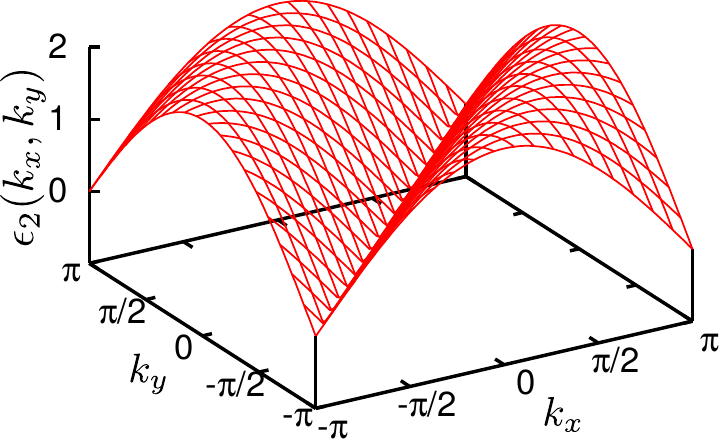}
\caption{(Color online.)~
A contour plot of the dispersion for 2D TFIM within average field approximation (AFA) is plotted for $J_1 = J_2 = \delta = 1$ and $h = 0$. The color code is for the energy dispersion value. For other $h$ values, the spectrum is gapped. In the two figures on upper row the dispersion is eq.({\ref{eq.e1}}). We take a cross-section on the $k_x - k_y$ plane along the dotted line, and present the dispersion as a function of $k_x$. This shows a Dirac-like (linear gap-closing) feature at the critical point. There exist two lines of such critical theories along $k_{y}=k_{x}\pm\pi$. In the lower row, the dispersion is eq.({\ref{eq.e2}}) is plotted for $J_1 = J_2 = \delta = 1$ and $h = 0$. For other $h$ values, the spectrum is gapped. It shows a Dirac-like (linear gap-closing) feature at the critical point along the $k_x = k_y$ line. We call these three critical lines the Weyl stripe metal (WStM) (see text for more).}
\label{fig:contour-ising}
\end{figure}

\end{widetext}

\par
In order to clarify the change in a topological quantum number across these transitions, we now calculate a topological index from these time-reversal symmetric points in the first quadrant of the B.Z.~\cite{fukane-2007}. For this, we write down the algebra for the full Parity operator $\hat{P}$ below. First, $\hat{P}$ commutes with the Hamiltonian, $\left[\hat{P},H \right ]=0$. Further, the identities satisfied by the full Parity operator $\hat{P}$ are 
\begin{eqnarray}
\hat{P} = \hat{\pi} \beta \\
\hat{P} \alpha_i \hat{P} = - \alpha_i \\
\hat{P} \beta \hat{P} = \beta
\end{eqnarray}
where $\hat{\pi}$ is the standard parity operator defines as ${\hat{\pi}}^\dagger x \hat{\pi} = -x $~,~${\hat{\pi}}^\dagger p \hat{\pi} = -p$, and $\alpha_{i}$, $\beta$ are as defined earlier. Thus, the full parity operator satisfies
\begin{eqnarray}
\hat{P} H(\Gamma_1) = H(-\Gamma_1 + \vec{K}) \hat{P}~,
\end{eqnarray} 
where $\Gamma_1 = (k_x, k_y)$ is a point on the 2D B.Z. and $\vec{K}$ is a reciprocal lattice vector. Thus, the full parity operation at the four time reversal invariant momenta ($\Gamma_{1}^{*}=(0,0),(0,\pi),(\pi,0),(\pi,\pi)$) is given by 
\begin{eqnarray}
\delta_{\Gamma_1^{*}} = \langle \psi_1  |\hat{P} |\psi_1\rangle|_{\Gamma_{1}^{*}}~,
\end{eqnarray} 
where the $|\psi_1 \rangle$ is ground state of the Hamiltonian Eq.(\ref{eq.hdirac2d}), and can be written as 
$$|\psi_1\rangle = {\frac{1}{\sqrt{\epsilon_1 (\epsilon_1 + B ) }}} \left[ \begin{array}{c}
A \\ 
0 \\ 
0 \\ 
\epsilon_1 + B 
\end{array} \right]~,$$
where $A$ and $B$ have been defined earlier, and $\hat{P}$ changes the momentum index by a negative sign. In this way, we obtain
\begin{eqnarray}
\delta_{(0,0)}     &=& sgn (h + \delta J_1 + J_2 ) \nonumber\\
\delta_{(\pi,0)}   &=& sgn (h - \delta J_1 + J_2) \nonumber\\
\delta_{(0,\pi)}   &=& sgn (h + \delta J_1 - J_2) \nonumber\\ 
\delta_{(\pi,\pi)} &=& sgn (h - \delta J_1 - J_2 )~.
\end{eqnarray}
\par
We thus calculate the Fu-Kane topological invariant~\cite{fukane-2007} at these TRS points as
\begin{eqnarray}
(-1)^{\Omega} &=& sgn (h + \delta J_1 + J_2 ) \times sgn (h - \delta J_1 + J_2) \nonumber \\
&& \times sgn (h + \delta J_1 - J_2) \nonumber \\
&& \times sgn (h - \delta J_1 - J_2 )
\end{eqnarray}
where $\Omega$ is a topological number. From this relation, it can be easily seen that there exist Lifshitz transitions (critical Weyl semimetals (Weyl SM)) that separate topologically trivial superconducting phases with $\Omega=0$ and topologically non-trivial (triplet pairing) superconducting phases with $\Omega=1$ at $h^{*}=\delta J_{1} + J_{2}$ as well as $h^{*}=-(\delta J_{1} + J_{2})$. Further, there is another Lifshitz transition between the two phases with $\Omega=1$ at $h^{*}=\delta J_{1} - J_{2}$. For the case of the 2D TFIM ($J_{1}=J_{2}\equiv J$, $\delta=1$), these are transitions at $h^{*}=\pm 2J$ and $h^{*}=0$ respectively. The former corresponds to the finite-temperature critical point of the 3D Ising model, while the latter to a quantum critical Weyl stripe metal (WStM) phase described in detail below. These transitions are shown in Fig.(\ref{fig:nu}) below. It is important to note that all three Lifshitz transitions are Lorentz invariant, i.e., possess dynamical exponent $z=1$ and an emergent $SU(2)$ symmetry at the transition (associated with massless Dirac fermions) where the topological quantum no. $\Omega$ is ill-defined. All of this is common to the transition we have studied earlier for the 1D TFIM, as well as appears to be in agreement with the nature of the transition in the 3D Ising model conjectured in Ref.(\cite{polyakov-book}). 
\par
Indeed, the line of critical points $h^{*}=\delta J_{1} + J_{2}$ obtained from the dispersion relations at these TRS points connects the critical point of the 2D TFIM theory ($J_{1}=J_{2}\equiv J$, $\delta=1$) to that of the 1D TFIM ($J_{1}\equiv J$, $J_{2}=0$, $\delta=1$). Similarly, the line of critical points $h^{*}=\delta J_{1} - J_{2}$ connects the critial point of the 1D TFIM ($J_{1}\equiv J$, $J_{2}=0$, $\delta=1$) to that of a quantum critical Weyl stripe metal (WStM) ($J_{1}=J_{2}\equiv J$, $\delta=1$). As seen in Fig.(\ref{fig:contour-ising}), the WStM possesses an open Fermi surface characterised by $k_{y}=k_{x}$, $k_{y}=k_{x}\pm\pi$, i.e., lines obtained by stacking effectively 1D Weyl semimetal theories along the three directions indicated above within the 2D B.Z. Such an open Fermi surface corresponds to the existence of stripe-like 1D (quantum critical) metallic systems in real space which are disconnected from one another. In this way, we see that the critical line $h^{*}=\delta J_{1} - J_{2}$ connects a set of 1D critical theories while the critical line $h^{*}=\delta J_{1} + J_{2}$ connects theories with dimensionality changing from 1 to 2. This is shown in Fig.(\ref{fig:nu}). Importantly, the latter critical line extends the holographic connection between the critical points of the 2D Ising model/1D TFIM and the 1D Ising model/qubit observed in subsection (\ref{HolographyinIsing}) to that between the 3D Ising model/2D TFIM and the 2D Ising model/1D TFIM. Following our discussion in section \ref{sec:dual} of the role played by duality in a Lifshitz transition, we are now in a position to be able to clarify the nature of the associated topological confinement-deconfinement transition of electric flux string loops in the gauge model reached via a duality transformation of the 2D TFIM (i.e., a theory written in terms of domain wall variables): the 2D quantum Ising lattice gauge theory~\cite{wegner-1971,kogut-1979,fradkin-susskind-1978,fradkin-book}. It will be interesting to investigate whether our approach can also capture the nature of the conjectured tricritical topological transition of the 2D Ising ($Z_{2}$) lattice gauge theory with a coupling to a dynamical Ising matter field (see Ref.(\cite{fradkin-book}) and references therein). 

\begin{figure}[hbtp] 
\centering
\includegraphics[width=0.48\textwidth]{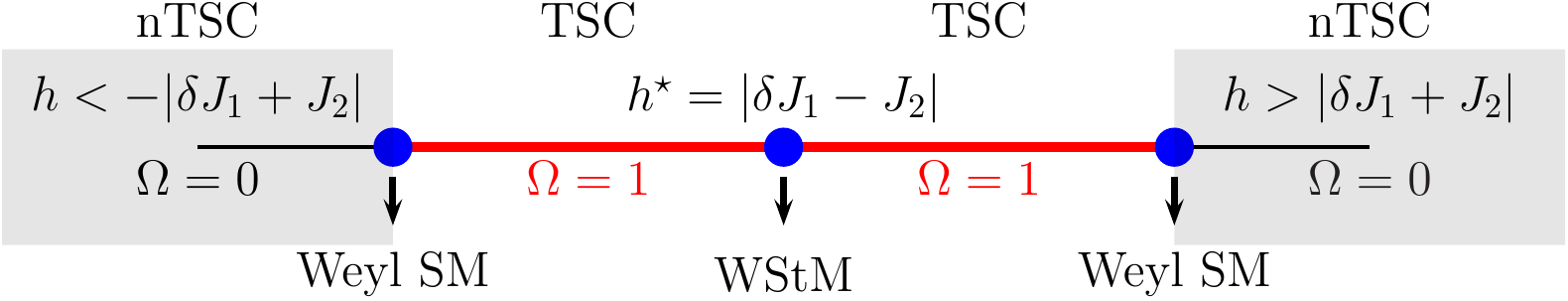}
\includegraphics[width=0.3\textwidth]{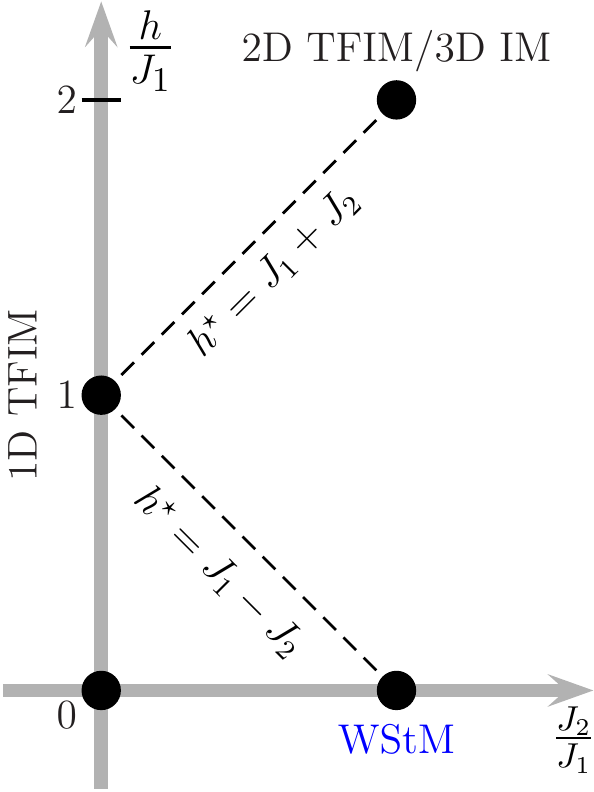}
\caption{(Color online.)~
(top panel) Lifshitz transitions seen via changes in the topological quantum number $\Omega$. Lifshitz transitions with Weyl semimetallic critical theories are found at $h^{*}=\pm | \delta J_1 + J_2 |$, separating non-topological superconductor phases (nTSC, $\Omega =0$) and topological superconductor phases (TSC, $\Omega = 1$). Another Lifshitz transition Weyl stripe metal (WStM) critical theory is observed at $h^{*}=|\delta J_1 - J_2 |$ separating to TSC phases with $\Omega=1$. (bottom panel) A plot of the critical fields for $\delta=1$ shows that there exists a line of Lifshitz critical theories that connect (a) the 2D TFIM and the 1D TFIM ($h^{*}=J_{1}+J_{2}$) and (b) the WStM to the 1D TFIM ($h^{*}=J_{1}-J_{2}$).
}
\label{fig:nu}
\end{figure}

For the sake of completeness, we end with a discussion of the critical exponents obtained for the 2D TFIM from the AFA. As discussed earlier, the dynamical exponent $z=1$ and the gap closes linearly with $|h-h^{*}|$, giving the gap exponent $y=1$ and the correlation length exponent $\nu=1$. We will now compute the specific heat exponent $\alpha$. For this, let us write the effective dispersion relation near one of the TRS symmetric point, lets say $(0,0)$. The total ground state energy is obtained from the dispersion by filling up all the negative energy single particle states and given by 
\begin{eqnarray}
E_{G} &=& - \sum_{\vec{k}} \left[ (h \pm (\delta J_1 + J_2))^2 + (J_1 k_x + \delta J_2 k_y)^2 \right]^{1/2}\nonumber\\
&=& \sum_{k_x , k_y} \sqrt{ g^2 + (J_1 k_x + \delta J_2 k_y)^2}
\end{eqnarray}
here $g$ is the variable which is the gap scale ($g = | h - (\delta J_1 + J_2 ) |$) of our problem at hand. We take double derivative of ground state energy $E_G$ per site, with respect to gap scale $g$ to get the specific heat as 
\begin{eqnarray}
C &=& - \frac{d^2 (E_G/L)}{dg^2} \nonumber\\
&=& \int^{k^\star}_{k_x} \int^{k^\star}_{k_y} \frac{ d k_x d k_y}{J \sqrt{ (g/J)^2 + (k_x + k_y)^2}} \nonumber\\
&\sim& \log \left( \frac{k^\star}{g} \right) \sim \log \left( \frac{k^\star}{|T -T_c|} \right) 
\end{eqnarray}
where we have taken $\delta = 1$ and $J_1 = J_2$ in equn.(\ref{eq:2dxyh}). From the relation for the specific heat $C(T) \sim \text{constant} ~ |T -T_c|^\alpha $, the $\log$ form of $C(T)$ again gives us $\alpha = 0$. In the absence of an exact solution for the 2D TFIM, we rely on a comparison with the values of the critical field $h^{*}=3.04J$ and exponents $(\alpha=0.134,\nu=0.622=y, z=1)$ obtained from various numerical approaches.~\cite{dejonghvanleeuwen-1998,riegerkawashima-1999, blotedeng-2002, evenblyvidal-2009}~While the same value of $z=1$ is obtained from the AFA, the disagreement with the other values, $\alpha=0,~\nu=1=y,~h^{*}=2J$, is apparent. This discrepancy is an outcome of the AFA and is expected to improve upon taking into consideration of the fluctuations of the C-S field around the AFA ground state~\cite{lopezrojofradkin-1994}. This is, however, beyond the scope of the present work. Nevertheless, as noted earlier, the AFA does meet the main goal of this work by capturing precisely the topological nature of the transition. 
\section{\label{sec:summary} Summary and Outlook}
\begin{figure}[!htb] 
\centering
\includegraphics[width=0.5\textwidth]{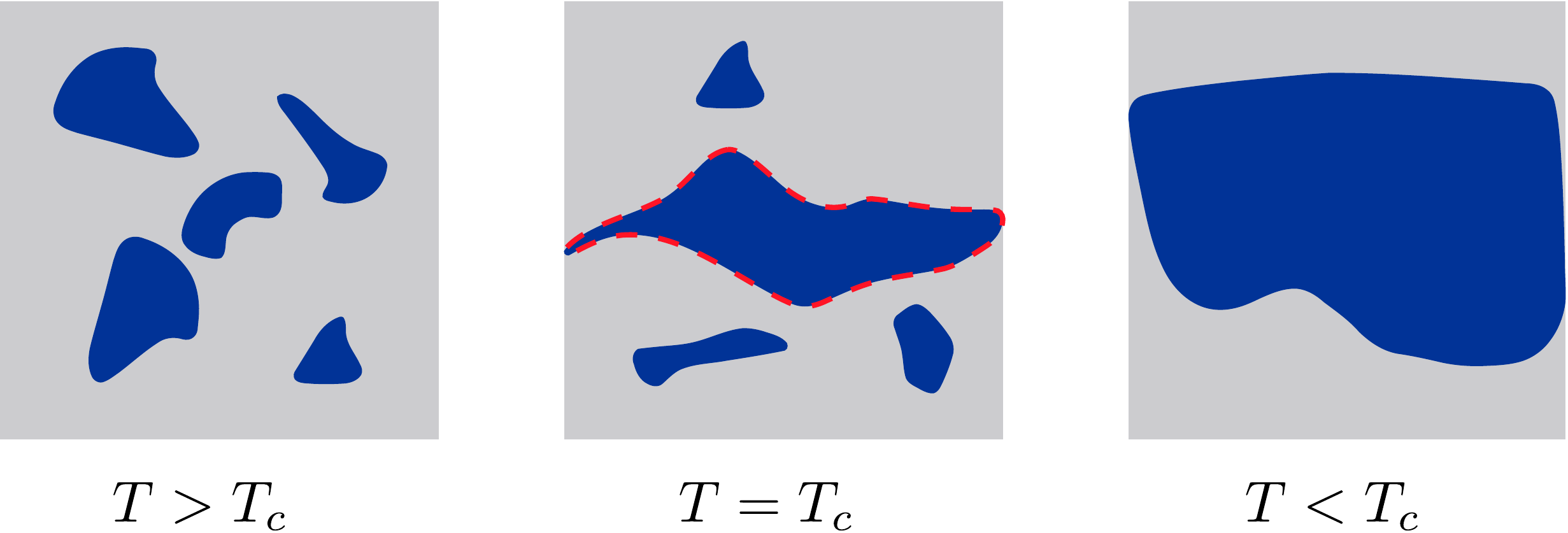}
\caption{(Color online.)~A representation of the evolution of ordered domains with varying temperatures is shown. The light shaded region is where spins are randomly oriented. Dark region is where spins are aligned to form a domain. At $T=T_c$ there is at least one domain (shown as a dashed red curve) which touches the boundary (and thus becomes sensitive to boundary conditions, i.e., topologically active). Below $T_c$ we have a big domain spanning the full system.
}
\label{fig:domain2d-ising}
\end{figure}
We begin by summarising our main findings. To begin with, in section \ref{sec:2d-classical-1d-QI}, by noting the fact that the Bogoliubov transformation for the $\pi$-mode in the quasi-particle eigenspectrum of the Jordan-Wigner fermionised 1D TFIM diverges at criticality, 
we identified that the linear gap-closing of the massive Dirac spectrum can be tracked via an effective Hamiltonian for this mode. This is the Hamiltonian of a single qubit in the presence of a field. In sections \ref{sec:lt}, \ref{sec:electric-field} and \ref{sec:dual}, various topological quantum nos. that characterise the Fermi-surface topology-changing Lifshitz transition are then computed from this effective Hamiltonian. Importantly, the vanishing effective Hamiltonian indicates a non-trivial topological quantum number called Wess-Zumino-Novikov-Witten term arising form the SU(2) nature of the massless Dirac spectrum at the critical point of the 1D TFIM. The ordered phase is characterised by a non-local fermion-parity topological invariant related to a $Z_{2}$ symmetry operation of the original spin problem. In this way, the algebraic divergences observed in the neighbourhood of a continuous transition in the GLW paradigm are replaced here by holographic relations (e.g., equns.(\ref{ChernAtiyahSinger}), (\ref{chernhall}), (\ref{volovikrelation}), (\ref{entholo}) etc.) that link a topological quantity at the critical point to other topological quantities in the ordered and disordered phases.
Further, measurable quantities like the specific heat $C$ are shown to be a function of a Berry connection arising from the $\pi$-mode Hamiltonian, while Pfeuty's end-spin correlation function is observed to be linked to the $Z_{2}$  topological invariant. The symmetry-protected topological (SPT) nature of the ground state in the ordered phase, as well as the entanglement content of the critical point and the phases it separates, is clarified. The duality of the 1D TFIM is captured through the holographic relations mentioned above. The bulk-boundary correspondence at the heart of this holography is captured via a Thouless charge-pumping mechanism which is equivalent to the Laughlin gedanken for the IQHE. We propose that similar topological transitions should be observed in various topological insulators and superconductors which are studied via Dirac operators. The transitions will be signalled very generally by the appearance of at least one zero eigenvalue of the Dirac mass matrix, i.e., an emergent SU(2) symmetry for massless Dirac fermions in the theory. The critical exponents $\nu, \alpha, y$ and $z$ are seen to arise naturally from these topological arguments; they satisfy the Josephson scaling relation as expected. We show that the thermodynamic limit needs to be taken such that the topological properties (i.e., sensitivity to boundary conditions) are manifest throughout. This is shown to be in agreement with the Yang-Lee theorem, and confirms the nature of the topological transition in the classical 2D Ising model. 
\par
The effects of a non-zero longitudinal field and interactions that scatter across the singular Fermi surface are treated within the renormalisation group (RG) formalism in section \ref{sec:rg-2d-1d-ising}, revealing a rich phase diagram in Fig.(\ref{fig:rg-tfim-hx}). The analysis shows that the critical point of the 1D TFIM can be destabilised by hedgehog topological excitations and flow under RG to the $SU(2)$-symmetric WZNW theory for the 1D spin-1/2 Heisenberg chain. Similarly, the flow from the WZNW theory to the finite magnetisation theories in the presence of non-zero longitudinal and transverse fields is caused by RG flows arising from meron topological excitations.
In section \ref{sec:univ}, we extend these ideas to various other Ising and related lattice gauge models. We show that the classical to quantum correspondence links the critical theories of Ising models in various dimensions holographically through the universal effective Hamiltonian that describes the Lifshitz transition of the 1D TFIM. For instance, the 2D transverse-field Ising model is shown to possess a $Z_{2}$ symmetry similar to the 1D model. Via a Jordan-Wigner fermionisation in 2D, we show that the fermionic model possess
a topological fermion-parity index equivalent to the $Z_{2}$ invariant of the original spin model. 
We show the existence of a Weyl-semimetal Lifshitz transition between the ordered phase (topological superconductor) and the disordered phase (non-Topological SC) in the 2D TFIM. This critical theory is connected to that of the 1D TFIM through a line of critical points, confirming their holographic relation. This also reveals the topological nature of the transition in the 3D Ising model and the 2D quantum Ising lattice gauge theory. We also find another novel Lifshitz transition of the 2D TFIM between two topological SC phases involving a gapless Weyl stripe metal. In the same manner, the $T=0$ topological transition of the 1D classical Ising model leads to similar conclusions for the 2D classical Ising lattice gauge theory, the Kitaev toric code and the Wen plaquette model. It is important to note that we have, thus far, not found any evidence for the existence of critical dimensions that can limit the validity of these statements.
\par
Our results for the quantum Ising models can be translated to their classical counterparts via the $T$-matrix formalism. First, they reveal that topological features are common to both quantum and classical Ising transitions and suggest ways in which macroscopic symmetry-protected topological quantum phenomena can be manifested in classical contexts. Certainly, the effects of dissipative environments as well as external influences that may spoil these topological features need a careful study. We do find, however, that dissipation from an Ohmic environment could instead secure the topological properties. Importantly, our discussions of the pathways that lead from the Lifshitz transitions discussed in the present work to the second-order transitions of the GLW paradigm suggest that the latter could be an experimentally observed reality emergent from the former. The notion of universality for such Lifshitz transitions in Ising models relates simply to the fact that they should possess (i) Lorentz invariance (i.e., dynamical exponent $z=1$) and (ii) an emergent $SU(2)$ symmetry, irrespective of dimensionality. We find that some critical exponents and scaling relations do exist for such non-GLW transitions. However, non-local order parameters, sensitivity to boundary conditions and holographic relations  between topological quantities at the transition, the ordered and disordered phases are equally important. It will be interesting to investigate whether such transitions exist in other symmetry groups as well~\cite{mukherjeejalallal}.
\par
Finally, what are the experimental signatures of such Lifshitz transitions in the classical model? Given that the Jordan-Wigner fermions of the TFIM correspond to domain wall excitations in the classical problem, we expect to observe spanning clusters precisely at criticality, i.e., domain walls that span the system from end to end and whose sensitivity to boundary conditions signals the topological transition. This is visualised in Fig.(\ref{fig:domain2d-ising}). It would be interesting to realise an equivalent of the Thouless charge pumping experiment as well. We expect that this should correspond to the periodic appearance and disappearance of such spanning domain walls as the critical point is crossed periodically and adiabatically. However, it is crucial that such an experiment be carried out in the absence of a longitudinal field that breaks the Ising symmetry, as well as any external influences that may quench the dynamics of the topological degrees of freedom in the bulk as well as the boundaries. It is to be expected that any success in such directions will spur further activity in the understanding of transitions that lie beyond the traditional GLW paradigm.
\section*{Acknowledgements}
The authors thank A. Mukherjee, S. Pal, V. Adak, R. K. Singh, A. Ghosh, B. Bansal, A. Taraphder, S. Rao for several enlightening discussions. We also thank J. Vidal, M. A. Martin-Delgado and J. H. H. Perk for valuable feedback.
S.J. acknowledges IISER-Kolkata for financial support through a postdoctoral fellowship. R. K. thanks the DST, Govt. of India for funding through an INSPIRE fellowship during his tenure at IISER-Kolkata as a student of the BS-MS program. S. L. thanks the DST, Govt. of India for funding through a Ramanujan Fellowship (2010-2015) during which a substantial part of this work was carried out. 
\bibliography{ref-2dising} 

\begin{thebibliography}{146}%
\makeatletter
\providecommand \@ifxundefined [1]{%
 \@ifx{#1\undefined}
}%
\providecommand \@ifnum [1]{%
 \ifnum #1\expandafter \@firstoftwo
 \else \expandafter \@secondoftwo
 \fi
}%
\providecommand \@ifx [1]{%
 \ifx #1\expandafter \@firstoftwo
 \else \expandafter \@secondoftwo
 \fi
}%
\providecommand \natexlab [1]{#1}%
\providecommand \enquote  [1]{``#1''}%
\providecommand \bibnamefont  [1]{#1}%
\providecommand \bibfnamefont [1]{#1}%
\providecommand \citenamefont [1]{#1}%
\providecommand \href@noop [0]{\@secondoftwo}%
\providecommand \href [0]{\begingroup \@sanitize@url \@href}%
\providecommand \@href[1]{\@@startlink{#1}\@@href}%
\providecommand \@@href[1]{\endgroup#1\@@endlink}%
\providecommand \@sanitize@url [0]{\catcode `\\12\catcode `\$12\catcode
  `\&12\catcode `\#12\catcode `\^12\catcode `\_12\catcode `\%12\relax}%
\providecommand \@@startlink[1]{}%
\providecommand \@@endlink[0]{}%
\providecommand \url  [0]{\begingroup\@sanitize@url \@url }%
\providecommand \@url [1]{\endgroup\@href {#1}{\urlprefix }}%
\providecommand \urlprefix  [0]{URL }%
\providecommand \Eprint [0]{\href }%
\providecommand \doibase [0]{http://dx.doi.org/}%
\providecommand \selectlanguage [0]{\@gobble}%
\providecommand \bibinfo  [0]{\@secondoftwo}%
\providecommand \bibfield  [0]{\@secondoftwo}%
\providecommand \translation [1]{[#1]}%
\providecommand \BibitemOpen [0]{}%
\providecommand \bibitemStop [0]{}%
\providecommand \bibitemNoStop [0]{.\EOS\space}%
\providecommand \EOS [0]{\spacefactor3000\relax}%
\providecommand \BibitemShut  [1]{\csname bibitem#1\endcsname}%
\let\auto@bib@innerbib\@empty
\bibitem [{\citenamefont {Landau}\ and\ \citenamefont
  {Lifshitz}(1958)}]{landaulifshitzstatphys}%
  \BibitemOpen
  \bibfield  {author} {\bibinfo {author} {\bibfnamefont {L.}~\bibnamefont
  {Landau}}\ and\ \bibinfo {author} {\bibfnamefont {E.~M.}\ \bibnamefont
  {Lifshitz}},\ }\href@noop {} {\emph {\bibinfo {title} {Statistical
  Physics}}}\ (\bibinfo  {publisher} {Pergamon Press, London},\ \bibinfo {year}
  {1958})\BibitemShut {NoStop}%
\bibitem [{\citenamefont {Goldenfeld}(1992)}]{goldenfeld-book}%
  \BibitemOpen
  \bibfield  {author} {\bibinfo {author} {\bibfnamefont {N.}~\bibnamefont
  {Goldenfeld}},\ }\href@noop {} {\emph {\bibinfo {title} {Lectures On Phase
  Transitions And The Renormalization Group}}}\ (\bibinfo  {publisher} {Perseus
  Books (The Advanced Book Program)},\ \bibinfo {year} {1992})\BibitemShut
  {NoStop}%
\bibitem [{\citenamefont {Wilson}(1975)}]{wilson-1975}%
  \BibitemOpen
  \bibfield  {author} {\bibinfo {author} {\bibfnamefont {K.}~\bibnamefont
  {Wilson}},\ }\href@noop {} {\bibfield  {journal} {\bibinfo  {journal} {Rev.
  Mod. Phys.}\ }\textbf {\bibinfo {volume} {47}},\ \bibinfo {pages} {733}
  (\bibinfo {year} {1975})}\BibitemShut {NoStop}%
\bibitem [{\citenamefont {Ising}(1925)}]{ising-1925}%
  \BibitemOpen
  \bibfield  {author} {\bibinfo {author} {\bibfnamefont {E.}~\bibnamefont
  {Ising}},\ }\href@noop {} {\bibfield  {journal} {\bibinfo  {journal} {Zeits.
  f{\"u}r Physik}\ }\textbf {\bibinfo {volume} {31}},\ \bibinfo {pages} {253}
  (\bibinfo {year} {1925})}\BibitemShut {NoStop}%
\bibitem [{\citenamefont {Onsager}(1944)}]{onsager-1944}%
  \BibitemOpen
  \bibfield  {author} {\bibinfo {author} {\bibfnamefont {L.}~\bibnamefont
  {Onsager}},\ }\href@noop {} {\bibfield  {journal} {\bibinfo  {journal}
  {Physical review}\ }\textbf {\bibinfo {volume} {65}},\ \bibinfo {pages} {117}
  (\bibinfo {year} {1944})}\BibitemShut {NoStop}%
\bibitem [{\citenamefont {Kaufman}(1949)}]{kaufman-1949}%
  \BibitemOpen
  \bibfield  {author} {\bibinfo {author} {\bibfnamefont {B.}~\bibnamefont
  {Kaufman}},\ }\href@noop {} {\bibfield  {journal} {\bibinfo  {journal}
  {Physical Review}\ }\textbf {\bibinfo {volume} {76}},\ \bibinfo {pages}
  {1232} (\bibinfo {year} {1949})}\BibitemShut {NoStop}%
\bibitem [{\citenamefont {Suzuki}(1976)}]{suzuki-1976}%
  \BibitemOpen
  \bibfield  {author} {\bibinfo {author} {\bibfnamefont {M.}~\bibnamefont
  {Suzuki}},\ }\href@noop {} {\bibfield  {journal} {\bibinfo  {journal} {Prog.
  Theor. Phys.}\ }\textbf {\bibinfo {volume} {56}},\ \bibinfo {pages} {1454}
  (\bibinfo {year} {1976})}\BibitemShut {NoStop}%
\bibitem [{\citenamefont {Fradkin}\ and\ \citenamefont
  {Susskind}(1978)}]{fradkin-susskind-1978}%
  \BibitemOpen
  \bibfield  {author} {\bibinfo {author} {\bibfnamefont {E.}~\bibnamefont
  {Fradkin}}\ and\ \bibinfo {author} {\bibfnamefont {L.}~\bibnamefont
  {Susskind}},\ }\href {\doibase 10.1103/PhysRevD.17.2637} {\bibfield
  {journal} {\bibinfo  {journal} {Phys. Rev. D}\ }\textbf {\bibinfo {volume}
  {17}},\ \bibinfo {pages} {2637} (\bibinfo {year} {1978})}\BibitemShut
  {NoStop}%
\bibitem [{\citenamefont {Schultz}\ \emph {et~al.}(1964)\citenamefont
  {Schultz}, \citenamefont {Mattis},\ and\ \citenamefont
  {Lieb}}]{schultz-1964}%
  \BibitemOpen
  \bibfield  {author} {\bibinfo {author} {\bibfnamefont {T.~D.}\ \bibnamefont
  {Schultz}}, \bibinfo {author} {\bibfnamefont {D.~C.}\ \bibnamefont {Mattis}},
  \ and\ \bibinfo {author} {\bibfnamefont {E.~H.}\ \bibnamefont {Lieb}},\
  }\href@noop {} {\bibfield  {journal} {\bibinfo  {journal} {Review of Modern
  Physics}\ }\textbf {\bibinfo {volume} {36}},\ \bibinfo {pages} {856}
  (\bibinfo {year} {1964})}\BibitemShut {NoStop}%
\bibitem [{\citenamefont {Elliot}\ \emph {et~al.}(1970)\citenamefont {Elliot},
  \citenamefont {Pfeuty},\ and\ \citenamefont {Wood}}]{elliot-1970}%
  \BibitemOpen
  \bibfield  {author} {\bibinfo {author} {\bibfnamefont {R.~J.}\ \bibnamefont
  {Elliot}}, \bibinfo {author} {\bibfnamefont {P.}~\bibnamefont {Pfeuty}}, \
  and\ \bibinfo {author} {\bibfnamefont {C.}~\bibnamefont {Wood}},\ }\href@noop
  {} {\bibfield  {journal} {\bibinfo  {journal} {Phys. Rev. Lett.}\ }\textbf
  {\bibinfo {volume} {36}},\ \bibinfo {pages} {856} (\bibinfo {year}
  {1970})}\BibitemShut {NoStop}%
\bibitem [{\citenamefont {Pfeuty}(1970)}]{pfeuty-1970}%
  \BibitemOpen
  \bibfield  {author} {\bibinfo {author} {\bibfnamefont {P.}~\bibnamefont
  {Pfeuty}},\ }\href@noop {} {\bibfield  {journal} {\bibinfo  {journal} {Annals
  of Physics}\ }\textbf {\bibinfo {volume} {57}},\ \bibinfo {pages} {79}
  (\bibinfo {year} {1970})}\BibitemShut {NoStop}%
\bibitem [{\citenamefont {Kramers}\ and\ \citenamefont
  {Wannier}(1941)}]{kramers-wannier-1941}%
  \BibitemOpen
  \bibfield  {author} {\bibinfo {author} {\bibfnamefont {H.}~\bibnamefont
  {Kramers}}\ and\ \bibinfo {author} {\bibfnamefont {G.}~\bibnamefont
  {Wannier}},\ }\href@noop {} {\bibfield  {journal} {\bibinfo  {journal} {Phys.
  Rev.}\ }\textbf {\bibinfo {volume} {60}},\ \bibinfo {pages} {252} (\bibinfo
  {year} {1941})}\BibitemShut {NoStop}%
\bibitem [{\citenamefont {Kadanoff}\ and\ \citenamefont
  {Ceva}(1971)}]{kadanoffceva-1971}%
  \BibitemOpen
  \bibfield  {author} {\bibinfo {author} {\bibfnamefont {L.~P.}\ \bibnamefont
  {Kadanoff}}\ and\ \bibinfo {author} {\bibfnamefont {H.}~\bibnamefont
  {Ceva}},\ }\href@noop {} {\bibfield  {journal} {\bibinfo  {journal} {Phys.
  Rev. B}\ }\textbf {\bibinfo {volume} {3}},\ \bibinfo {pages} {3918} (\bibinfo
  {year} {1971})}\BibitemShut {NoStop}%
\bibitem [{\citenamefont {Sachdev}(2011)}]{sachdev-book}%
  \BibitemOpen
  \bibfield  {author} {\bibinfo {author} {\bibfnamefont {S.}~\bibnamefont
  {Sachdev}},\ }\href@noop {} {\emph {\bibinfo {title} {Quantum Phase
  Transitions}}}\ (\bibinfo  {publisher} {Second Edition, Cambridge University
  Press},\ \bibinfo {year} {2011})\BibitemShut {NoStop}%
\bibitem [{\citenamefont {Peierls}(1936)}]{peierls-1936}%
  \BibitemOpen
  \bibfield  {author} {\bibinfo {author} {\bibfnamefont {R.}~\bibnamefont
  {Peierls}},\ }\href@noop {} {\bibfield  {journal} {\bibinfo  {journal} {Math.
  Proc. of the Cambridge Phil. Soc.}\ }\textbf {\bibinfo {volume} {32}},\
  \bibinfo {pages} {477} (\bibinfo {year} {1936})}\BibitemShut {NoStop}%
\bibitem [{\citenamefont {M{\"u}ller-Hartmann}\ and\ \citenamefont
  {Zittartz}(1977)}]{mullerhartmannzittartz-1977}%
  \BibitemOpen
  \bibfield  {author} {\bibinfo {author} {\bibfnamefont {E.}~\bibnamefont
  {M{\"u}ller-Hartmann}}\ and\ \bibinfo {author} {\bibfnamefont
  {J.}~\bibnamefont {Zittartz}},\ }\href {\doibase 10.1007/BF01325537}
  {\bibfield  {journal} {\bibinfo  {journal} {Zeitschrift f{\"u}r Physik B
  Condensed Matter}\ }\textbf {\bibinfo {volume} {27}},\ \bibinfo {pages} {261}
  (\bibinfo {year} {1977})}\BibitemShut {NoStop}%
\bibitem [{\citenamefont {Kasteleyn}(1961)}]{kasteleyn1961}%
  \BibitemOpen
  \bibfield  {author} {\bibinfo {author} {\bibfnamefont {P.~W.}\ \bibnamefont
  {Kasteleyn}},\ }\href@noop {} {\bibfield  {journal} {\bibinfo  {journal}
  {Physica}\ }\textbf {\bibinfo {volume} {27}},\ \bibinfo {pages} {1209}
  (\bibinfo {year} {1961})}\BibitemShut {NoStop}%
\bibitem [{\citenamefont {Yang}(1952)}]{yang-1952}%
  \BibitemOpen
  \bibfield  {author} {\bibinfo {author} {\bibfnamefont {C.~N.}\ \bibnamefont
  {Yang}},\ }\href@noop {} {\bibfield  {journal} {\bibinfo  {journal} {Physical
  Review}\ }\textbf {\bibinfo {volume} {85}},\ \bibinfo {pages} {809} (\bibinfo
  {year} {1952})}\BibitemShut {NoStop}%
\bibitem [{\citenamefont {Wu}(1966)}]{wu-1966}%
  \BibitemOpen
  \bibfield  {author} {\bibinfo {author} {\bibfnamefont {T.~T.}\ \bibnamefont
  {Wu}},\ }\href@noop {} {\bibfield  {journal} {\bibinfo  {journal} {Phys.
  Rev.}\ }\textbf {\bibinfo {volume} {149}},\ \bibinfo {pages} {380} (\bibinfo
  {year} {1966})}\BibitemShut {NoStop}%
\bibitem [{\citenamefont {McCoy}(1968)}]{mccoy-1968}%
  \BibitemOpen
  \bibfield  {author} {\bibinfo {author} {\bibfnamefont {B.~M.}\ \bibnamefont
  {McCoy}},\ }\href@noop {} {\bibfield  {journal} {\bibinfo  {journal} {Phys.
  Rev.}\ }\textbf {\bibinfo {volume} {173}},\ \bibinfo {pages} {531} (\bibinfo
  {year} {1968})}\BibitemShut {NoStop}%
\bibitem [{\citenamefont {Lifshitz}(1960)}]{lifshitz-1960}%
  \BibitemOpen
  \bibfield  {author} {\bibinfo {author} {\bibfnamefont {I.~M.}\ \bibnamefont
  {Lifshitz}},\ }\href@noop {} {\bibfield  {journal} {\bibinfo  {journal} {Sov.
  Phys. JETP}\ }\textbf {\bibinfo {volume} {11}},\ \bibinfo {pages} {1130}
  (\bibinfo {year} {1960})}\BibitemShut {NoStop}%
\bibitem [{\citenamefont {Ruelle}(1969)}]{ruelle-book}%
  \BibitemOpen
  \bibfield  {author} {\bibinfo {author} {\bibfnamefont {D.}~\bibnamefont
  {Ruelle}},\ }\href@noop {} {\emph {\bibinfo {title} {Statistical Mechanics:
  rigorous results.}}}\ (\bibinfo  {publisher} {W. A. Benjamin},\ \bibinfo
  {year} {1969})\BibitemShut {NoStop}%
\bibitem [{\citenamefont {Kohn}(1964)}]{kohn-1964}%
  \BibitemOpen
  \bibfield  {author} {\bibinfo {author} {\bibfnamefont {W.}~\bibnamefont
  {Kohn}},\ }\href@noop {} {\bibfield  {journal} {\bibinfo  {journal} {Phys.
  Rev.}\ }\textbf {\bibinfo {volume} {\bf 133}},\ \bibinfo {pages} {A171}
  (\bibinfo {year} {1964})}\BibitemShut {NoStop}%
\bibitem [{\citenamefont {Edwards}\ and\ \citenamefont
  {Thouless}(1972)}]{thoulessedwards-1972}%
  \BibitemOpen
  \bibfield  {author} {\bibinfo {author} {\bibfnamefont {J.~T.}\ \bibnamefont
  {Edwards}}\ and\ \bibinfo {author} {\bibfnamefont {D.~J.}\ \bibnamefont
  {Thouless}},\ }\href@noop {} {\bibfield  {journal} {\bibinfo  {journal} {J.
  Phys. C: Solid State Phys.}\ }\textbf {\bibinfo {volume} {5}},\ \bibinfo
  {pages} {807} (\bibinfo {year} {1972})}\BibitemShut {NoStop}%
\bibitem [{\citenamefont {Fradkin}(2013)}]{fradkin-book}%
  \BibitemOpen
  \bibfield  {author} {\bibinfo {author} {\bibfnamefont {E.}~\bibnamefont
  {Fradkin}},\ }\href@noop {} {\emph {\bibinfo {title} {Field theories of
  Condensed Matter Physics}}}\ (\bibinfo  {publisher} {Cambridge University
  Press},\ \bibinfo {year} {2013})\BibitemShut {NoStop}%
\bibitem [{\citenamefont {Kogut}(1979)}]{kogut-1979}%
  \BibitemOpen
  \bibfield  {author} {\bibinfo {author} {\bibfnamefont {J.~B.}\ \bibnamefont
  {Kogut}},\ }\href@noop {} {\bibfield  {journal} {\bibinfo  {journal} {Review
  of Modern Physics}\ }\textbf {\bibinfo {volume} {51}},\ \bibinfo {pages}
  {659} (\bibinfo {year} {1979})}\BibitemShut {NoStop}%
\bibitem [{\citenamefont {Emch}\ and\ \citenamefont {Liu}(2002)}]{emch-book}%
  \BibitemOpen
  \bibfield  {author} {\bibinfo {author} {\bibfnamefont {G.~G.}\ \bibnamefont
  {Emch}}\ and\ \bibinfo {author} {\bibfnamefont {C.}~\bibnamefont {Liu}},\
  }\href@noop {} {\emph {\bibinfo {title} {The Logic of Thermostatistical
  Physics}}}\ (\bibinfo  {publisher} {Springer},\ \bibinfo {year}
  {2002})\BibitemShut {NoStop}%
\bibitem [{\citenamefont {Kitaev}(2009)}]{kitaevperiodictable}%
  \BibitemOpen
  \bibfield  {author} {\bibinfo {author} {\bibfnamefont {A.}~\bibnamefont
  {Kitaev}},\ }\href@noop {} {\bibfield  {journal} {\bibinfo  {journal} {AIP.
  Conf. Proc.}\ }\textbf {\bibinfo {volume} {\bf 1134}},\ \bibinfo {pages} {22}
  (\bibinfo {year} {2009})}\BibitemShut {NoStop}%
\bibitem [{\citenamefont {Bender}\ and\ \citenamefont
  {Orszag}(1999)}]{benderorszag}%
  \BibitemOpen
  \bibfield  {author} {\bibinfo {author} {\bibfnamefont {C.~M.}\ \bibnamefont
  {Bender}}\ and\ \bibinfo {author} {\bibfnamefont {S.~A.}\ \bibnamefont
  {Orszag}},\ }\href@noop {} {\emph {\bibinfo {title} {Advanced Mathematical
  Methods for Scientists and Engineers}}}\ (\bibinfo  {publisher} {Springer},\
  \bibinfo {year} {1999})\BibitemShut {NoStop}%
\bibitem [{\citenamefont {Rotter}(2009)}]{rotter-2009}%
  \BibitemOpen
  \bibfield  {author} {\bibinfo {author} {\bibfnamefont {I.}~\bibnamefont
  {Rotter}},\ }\href@noop {} {\bibfield  {journal} {\bibinfo  {journal} {J.
  Phys. A}\ }\textbf {\bibinfo {volume} {\bf 42}},\ \bibinfo {pages} {153001}
  (\bibinfo {year} {2009})}\BibitemShut {NoStop}%
\bibitem [{\citenamefont {Moiseyev}(2011)}]{moiseyev-book}%
  \BibitemOpen
  \bibfield  {author} {\bibinfo {author} {\bibfnamefont {N.}~\bibnamefont
  {Moiseyev}},\ }\href@noop {} {\emph {\bibinfo {title} {Non-Hermitian Quantum
  Mechanics}}}\ (\bibinfo  {publisher} {Cambridge University Press},\ \bibinfo
  {year} {2011})\BibitemShut {NoStop}%
\bibitem [{\citenamefont {Chandler}(1987)}]{chandler-book}%
  \BibitemOpen
  \bibfield  {author} {\bibinfo {author} {\bibfnamefont {D.}~\bibnamefont
  {Chandler}},\ }\href@noop {} {\emph {\bibinfo {title} {Introduction to Modern
  Statistical mechanics}}}\ (\bibinfo  {publisher} {Oxford University Press},\
  \bibinfo {year} {1987})\BibitemShut {NoStop}%
\bibitem [{\citenamefont {Perk}\ and\ \citenamefont
  {Capel}(1977)}]{perkcapel-1977}%
  \BibitemOpen
  \bibfield  {author} {\bibinfo {author} {\bibfnamefont {J.~H.~H.}\
  \bibnamefont {Perk}}\ and\ \bibinfo {author} {\bibfnamefont {H.~W.}\
  \bibnamefont {Capel}},\ }\href@noop {} {\bibfield  {journal} {\bibinfo
  {journal} {Physica}\ }\textbf {\bibinfo {volume} {89A}},\ \bibinfo {pages}
  {265} (\bibinfo {year} {1977})}\BibitemShut {NoStop}%
\bibitem [{\citenamefont {du~Croo~de Jongh}\ and\ \citenamefont {van
  Leeuwen}(1998)}]{dejonghvanleeuwen-1998}%
  \BibitemOpen
  \bibfield  {author} {\bibinfo {author} {\bibfnamefont {M.~S.~L.}\
  \bibnamefont {du~Croo~de Jongh}}\ and\ \bibinfo {author} {\bibfnamefont
  {J.~M.~J.}\ \bibnamefont {van Leeuwen}},\ }\href@noop {} {\bibfield
  {journal} {\bibinfo  {journal} {Phys. Rev. B}\ }\textbf {\bibinfo {volume}
  {57}},\ \bibinfo {pages} {8494} (\bibinfo {year} {1998})}\BibitemShut
  {NoStop}%
\bibitem [{\citenamefont {Anderson}(1958)}]{anderson-1958}%
  \BibitemOpen
  \bibfield  {author} {\bibinfo {author} {\bibfnamefont {P.~W.}\ \bibnamefont
  {Anderson}},\ }\href@noop {} {\bibfield  {journal} {\bibinfo  {journal}
  {Phys. Rev.}\ }\textbf {\bibinfo {volume} {110}},\ \bibinfo {pages} {827}
  (\bibinfo {year} {1958})}\BibitemShut {NoStop}%
\bibitem [{\citenamefont {Su}\ \emph {et~al.}(1979)\citenamefont {Su},
  \citenamefont {Schrieffer},\ and\ \citenamefont {Heeger}}]{ssh-1979}%
  \BibitemOpen
  \bibfield  {author} {\bibinfo {author} {\bibfnamefont {W.~P.}\ \bibnamefont
  {Su}}, \bibinfo {author} {\bibfnamefont {J.~R.}\ \bibnamefont {Schrieffer}},
  \ and\ \bibinfo {author} {\bibfnamefont {A.~J.}\ \bibnamefont {Heeger}},\
  }\href@noop {} {\bibfield  {journal} {\bibinfo  {journal} {Phys. Rev. Lett.}\
  }\textbf {\bibinfo {volume} {42}},\ \bibinfo {pages} {1698} (\bibinfo {year}
  {1979})}\BibitemShut {NoStop}%
\bibitem [{\citenamefont {Varney}\ \emph {et~al.}(2011)\citenamefont {Varney},
  \citenamefont {Sun}, \citenamefont {Rigol},\ and\ \citenamefont
  {Galitski}}]{varney-2011}%
  \BibitemOpen
  \bibfield  {author} {\bibinfo {author} {\bibfnamefont {C.~N.}\ \bibnamefont
  {Varney}}, \bibinfo {author} {\bibfnamefont {K.}~\bibnamefont {Sun}},
  \bibinfo {author} {\bibfnamefont {M.}~\bibnamefont {Rigol}}, \ and\ \bibinfo
  {author} {\bibfnamefont {V.}~\bibnamefont {Galitski}},\ }\href@noop {}
  {\bibfield  {journal} {\bibinfo  {journal} {Phys. Rev. B}\ }\textbf {\bibinfo
  {volume} {84}},\ \bibinfo {pages} {241105(R)} (\bibinfo {year}
  {2011})}\BibitemShut {NoStop}%
\bibitem [{\citenamefont {Zak}(1989)}]{zak-1989}%
  \BibitemOpen
  \bibfield  {author} {\bibinfo {author} {\bibfnamefont {J.}~\bibnamefont
  {Zak}},\ }\href@noop {} {\bibfield  {journal} {\bibinfo  {journal} {Phys.
  Rev. Lett.}\ }\textbf {\bibinfo {volume} {62}},\ \bibinfo {pages} {2747}
  (\bibinfo {year} {1989})}\BibitemShut {NoStop}%
\bibitem [{\citenamefont {Shen}(2012)}]{sqshen}%
  \BibitemOpen
  \bibfield  {author} {\bibinfo {author} {\bibfnamefont {S.-Q.}\ \bibnamefont
  {Shen}},\ }\href@noop {} {\emph {\bibinfo {title} {Topological Insulators:
  Dirac Equation in Condensed Matters}}}\ (\bibinfo  {publisher} {Springer},\
  \bibinfo {year} {2012})\BibitemShut {NoStop}%
\bibitem [{\citenamefont {McGreevy}(2015)}]{mcgreevylectures}%
  \BibitemOpen
  \bibfield  {author} {\bibinfo {author} {\bibnamefont {McGreevy}},\
  }\href@noop {} {\enquote {\bibinfo {title} {Where do quantum field theories
  come from?}}\ } (\bibinfo {year} {2015}),\ \bibinfo {note} {lecture
  notes}\BibitemShut {NoStop}%
\bibitem [{\citenamefont {Oshikawa}\ \emph {et~al.}(1997)\citenamefont
  {Oshikawa}, \citenamefont {Yamanaka},\ and\ \citenamefont
  {Affleck}}]{oya-1997}%
  \BibitemOpen
  \bibfield  {author} {\bibinfo {author} {\bibfnamefont {M.}~\bibnamefont
  {Oshikawa}}, \bibinfo {author} {\bibfnamefont {M.}~\bibnamefont {Yamanaka}},
  \ and\ \bibinfo {author} {\bibfnamefont {I.}~\bibnamefont {Affleck}},\
  }\href@noop {} {\bibfield  {journal} {\bibinfo  {journal} {Phys. Rev. Lett.}\
  }\textbf {\bibinfo {volume} {78}},\ \bibinfo {pages} {1984} (\bibinfo {year}
  {1997})}\BibitemShut {NoStop}%
\bibitem [{\citenamefont {Niu}\ \emph {et~al.}(2012)\citenamefont {Niu},
  \citenamefont {Chung}, \citenamefont {Hsu}, \citenamefont {Mandal},
  \citenamefont {Raghu},\ and\ \citenamefont {Chakravarty}}]{niu-2012}%
  \BibitemOpen
  \bibfield  {author} {\bibinfo {author} {\bibfnamefont {Y.}~\bibnamefont
  {Niu}}, \bibinfo {author} {\bibfnamefont {S.~B.}\ \bibnamefont {Chung}},
  \bibinfo {author} {\bibfnamefont {C.-H.}\ \bibnamefont {Hsu}}, \bibinfo
  {author} {\bibfnamefont {I.}~\bibnamefont {Mandal}}, \bibinfo {author}
  {\bibfnamefont {S.}~\bibnamefont {Raghu}}, \ and\ \bibinfo {author}
  {\bibfnamefont {S.}~\bibnamefont {Chakravarty}},\ }\href@noop {} {\bibfield
  {journal} {\bibinfo  {journal} {Phys. Rev. B}\ }\textbf {\bibinfo {volume}
  {85}},\ \bibinfo {pages} {035110} (\bibinfo {year} {2012})}\BibitemShut
  {NoStop}%
\bibitem [{\citenamefont {Zhang}\ and\ \citenamefont
  {Song}(2015)}]{zhangsong-2015}%
  \BibitemOpen
  \bibfield  {author} {\bibinfo {author} {\bibfnamefont {G.}~\bibnamefont
  {Zhang}}\ and\ \bibinfo {author} {\bibfnamefont {Z.}~\bibnamefont {Song}},\
  }\href@noop {} {\bibfield  {journal} {\bibinfo  {journal} {Phys. Rev. Lett.}\
  }\textbf {\bibinfo {volume} {115}},\ \bibinfo {pages} {177204} (\bibinfo
  {year} {2015})}\BibitemShut {NoStop}%
\bibitem [{\citenamefont {deGottardi}\ \emph {et~al.}(2011)\citenamefont
  {deGottardi}, \citenamefont {Sen},\ and\ \citenamefont
  {Vishveshwara}}]{degottardi-2011}%
  \BibitemOpen
  \bibfield  {author} {\bibinfo {author} {\bibfnamefont {W.}~\bibnamefont
  {deGottardi}}, \bibinfo {author} {\bibfnamefont {D.}~\bibnamefont {Sen}}, \
  and\ \bibinfo {author} {\bibfnamefont {S.}~\bibnamefont {Vishveshwara}},\
  }\href@noop {} {\bibfield  {journal} {\bibinfo  {journal} {New Journal of
  Physics}\ }\textbf {\bibinfo {volume} {13}},\ \bibinfo {pages} {065028}
  (\bibinfo {year} {2011})}\BibitemShut {NoStop}%
\bibitem [{\citenamefont {Feng}\ \emph {et~al.}(2007)\citenamefont {Feng},
  \citenamefont {Zhang},\ and\ \citenamefont {Xiang}}]{feng-2007}%
  \BibitemOpen
  \bibfield  {author} {\bibinfo {author} {\bibfnamefont {X.-Y.}\ \bibnamefont
  {Feng}}, \bibinfo {author} {\bibfnamefont {G.-M.}\ \bibnamefont {Zhang}}, \
  and\ \bibinfo {author} {\bibfnamefont {T.}~\bibnamefont {Xiang}},\
  }\href@noop {} {\bibfield  {journal} {\bibinfo  {journal} {Phys. Rev. Lett.}\
  }\textbf {\bibinfo {volume} {98}},\ \bibinfo {pages} {087204} (\bibinfo
  {year} {2007})}\BibitemShut {NoStop}%
\bibitem [{\citenamefont {Ferrell}(1973)}]{ferrell-1973}%
  \BibitemOpen
  \bibfield  {author} {\bibinfo {author} {\bibfnamefont {R.~A.}\ \bibnamefont
  {Ferrell}},\ }\href@noop {} {\bibfield  {journal} {\bibinfo  {journal}
  {Journal of Statistical Physics}\ }\textbf {\bibinfo {volume} {8}},\ \bibinfo
  {pages} {265} (\bibinfo {year} {1973})}\BibitemShut {NoStop}%
\bibitem [{\citenamefont {Ferrell}(1981)}]{ferrell-1981}%
  \BibitemOpen
  \bibfield  {author} {\bibinfo {author} {\bibfnamefont {R.~A.}\ \bibnamefont
  {Ferrell}},\ }\href@noop {} {\bibfield  {journal} {\bibinfo  {journal}
  {Journal of Statistical Physics}\ }\textbf {\bibinfo {volume} {25}},\
  \bibinfo {pages} {361} (\bibinfo {year} {1981})}\BibitemShut {NoStop}%
\bibitem [{\citenamefont {Yang}\ and\ \citenamefont {Lee}(1952)}]{yang-lee}%
  \BibitemOpen
  \bibfield  {author} {\bibinfo {author} {\bibfnamefont {C.~N.}\ \bibnamefont
  {Yang}}\ and\ \bibinfo {author} {\bibfnamefont {T.~D.}\ \bibnamefont {Lee}},\
  }\href {\doibase 10.1103/PhysRev.87.404} {\bibfield  {journal} {\bibinfo
  {journal} {Phys. Rev.}\ }\textbf {\bibinfo {volume} {87}},\ \bibinfo {pages}
  {404 and 410} (\bibinfo {year} {1952})}\BibitemShut {NoStop}%
\bibitem [{\citenamefont {Guralnik}\ and\ \citenamefont
  {Guralnik}(2010)}]{guralnik:0710.1256}%
  \BibitemOpen
  \bibfield  {author} {\bibinfo {author} {\bibfnamefont {G.}~\bibnamefont
  {Guralnik}}\ and\ \bibinfo {author} {\bibfnamefont {Z.}~\bibnamefont
  {Guralnik}},\ }\href {\doibase http://dx.doi.org/10.1016/j.aop.2010.06.001}
  {\bibfield  {journal} {\bibinfo  {journal} {Annals of Physics}\ }\textbf
  {\bibinfo {volume} {325}},\ \bibinfo {pages} {2486 } (\bibinfo {year}
  {2010})}\BibitemShut {NoStop}%
\bibitem [{\citenamefont {Kortman}\ and\ \citenamefont
  {Griffiths}(1971)}]{kortman-1971}%
  \BibitemOpen
  \bibfield  {author} {\bibinfo {author} {\bibfnamefont {P.~J.}\ \bibnamefont
  {Kortman}}\ and\ \bibinfo {author} {\bibfnamefont {R.~B.}\ \bibnamefont
  {Griffiths}},\ }\href@noop {} {\bibfield  {journal} {\bibinfo  {journal}
  {Phys. Rev. Lett.}\ }\textbf {\bibinfo {volume} {\bf 27}},\ \bibinfo {pages}
  {1439} (\bibinfo {year} {1971})}\BibitemShut {NoStop}%
\bibitem [{\citenamefont {Fisher}(1978)}]{fisher-1978}%
  \BibitemOpen
  \bibfield  {author} {\bibinfo {author} {\bibfnamefont {M.~E.}\ \bibnamefont
  {Fisher}},\ }\href@noop {} {\bibfield  {journal} {\bibinfo  {journal} {Phys.
  Rev. Lett.}\ }\textbf {\bibinfo {volume} {\bf 40}},\ \bibinfo {pages} {1610}
  (\bibinfo {year} {1978})}\BibitemShut {NoStop}%
\bibitem [{\citenamefont {Ryu}\ \emph {et~al.}(2010)\citenamefont {Ryu},
  \citenamefont {Schnyder}, \citenamefont {Furusaki},\ and\ \citenamefont
  {Ludwig}}]{ryuschnyderfurusakiludwig}%
  \BibitemOpen
  \bibfield  {author} {\bibinfo {author} {\bibfnamefont {S.}~\bibnamefont
  {Ryu}}, \bibinfo {author} {\bibfnamefont {A.~P.}\ \bibnamefont {Schnyder}},
  \bibinfo {author} {\bibfnamefont {A.}~\bibnamefont {Furusaki}}, \ and\
  \bibinfo {author} {\bibfnamefont {A.~W.~W.}\ \bibnamefont {Ludwig}},\
  }\href@noop {} {\bibfield  {journal} {\bibinfo  {journal} {New J. Phys.}\
  }\textbf {\bibinfo {volume} {\bf 12}},\ \bibinfo {pages} {065010} (\bibinfo
  {year} {2010})}\BibitemShut {NoStop}%
\bibitem [{\citenamefont {LeClair}\ and\ \citenamefont
  {Bernard}(2012)}]{leclairbernard}%
  \BibitemOpen
  \bibfield  {author} {\bibinfo {author} {\bibfnamefont {A.}~\bibnamefont
  {LeClair}}\ and\ \bibinfo {author} {\bibfnamefont {D.}~\bibnamefont
  {Bernard}},\ }\href@noop {} {\bibfield  {journal} {\bibinfo  {journal} {J.
  Phys. A: Math. Theor.}\ }\textbf {\bibinfo {volume} {\bf 45}},\ \bibinfo
  {pages} {435203} (\bibinfo {year} {2012})}\BibitemShut {NoStop}%
\bibitem [{\citenamefont {Fu}\ and\ \citenamefont {Kane}(2009)}]{fukaneJJ}%
  \BibitemOpen
  \bibfield  {author} {\bibinfo {author} {\bibfnamefont {L.}~\bibnamefont
  {Fu}}\ and\ \bibinfo {author} {\bibfnamefont {C.~L.}\ \bibnamefont {Kane}},\
  }\href@noop {} {\bibfield  {journal} {\bibinfo  {journal} {Phys. Rev. B}\
  }\textbf {\bibinfo {volume} {\bf 79}},\ \bibinfo {pages} {161408(R)}
  (\bibinfo {year} {2009})}\BibitemShut {NoStop}%
\bibitem [{\citenamefont {Mohanta}\ \emph {et~al.}(2014)\citenamefont
  {Mohanta}, \citenamefont {Bandopadhyay}, \citenamefont {Lal},\ and\
  \citenamefont {Taraphder}}]{arxiv:1407.6539}%
  \BibitemOpen
  \bibfield  {author} {\bibinfo {author} {\bibfnamefont {N.}~\bibnamefont
  {Mohanta}}, \bibinfo {author} {\bibfnamefont {S.}~\bibnamefont
  {Bandopadhyay}}, \bibinfo {author} {\bibfnamefont {S.}~\bibnamefont {Lal}}, \
  and\ \bibinfo {author} {\bibfnamefont {A.}~\bibnamefont {Taraphder}},\ }\href
  {http://arxiv.org/abs/1407.6539} {\enquote {\bibinfo {title} {Emergent spin
  hall phase at a lifshitz transition from competing orders},}\ } (\bibinfo
  {year} {2014}),\ \bibinfo {note} {arxiv:1407.6539}\BibitemShut {NoStop}%
\bibitem [{\citenamefont {Stone}(1985)}]{stone-1985}%
  \BibitemOpen
  \bibfield  {author} {\bibinfo {author} {\bibfnamefont {M.}~\bibnamefont
  {Stone}},\ }\href@noop {} {\bibfield  {journal} {\bibinfo  {journal} {Phys.
  Rev. B}\ }\textbf {\bibinfo {volume} {\bf 31}},\ \bibinfo {pages} {6112}
  (\bibinfo {year} {1985})}\BibitemShut {NoStop}%
\bibitem [{\citenamefont {Lee}\ \emph {et~al.}(2007)\citenamefont {Lee},
  \citenamefont {Zhang},\ and\ \citenamefont {Xiang}}]{dhlee-2007}%
  \BibitemOpen
  \bibfield  {author} {\bibinfo {author} {\bibfnamefont {D.-H.}\ \bibnamefont
  {Lee}}, \bibinfo {author} {\bibfnamefont {G.-M.}\ \bibnamefont {Zhang}}, \
  and\ \bibinfo {author} {\bibfnamefont {T.}~\bibnamefont {Xiang}},\
  }\href@noop {} {\bibfield  {journal} {\bibinfo  {journal} {Phys. Rev. Lett.}\
  }\textbf {\bibinfo {volume} {99}},\ \bibinfo {pages} {196805} (\bibinfo
  {year} {2007})}\BibitemShut {NoStop}%
\bibitem [{\citenamefont {Atiyah}\ and\ \citenamefont
  {Singer}(1968{\natexlab{a}})}]{atiyahsingera}%
  \BibitemOpen
  \bibfield  {author} {\bibinfo {author} {\bibfnamefont {M.~F.}\ \bibnamefont
  {Atiyah}}\ and\ \bibinfo {author} {\bibfnamefont {I.~M.}\ \bibnamefont
  {Singer}},\ }\href@noop {} {\bibfield  {journal} {\bibinfo  {journal} {Ann.
  Math.}\ }\textbf {\bibinfo {volume} {\bf 87}},\ \bibinfo {pages} {485}
  (\bibinfo {year} {1968}{\natexlab{a}})}\BibitemShut {NoStop}%
\bibitem [{\citenamefont {Atiyah}\ and\ \citenamefont
  {Singer}(1968{\natexlab{b}})}]{atiyahsingerb}%
  \BibitemOpen
  \bibfield  {author} {\bibinfo {author} {\bibfnamefont {M.~F.}\ \bibnamefont
  {Atiyah}}\ and\ \bibinfo {author} {\bibfnamefont {I.~M.}\ \bibnamefont
  {Singer}},\ }\href@noop {} {\bibfield  {journal} {\bibinfo  {journal} {Ann.
  Math.}\ }\textbf {\bibinfo {volume} {\bf 87}},\ \bibinfo {pages} {546}
  (\bibinfo {year} {1968}{\natexlab{b}})}\BibitemShut {NoStop}%
\bibitem [{\citenamefont {Akhoury}\ and\ \citenamefont
  {Comtet}(1986)}]{akhourycomtet-1986}%
  \BibitemOpen
  \bibfield  {author} {\bibinfo {author} {\bibfnamefont {R.}~\bibnamefont
  {Akhoury}}\ and\ \bibinfo {author} {\bibfnamefont {A.}~\bibnamefont
  {Comtet}},\ }\href@noop {} {\bibfield  {journal} {\bibinfo  {journal} {Ann.
  Phys.}\ }\textbf {\bibinfo {volume} {172}},\ \bibinfo {pages} {245} (\bibinfo
  {year} {1986})}\BibitemShut {NoStop}%
\bibitem [{\citenamefont {Kubo}(1957)}]{kubo-1957}%
  \BibitemOpen
  \bibfield  {author} {\bibinfo {author} {\bibfnamefont {R.}~\bibnamefont
  {Kubo}},\ }\href@noop {} {\bibfield  {journal} {\bibinfo  {journal} {J. Phys.
  Soc. Jpn.}\ }\textbf {\bibinfo {volume} {12}},\ \bibinfo {pages} {570}
  (\bibinfo {year} {1957})}\BibitemShut {NoStop}%
\bibitem [{\citenamefont {Martin}\ and\ \citenamefont
  {Schwinger}(1959)}]{martinschwinger-1959}%
  \BibitemOpen
  \bibfield  {author} {\bibinfo {author} {\bibfnamefont {P.~C.}\ \bibnamefont
  {Martin}}\ and\ \bibinfo {author} {\bibfnamefont {J.}~\bibnamefont
  {Schwinger}},\ }\href@noop {} {\bibfield  {journal} {\bibinfo  {journal}
  {Phys. Rev.}\ }\textbf {\bibinfo {volume} {115}},\ \bibinfo {pages} {1342}
  (\bibinfo {year} {1959})}\BibitemShut {NoStop}%
\bibitem [{\citenamefont {Perk}\ and\ \citenamefont
  {Au-Yang}(2009)}]{perkyang-2009}%
  \BibitemOpen
  \bibfield  {author} {\bibinfo {author} {\bibfnamefont {J.~H.~H.}\
  \bibnamefont {Perk}}\ and\ \bibinfo {author} {\bibfnamefont {H.}~\bibnamefont
  {Au-Yang}},\ }\href@noop {} {\bibfield  {journal} {\bibinfo  {journal} {J.
  Stat. Phys.}\ }\textbf {\bibinfo {volume} {135}},\ \bibinfo {pages} {599}
  (\bibinfo {year} {2009})}\BibitemShut {NoStop}%
\bibitem [{\citenamefont {Kitaev}(2001)}]{kitaevmajoranachain}%
  \BibitemOpen
  \bibfield  {author} {\bibinfo {author} {\bibfnamefont {A.}~\bibnamefont
  {Kitaev}},\ }\href@noop {} {\bibfield  {journal} {\bibinfo  {journal} {Phys.
  Usp.}\ }\textbf {\bibinfo {volume} {\bf 44}},\ \bibinfo {pages} {131}
  (\bibinfo {year} {2001})}\BibitemShut {NoStop}%
\bibitem [{\citenamefont {Bogolubov}\ and\ \citenamefont
  {Bogolubov~Jr.}(2010)}]{bogoliubov-book}%
  \BibitemOpen
  \bibfield  {author} {\bibinfo {author} {\bibfnamefont {N.~N.}\ \bibnamefont
  {Bogolubov}}\ and\ \bibinfo {author} {\bibfnamefont {N.~N.}\ \bibnamefont
  {Bogolubov~Jr.}},\ }\href@noop {} {\emph {\bibinfo {title} {Introduction to
  Quantum Statistical Mechanics}}}\ (\bibinfo  {publisher} {Second Edition,
  World Scientific},\ \bibinfo {year} {2010})\BibitemShut {NoStop}%
\bibitem [{\citenamefont {Scalapino}\ \emph {et~al.}(1993)\citenamefont
  {Scalapino}, \citenamefont {White},\ and\ \citenamefont {Zhang}}]{swz-1993}%
  \BibitemOpen
  \bibfield  {author} {\bibinfo {author} {\bibfnamefont {D.~J.}\ \bibnamefont
  {Scalapino}}, \bibinfo {author} {\bibfnamefont {S.~R.}\ \bibnamefont
  {White}}, \ and\ \bibinfo {author} {\bibfnamefont {S.~C.}\ \bibnamefont
  {Zhang}},\ }\href@noop {} {\bibfield  {journal} {\bibinfo  {journal} {Phys.
  Rev. B}\ }\textbf {\bibinfo {volume} {\bf 47}},\ \bibinfo {pages} {7995}
  (\bibinfo {year} {1993})}\BibitemShut {NoStop}%
\bibitem [{\citenamefont {Leggett}\ \emph {et~al.}(1987)\citenamefont
  {Leggett}, \citenamefont {Chakravarty}, \citenamefont {Dorsey}, \citenamefont
  {Fisher}, \citenamefont {Garg},\ and\ \citenamefont
  {Zwerger}}]{leggettreview-1987}%
  \BibitemOpen
  \bibfield  {author} {\bibinfo {author} {\bibfnamefont {A.~J.}\ \bibnamefont
  {Leggett}}, \bibinfo {author} {\bibfnamefont {S.}~\bibnamefont
  {Chakravarty}}, \bibinfo {author} {\bibfnamefont {A.~T.}\ \bibnamefont
  {Dorsey}}, \bibinfo {author} {\bibfnamefont {M.~P.~A.}\ \bibnamefont
  {Fisher}}, \bibinfo {author} {\bibfnamefont {A.}~\bibnamefont {Garg}}, \ and\
  \bibinfo {author} {\bibfnamefont {W.}~\bibnamefont {Zwerger}},\ }\href@noop
  {} {\bibfield  {journal} {\bibinfo  {journal} {Rev. Mod. Phys.}\ }\textbf
  {\bibinfo {volume} {59}},\ \bibinfo {pages} {4} (\bibinfo {year}
  {1987})}\BibitemShut {NoStop}%
\bibitem [{\citenamefont {Chakravarty}(1982)}]{chakravarty-1982}%
  \BibitemOpen
  \bibfield  {author} {\bibinfo {author} {\bibfnamefont {S.}~\bibnamefont
  {Chakravarty}},\ }\href@noop {} {\bibfield  {journal} {\bibinfo  {journal}
  {Phys. Rev. Lett.}\ }\textbf {\bibinfo {volume} {49}},\ \bibinfo {pages}
  {681} (\bibinfo {year} {1982})}\BibitemShut {NoStop}%
\bibitem [{\citenamefont {Bray}\ and\ \citenamefont
  {Moore}(1982)}]{braymoore-1982}%
  \BibitemOpen
  \bibfield  {author} {\bibinfo {author} {\bibfnamefont {A.~J.}\ \bibnamefont
  {Bray}}\ and\ \bibinfo {author} {\bibfnamefont {M.~A.}\ \bibnamefont
  {Moore}},\ }\href@noop {} {\bibfield  {journal} {\bibinfo  {journal} {Phys.
  Rev. Lett.}\ }\textbf {\bibinfo {volume} {49}},\ \bibinfo {pages} {1545}
  (\bibinfo {year} {1982})}\BibitemShut {NoStop}%
\bibitem [{\citenamefont {Berezinskii}(1971)}]{BKT1}%
  \BibitemOpen
  \bibfield  {author} {\bibinfo {author} {\bibfnamefont {V.~L.}\ \bibnamefont
  {Berezinskii}},\ }\href@noop {} {\bibfield  {journal} {\bibinfo  {journal}
  {Sov. Phys. JETP}\ }\textbf {\bibinfo {volume} {\bf 32}},\ \bibinfo {pages}
  {493} (\bibinfo {year} {1971})}\BibitemShut {NoStop}%
\bibitem [{\citenamefont {Kosterlitz}\ and\ \citenamefont
  {Thouless}(1973)}]{BKT2}%
  \BibitemOpen
  \bibfield  {author} {\bibinfo {author} {\bibfnamefont {J.~M.}\ \bibnamefont
  {Kosterlitz}}\ and\ \bibinfo {author} {\bibfnamefont {D.~J.}\ \bibnamefont
  {Thouless}},\ }\href@noop {} {\bibfield  {journal} {\bibinfo  {journal} {J.
  Phys. C}\ }\textbf {\bibinfo {volume} {\bf 6}},\ \bibinfo {pages} {1181}
  (\bibinfo {year} {1973})}\BibitemShut {NoStop}%
\bibitem [{\citenamefont {Bardyn}\ \emph {et~al.}(2013)\citenamefont {Bardyn},
  \citenamefont {Baranov}, \citenamefont {Kraus}, \citenamefont {Rico},
  \citenamefont {Imam\u{o}glu}, \citenamefont {Zoller},\ and\ \citenamefont
  {Diehl}}]{bardyn-2013}%
  \BibitemOpen
  \bibfield  {author} {\bibinfo {author} {\bibfnamefont {C.-E.}\ \bibnamefont
  {Bardyn}}, \bibinfo {author} {\bibfnamefont {M.~A.}\ \bibnamefont {Baranov}},
  \bibinfo {author} {\bibfnamefont {C.~V.}\ \bibnamefont {Kraus}}, \bibinfo
  {author} {\bibfnamefont {E.}~\bibnamefont {Rico}}, \bibinfo {author}
  {\bibfnamefont {A.}~\bibnamefont {Imam\u{o}glu}}, \bibinfo {author}
  {\bibfnamefont {P.}~\bibnamefont {Zoller}}, \ and\ \bibinfo {author}
  {\bibfnamefont {S.}~\bibnamefont {Diehl}},\ }\href@noop {} {\bibfield
  {journal} {\bibinfo  {journal} {New Journal of Physics}\ }\textbf {\bibinfo
  {volume} {15}},\ \bibinfo {pages} {085001} (\bibinfo {year}
  {2013})}\BibitemShut {NoStop}%
\bibitem [{\citenamefont {Manton}(1985)}]{manton-1985}%
  \BibitemOpen
  \bibfield  {author} {\bibinfo {author} {\bibfnamefont {N.~S.}\ \bibnamefont
  {Manton}},\ }\href@noop {} {\bibfield  {journal} {\bibinfo  {journal} {Annals
  of Physics}\ }\textbf {\bibinfo {volume} {159}},\ \bibinfo {pages} {220}
  (\bibinfo {year} {1985})}\BibitemShut {NoStop}%
\bibitem [{\citenamefont {Thacker}(2014)}]{thacker-2014}%
  \BibitemOpen
  \bibfield  {author} {\bibinfo {author} {\bibfnamefont {H.~B.}\ \bibnamefont
  {Thacker}},\ }\href {\doibase 10.1103/PhysRevD.89.125011} {\bibfield
  {journal} {\bibinfo  {journal} {Phys. Rev. D}\ }\textbf {\bibinfo {volume}
  {89}},\ \bibinfo {pages} {125011} (\bibinfo {year} {2014})}\BibitemShut
  {NoStop}%
\bibitem [{\citenamefont {Thouless}\ \emph {et~al.}(1982)\citenamefont
  {Thouless}, \citenamefont {Kohmoto}, \citenamefont {Nightingale},\ and\
  \citenamefont {den Nijs}}]{thouless}%
  \BibitemOpen
  \bibfield  {author} {\bibinfo {author} {\bibfnamefont {D.~J.}\ \bibnamefont
  {Thouless}}, \bibinfo {author} {\bibfnamefont {M.}~\bibnamefont {Kohmoto}},
  \bibinfo {author} {\bibfnamefont {M.~P.}\ \bibnamefont {Nightingale}}, \ and\
  \bibinfo {author} {\bibfnamefont {M.}~\bibnamefont {den Nijs}},\ }\href
  {\doibase 10.1103/PhysRevLett.49.405} {\bibfield  {journal} {\bibinfo
  {journal} {Phys. Rev. Lett.}\ }\textbf {\bibinfo {volume} {49}},\ \bibinfo
  {pages} {405} (\bibinfo {year} {1982})}\BibitemShut {NoStop}%
\bibitem [{\citenamefont {Cayssol}\ \emph {et~al.}(2013)\citenamefont
  {Cayssol}, \citenamefont {Dora}, \citenamefont {Simon},\ and\ \citenamefont
  {Moessner}}]{cayssol-2013}%
  \BibitemOpen
  \bibfield  {author} {\bibinfo {author} {\bibfnamefont {J.}~\bibnamefont
  {Cayssol}}, \bibinfo {author} {\bibfnamefont {B.}~\bibnamefont {Dora}},
  \bibinfo {author} {\bibfnamefont {F.}~\bibnamefont {Simon}}, \ and\ \bibinfo
  {author} {\bibfnamefont {R.}~\bibnamefont {Moessner}},\ }\href@noop {}
  {\bibfield  {journal} {\bibinfo  {journal} {Phys. Stat. Solidi RRL}\ }\textbf
  {\bibinfo {volume} {7}},\ \bibinfo {pages} {101} (\bibinfo {year}
  {2013})}\BibitemShut {NoStop}%
\bibitem [{\citenamefont {Laughlin}(1981)}]{laughlin-1981}%
  \BibitemOpen
  \bibfield  {author} {\bibinfo {author} {\bibfnamefont {R.~B.}\ \bibnamefont
  {Laughlin}},\ }\href@noop {} {\bibfield  {journal} {\bibinfo  {journal}
  {Phys. Rev. B}\ }\textbf {\bibinfo {volume} {23}},\ \bibinfo {pages} {5632}
  (\bibinfo {year} {1981})}\BibitemShut {NoStop}%
\bibitem [{\citenamefont {Halperin}(1982)}]{halperin-1982}%
  \BibitemOpen
  \bibfield  {author} {\bibinfo {author} {\bibfnamefont {B.~I.}\ \bibnamefont
  {Halperin}},\ }\href@noop {} {\bibfield  {journal} {\bibinfo  {journal}
  {Phys. Rev. B}\ }\textbf {\bibinfo {volume} {25}},\ \bibinfo {pages} {2185}
  (\bibinfo {year} {1982})}\BibitemShut {NoStop}%
\bibitem [{\citenamefont {Tao}\ and\ \citenamefont
  {Haldane}(1986)}]{tao-haldane}%
  \BibitemOpen
  \bibfield  {author} {\bibinfo {author} {\bibfnamefont {R.}~\bibnamefont
  {Tao}}\ and\ \bibinfo {author} {\bibfnamefont {F.~D.~M.}\ \bibnamefont
  {Haldane}},\ }\href@noop {} {\bibfield  {journal} {\bibinfo  {journal} {Phys.
  Rev. B}\ }\textbf {\bibinfo {volume} {33}},\ \bibinfo {pages} {3844}
  (\bibinfo {year} {1986})}\BibitemShut {NoStop}%
\bibitem [{\citenamefont {Stone}(1991)}]{stone-1991}%
  \BibitemOpen
  \bibfield  {author} {\bibinfo {author} {\bibfnamefont {M.}~\bibnamefont
  {Stone}},\ }\href@noop {} {\bibfield  {journal} {\bibinfo  {journal} {Annals
  of Physics}\ }\textbf {\bibinfo {volume} {207}},\ \bibinfo {pages} {38}
  (\bibinfo {year} {1991})}\BibitemShut {NoStop}%
\bibitem [{\citenamefont {Callan}\ and\ \citenamefont
  {Harvey}(1985)}]{callanharvey-1985}%
  \BibitemOpen
  \bibfield  {author} {\bibinfo {author} {\bibfnamefont {C.~G.}\ \bibnamefont
  {Callan}, \bibfnamefont {Jr.}}\ and\ \bibinfo {author} {\bibfnamefont
  {J.~A.}\ \bibnamefont {Harvey}},\ }\href@noop {} {\bibfield  {journal}
  {\bibinfo  {journal} {Nucl. Phys. B}\ }\textbf {\bibinfo {volume} {250}},\
  \bibinfo {pages} {427} (\bibinfo {year} {1985})}\BibitemShut {NoStop}%
\bibitem [{\citenamefont {Hatsugai}(1993)}]{hatsugai}%
  \BibitemOpen
  \bibfield  {author} {\bibinfo {author} {\bibfnamefont {Y.}~\bibnamefont
  {Hatsugai}},\ }\href@noop {} {\bibfield  {journal} {\bibinfo  {journal}
  {Phys. Rev. B}\ }\textbf {\bibinfo {volume} {\bf 48}},\ \bibinfo {pages}
  {11851} (\bibinfo {year} {1993})}\BibitemShut {NoStop}%
\bibitem [{\citenamefont {Ishikawa}\ and\ \citenamefont
  {Matsuyama}(1987)}]{ishikawamatsuyama}%
  \BibitemOpen
  \bibfield  {author} {\bibinfo {author} {\bibfnamefont {K.}~\bibnamefont
  {Ishikawa}}\ and\ \bibinfo {author} {\bibfnamefont {T.}~\bibnamefont
  {Matsuyama}},\ }\href@noop {} {\bibfield  {journal} {\bibinfo  {journal}
  {Nucl. Phys. B}\ }\textbf {\bibinfo {volume} {\bf 280}},\ \bibinfo {pages}
  {523} (\bibinfo {year} {1987})}\BibitemShut {NoStop}%
\bibitem [{\citenamefont {Wegner}(1994)}]{wegner-1994}%
  \BibitemOpen
  \bibfield  {author} {\bibinfo {author} {\bibfnamefont {F.}~\bibnamefont
  {Wegner}},\ }\href@noop {} {\bibfield  {journal} {\bibinfo  {journal} {Ann.
  Phys. (Leipzig)}\ }\textbf {\bibinfo {volume} {3}},\ \bibinfo {pages} {77}
  (\bibinfo {year} {1994})}\BibitemShut {NoStop}%
\bibitem [{\citenamefont {Kehrein}(2006)}]{kehrein-book}%
  \BibitemOpen
  \bibfield  {author} {\bibinfo {author} {\bibfnamefont {S.}~\bibnamefont
  {Kehrein}},\ }\href@noop {} {\emph {\bibinfo {title} {The Flow Equation
  Approach to Many-Particle Systems}}}\ (\bibinfo  {publisher} {Springer
  Verlag},\ \bibinfo {year} {2006})\BibitemShut {NoStop}%
\bibitem [{\citenamefont {Correa}\ and\ \citenamefont
  {Plyushchay}(2007)}]{correaplyushchay-2007}%
  \BibitemOpen
  \bibfield  {author} {\bibinfo {author} {\bibfnamefont {F.}~\bibnamefont
  {Correa}}\ and\ \bibinfo {author} {\bibfnamefont {M.~S.}\ \bibnamefont
  {Plyushchay}},\ }\href@noop {} {\bibfield  {journal} {\bibinfo  {journal}
  {Ann. Phys.}\ }\textbf {\bibinfo {volume} {322}},\ \bibinfo {pages} {2493}
  (\bibinfo {year} {2007})}\BibitemShut {NoStop}%
\bibitem [{\citenamefont {Rau}(2004)}]{rau-2004}%
  \BibitemOpen
  \bibfield  {author} {\bibinfo {author} {\bibfnamefont {A.~R.~P.}\
  \bibnamefont {Rau}},\ }\href@noop {} {\bibfield  {journal} {\bibinfo
  {journal} {J. Phys. A: Math. Gen.}\ }\textbf {\bibinfo {volume} {37}},\
  \bibinfo {pages} {10421} (\bibinfo {year} {2004})}\BibitemShut {NoStop}%
\bibitem [{\citenamefont {Witten}(1982)}]{witten-1982}%
  \BibitemOpen
  \bibfield  {author} {\bibinfo {author} {\bibfnamefont {E.}~\bibnamefont
  {Witten}},\ }\href@noop {} {\bibfield  {journal} {\bibinfo  {journal} {Nucl.
  Phys. B}\ }\textbf {\bibinfo {volume} {202}},\ \bibinfo {pages} {253}
  (\bibinfo {year} {1982})}\BibitemShut {NoStop}%
\bibitem [{\citenamefont {Bolle}\ \emph {et~al.}(1987)\citenamefont {Bolle},
  \citenamefont {Gesztesy}, \citenamefont {Grosse},\ and\ \citenamefont
  {Simon}}]{bollesimon-1987}%
  \BibitemOpen
  \bibfield  {author} {\bibinfo {author} {\bibfnamefont {D.}~\bibnamefont
  {Bolle}}, \bibinfo {author} {\bibfnamefont {F.}~\bibnamefont {Gesztesy}},
  \bibinfo {author} {\bibfnamefont {H.}~\bibnamefont {Grosse}}, \ and\ \bibinfo
  {author} {\bibfnamefont {B.}~\bibnamefont {Simon}},\ }\href@noop {}
  {\bibfield  {journal} {\bibinfo  {journal} {J. Math. Phys.}\ }\textbf
  {\bibinfo {volume} {28}},\ \bibinfo {pages} {1512} (\bibinfo {year}
  {1987})}\BibitemShut {NoStop}%
\bibitem [{\citenamefont {Lewis}\ and\ \citenamefont
  {Sisson}(1975)}]{lewississon-1975}%
  \BibitemOpen
  \bibfield  {author} {\bibinfo {author} {\bibfnamefont {J.~T.}\ \bibnamefont
  {Lewis}}\ and\ \bibinfo {author} {\bibfnamefont {P.~N.~M.}\ \bibnamefont
  {Sisson}},\ }\href@noop {} {\bibfield  {journal} {\bibinfo  {journal}
  {Commun. Math. Phys.}\ }\textbf {\bibinfo {volume} {44}},\ \bibinfo {pages}
  {279} (\bibinfo {year} {1975})}\BibitemShut {NoStop}%
\bibitem [{\citenamefont {Asorey}\ \emph {et~al.}(1983)\citenamefont {Asorey},
  \citenamefont {Esteve},\ and\ \citenamefont {Pacheco}}]{asorey-1983}%
  \BibitemOpen
  \bibfield  {author} {\bibinfo {author} {\bibfnamefont {M.}~\bibnamefont
  {Asorey}}, \bibinfo {author} {\bibfnamefont {J.~G.}\ \bibnamefont {Esteve}},
  \ and\ \bibinfo {author} {\bibfnamefont {A.~F.}\ \bibnamefont {Pacheco}},\
  }\href@noop {} {\bibfield  {journal} {\bibinfo  {journal} {Phys. Rev. D}\
  }\textbf {\bibinfo {volume} {27}},\ \bibinfo {pages} {1852} (\bibinfo {year}
  {1983})}\BibitemShut {NoStop}%
\bibitem [{\citenamefont {Rajaraman}(1982)}]{rajaraman}%
  \BibitemOpen
  \bibfield  {author} {\bibinfo {author} {\bibfnamefont {R.}~\bibnamefont
  {Rajaraman}},\ }\href@noop {} {\emph {\bibinfo {title} {Solitons and
  Instantons: An Introduction to Solitons and Instantons in Quantum Field
  Theory}}}\ (\bibinfo  {publisher} {North-Holland},\ \bibinfo {year}
  {1982})\BibitemShut {NoStop}%
\bibitem [{\citenamefont {Apenko}(2008)}]{apenko-2008}%
  \BibitemOpen
  \bibfield  {author} {\bibinfo {author} {\bibfnamefont {S.~M.}\ \bibnamefont
  {Apenko}},\ }\href@noop {} {\bibfield  {journal} {\bibinfo  {journal} {J.
  Phys. A: Math. Theor.}\ }\textbf {\bibinfo {volume} {41}},\ \bibinfo {pages}
  {315301} (\bibinfo {year} {2008})}\BibitemShut {NoStop}%
\bibitem [{\citenamefont {Bulgadaev}(2006)}]{bulgadaev-2006}%
  \BibitemOpen
  \bibfield  {author} {\bibinfo {author} {\bibfnamefont {S.~A.}\ \bibnamefont
  {Bulgadaev}},\ }\href@noop {} {\bibfield  {journal} {\bibinfo  {journal}
  {JETP Letters}\ }\textbf {\bibinfo {volume} {83}},\ \bibinfo {pages} {563}
  (\bibinfo {year} {2006})}\BibitemShut {NoStop}%
\bibitem [{\citenamefont {Bogachek}\ \emph {et~al.}(1990)\citenamefont
  {Bogachek}, \citenamefont {Krive}, \citenamefont {Kulik},\ and\ \citenamefont
  {Rozhavsky}}]{bogachek-1990}%
  \BibitemOpen
  \bibfield  {author} {\bibinfo {author} {\bibfnamefont {E.~N.}\ \bibnamefont
  {Bogachek}}, \bibinfo {author} {\bibfnamefont {I.~V.}\ \bibnamefont {Krive}},
  \bibinfo {author} {\bibfnamefont {I.~O.}\ \bibnamefont {Kulik}}, \ and\
  \bibinfo {author} {\bibfnamefont {A.~S.}\ \bibnamefont {Rozhavsky}},\
  }\href@noop {} {\bibfield  {journal} {\bibinfo  {journal} {Phys. Rev. B}\
  }\textbf {\bibinfo {volume} {42}},\ \bibinfo {pages} {7614} (\bibinfo {year}
  {1990})}\BibitemShut {NoStop}%
\bibitem [{\citenamefont {Bermudez}\ \emph {et~al.}(2010)\citenamefont
  {Bermudez}, \citenamefont {Amico},\ and\ \citenamefont
  {Martin-Delgado}}]{martindelgado-2010}%
  \BibitemOpen
  \bibfield  {author} {\bibinfo {author} {\bibfnamefont {A.}~\bibnamefont
  {Bermudez}}, \bibinfo {author} {\bibfnamefont {L.}~\bibnamefont {Amico}}, \
  and\ \bibinfo {author} {\bibfnamefont {M.~A.}\ \bibnamefont
  {Martin-Delgado}},\ }\href@noop {} {\bibfield  {journal} {\bibinfo  {journal}
  {New J. Phys.}\ }\textbf {\bibinfo {volume} {12}},\ \bibinfo {pages} {055014}
  (\bibinfo {year} {2010})}\BibitemShut {NoStop}%
\bibitem [{\citenamefont {Kibble}(1976)}]{kibble-1976}%
  \BibitemOpen
  \bibfield  {author} {\bibinfo {author} {\bibfnamefont {T.~W.~B.}\
  \bibnamefont {Kibble}},\ }\href@noop {} {\bibfield  {journal} {\bibinfo
  {journal} {J. Phys. A}\ }\textbf {\bibinfo {volume} {9}},\ \bibinfo {pages}
  {1387} (\bibinfo {year} {1976})}\BibitemShut {NoStop}%
\bibitem [{\citenamefont {Zurek}(1985)}]{zurek-1985}%
  \BibitemOpen
  \bibfield  {author} {\bibinfo {author} {\bibfnamefont {W.~H.}\ \bibnamefont
  {Zurek}},\ }\href@noop {} {\bibfield  {journal} {\bibinfo  {journal} {Nature
  (London)}\ }\textbf {\bibinfo {volume} {317}},\ \bibinfo {pages} {505}
  (\bibinfo {year} {1985})}\BibitemShut {NoStop}%
\bibitem [{\citenamefont {Zeng}\ and\ \citenamefont {Wen}(2015)}]{zengwen}%
  \BibitemOpen
  \bibfield  {author} {\bibinfo {author} {\bibfnamefont {B.}~\bibnamefont
  {Zeng}}\ and\ \bibinfo {author} {\bibfnamefont {X.-G.}\ \bibnamefont {Wen}},\
  }\href {\doibase 10.1103/PhysRevB.91.125121} {\bibfield  {journal} {\bibinfo
  {journal} {Phys. Rev. B}\ }\textbf {\bibinfo {volume} {91}},\ \bibinfo
  {pages} {125121} (\bibinfo {year} {2015})}\BibitemShut {NoStop}%
\bibitem [{\citenamefont {Amico}\ \emph {et~al.}(2008)\citenamefont {Amico},
  \citenamefont {Fazio}, \citenamefont {Osterloh},\ and\ \citenamefont
  {Vedral}}]{amicovedral-2008}%
  \BibitemOpen
  \bibfield  {author} {\bibinfo {author} {\bibfnamefont {L.}~\bibnamefont
  {Amico}}, \bibinfo {author} {\bibfnamefont {R.}~\bibnamefont {Fazio}},
  \bibinfo {author} {\bibfnamefont {A.}~\bibnamefont {Osterloh}}, \ and\
  \bibinfo {author} {\bibfnamefont {V.}~\bibnamefont {Vedral}},\ }\href@noop {}
  {\bibfield  {journal} {\bibinfo  {journal} {Rev. Mod. Phys.}\ }\textbf
  {\bibinfo {volume} {80}},\ \bibinfo {pages} {517} (\bibinfo {year}
  {2008})}\BibitemShut {NoStop}%
\bibitem [{\citenamefont {Ryu}\ and\ \citenamefont
  {Hatsugai}(2006)}]{ryuhatsugai-2006}%
  \BibitemOpen
  \bibfield  {author} {\bibinfo {author} {\bibfnamefont {S.}~\bibnamefont
  {Ryu}}\ and\ \bibinfo {author} {\bibfnamefont {Y.}~\bibnamefont {Hatsugai}},\
  }\href@noop {} {\bibfield  {journal} {\bibinfo  {journal} {Rev. Mod. Phys.}\
  }\textbf {\bibinfo {volume} {73}},\ \bibinfo {pages} {245115} (\bibinfo
  {year} {2006})}\BibitemShut {NoStop}%
\bibitem [{\citenamefont {Wootters}(2001)}]{wootters-2001}%
  \BibitemOpen
  \bibfield  {author} {\bibinfo {author} {\bibfnamefont {W.~K.}\ \bibnamefont
  {Wootters}},\ }\href@noop {} {\bibfield  {journal} {\bibinfo  {journal}
  {Quant. Inf. and Comp.}\ }\textbf {\bibinfo {volume} {1}},\ \bibinfo {pages}
  {27} (\bibinfo {year} {2001})}\BibitemShut {NoStop}%
\bibitem [{\citenamefont {Swingle}(2010)}]{swingle-2010}%
  \BibitemOpen
  \bibfield  {author} {\bibinfo {author} {\bibfnamefont {B.}~\bibnamefont
  {Swingle}},\ }\href@noop {} {\bibfield  {journal} {\bibinfo  {journal} {Phys.
  Rev. Lett.}\ }\textbf {\bibinfo {volume} {105}},\ \bibinfo {pages} {050502}
  (\bibinfo {year} {2010})}\BibitemShut {NoStop}%
\bibitem [{\citenamefont {Calabrese}\ and\ \citenamefont
  {Cardy}(2004)}]{calabresecardy-2004}%
  \BibitemOpen
  \bibfield  {author} {\bibinfo {author} {\bibfnamefont {P.}~\bibnamefont
  {Calabrese}}\ and\ \bibinfo {author} {\bibfnamefont {J.}~\bibnamefont
  {Cardy}},\ }\href@noop {} {\bibfield  {journal} {\bibinfo  {journal} {J.
  Stat. Mech.: Theor. Exp.}\ }\textbf {\bibinfo {volume} {04}},\ \bibinfo
  {pages} {P06002} (\bibinfo {year} {2004})}\BibitemShut {NoStop}%
\bibitem [{\citenamefont {Zamolodchikov}(1988)}]{zamolodchikov}%
  \BibitemOpen
  \bibfield  {author} {\bibinfo {author} {\bibfnamefont {A.}~\bibnamefont
  {Zamolodchikov}},\ }\href@noop {} {\bibfield  {journal} {\bibinfo  {journal}
  {Int. J. Mod. Phys. A}\ }\textbf {\bibinfo {volume} {3}},\ \bibinfo {pages}
  {743} (\bibinfo {year} {1988})}\BibitemShut {NoStop}%
\bibitem [{\citenamefont {Bander}\ and\ \citenamefont
  {Itzykson}(1977)}]{banderitzykson}%
  \BibitemOpen
  \bibfield  {author} {\bibinfo {author} {\bibfnamefont {M.}~\bibnamefont
  {Bander}}\ and\ \bibinfo {author} {\bibfnamefont {C.}~\bibnamefont
  {Itzykson}},\ }\href@noop {} {\bibfield  {journal} {\bibinfo  {journal}
  {Phys. Rev. D}\ }\textbf {\bibinfo {volume} {\bf 15}},\ \bibinfo {pages}
  {463} (\bibinfo {year} {1977})}\BibitemShut {NoStop}%
\bibitem [{\citenamefont {Zuber}\ and\ \citenamefont
  {Itzykson}(1977)}]{zuberitzykson}%
  \BibitemOpen
  \bibfield  {author} {\bibinfo {author} {\bibfnamefont {J.~B.}\ \bibnamefont
  {Zuber}}\ and\ \bibinfo {author} {\bibfnamefont {C.}~\bibnamefont
  {Itzykson}},\ }\href@noop {} {\bibfield  {journal} {\bibinfo  {journal}
  {Phys. Rev. D}\ }\textbf {\bibinfo {volume} {\bf 15}},\ \bibinfo {pages}
  {2875} (\bibinfo {year} {1977})}\BibitemShut {NoStop}%
\bibitem [{\citenamefont {Witten}(1984)}]{witten}%
  \BibitemOpen
  \bibfield  {author} {\bibinfo {author} {\bibfnamefont {E.}~\bibnamefont
  {Witten}},\ }\href@noop {} {\bibfield  {journal} {\bibinfo  {journal}
  {Commun. Math. Phys.}\ }\textbf {\bibinfo {volume} {92}},\ \bibinfo {pages}
  {455} (\bibinfo {year} {1984})}\BibitemShut {NoStop}%
\bibitem [{\citenamefont {Knizhnik}\ and\ \citenamefont
  {Zamolodchikov}(1984)}]{knizhnik}%
  \BibitemOpen
  \bibfield  {author} {\bibinfo {author} {\bibfnamefont {V.}~\bibnamefont
  {Knizhnik}}\ and\ \bibinfo {author} {\bibfnamefont {A.}~\bibnamefont
  {Zamolodchikov}},\ }\href {\doibase
  http://dx.doi.org/10.1016/0550-3213(84)90374-2} {\bibfield  {journal}
  {\bibinfo  {journal} {Nuclear Physics B}\ }\textbf {\bibinfo {volume}
  {247}},\ \bibinfo {pages} {83 } (\bibinfo {year} {1984})}\BibitemShut
  {NoStop}%
\bibitem [{\citenamefont {Giamarchi}(2003)}]{giamarchi}%
  \BibitemOpen
  \bibfield  {author} {\bibinfo {author} {\bibfnamefont {T.}~\bibnamefont
  {Giamarchi}},\ }\href@noop {} {\emph {\bibinfo {title} {Quantum Physics in
  One Dimension}}}\ (\bibinfo  {publisher} {Oxford Science Publications},\
  \bibinfo {year} {2003})\BibitemShut {NoStop}%
\bibitem [{\citenamefont {Affleck}(1986)}]{affleck}%
  \BibitemOpen
  \bibfield  {author} {\bibinfo {author} {\bibfnamefont {I.}~\bibnamefont
  {Affleck}},\ }\href {\doibase 10.1103/PhysRevLett.56.408} {\bibfield
  {journal} {\bibinfo  {journal} {Phys. Rev. Lett.}\ }\textbf {\bibinfo
  {volume} {56}},\ \bibinfo {pages} {408} (\bibinfo {year} {1986})}\BibitemShut
  {NoStop}%
\bibitem [{\citenamefont {Tanaka}\ \emph {et~al.}(2009)\citenamefont {Tanaka},
  \citenamefont {Totsuka},\ and\ \citenamefont {Hu}}]{tanakatotsukahu}%
  \BibitemOpen
  \bibfield  {author} {\bibinfo {author} {\bibfnamefont {A.}~\bibnamefont
  {Tanaka}}, \bibinfo {author} {\bibfnamefont {K.}~\bibnamefont {Totsuka}}, \
  and\ \bibinfo {author} {\bibfnamefont {X.}~\bibnamefont {Hu}},\ }\href
  {\doibase 10.1103/PhysRevB.79.064412} {\bibfield  {journal} {\bibinfo
  {journal} {Phys. Rev. B}\ }\textbf {\bibinfo {volume} {79}},\ \bibinfo
  {pages} {064412} (\bibinfo {year} {2009})}\BibitemShut {NoStop}%
\bibitem [{\citenamefont {Horovitz}\ \emph {et~al.}(1983)\citenamefont
  {Horovitz}, \citenamefont {Bohr}, \citenamefont {Kosterlitz},\ and\
  \citenamefont {Schulz}}]{horovitz}%
  \BibitemOpen
  \bibfield  {author} {\bibinfo {author} {\bibfnamefont {B.}~\bibnamefont
  {Horovitz}}, \bibinfo {author} {\bibfnamefont {T.}~\bibnamefont {Bohr}},
  \bibinfo {author} {\bibfnamefont {J.~M.}\ \bibnamefont {Kosterlitz}}, \ and\
  \bibinfo {author} {\bibfnamefont {H.~J.}\ \bibnamefont {Schulz}},\ }\href
  {\doibase 10.1103/PhysRevB.28.6596} {\bibfield  {journal} {\bibinfo
  {journal} {Phys. Rev. B}\ }\textbf {\bibinfo {volume} {28}},\ \bibinfo
  {pages} {6596} (\bibinfo {year} {1983})}\BibitemShut {NoStop}%
\bibitem [{\citenamefont {Nussinov}\ and\ \citenamefont
  {Ortiz}(2009)}]{nussinovortiz}%
  \BibitemOpen
  \bibfield  {author} {\bibinfo {author} {\bibfnamefont {Z.}~\bibnamefont
  {Nussinov}}\ and\ \bibinfo {author} {\bibfnamefont {G.}~\bibnamefont
  {Ortiz}},\ }\href@noop {} {\bibfield  {journal} {\bibinfo  {journal} {Ann.
  Phys.}\ }\textbf {\bibinfo {volume} {\bf 324}},\ \bibinfo {pages} {977}
  (\bibinfo {year} {2009})}\BibitemShut {NoStop}%
\bibitem [{\citenamefont {Kitaev}(2003)}]{kitaevtoriccode}%
  \BibitemOpen
  \bibfield  {author} {\bibinfo {author} {\bibfnamefont {A.~Y.}\ \bibnamefont
  {Kitaev}},\ }\href@noop {} {\bibfield  {journal} {\bibinfo  {journal} {Ann.
  Phys.}\ }\textbf {\bibinfo {volume} {\bf 302}},\ \bibinfo {pages} {2}
  (\bibinfo {year} {2003})}\BibitemShut {NoStop}%
\bibitem [{\citenamefont {Wen}(2003)}]{wenplaquettemodel}%
  \BibitemOpen
  \bibfield  {author} {\bibinfo {author} {\bibfnamefont {X.-G.}\ \bibnamefont
  {Wen}},\ }\href@noop {} {\bibfield  {journal} {\bibinfo  {journal} {Phys.
  Rev. Lett.}\ }\textbf {\bibinfo {volume} {\bf 90}},\ \bibinfo {pages}
  {016803} (\bibinfo {year} {2003})}\BibitemShut {NoStop}%
\bibitem [{\citenamefont {Eggarter}(1974)}]{eggarter-1974}%
  \BibitemOpen
  \bibfield  {author} {\bibinfo {author} {\bibfnamefont {T.~P.}\ \bibnamefont
  {Eggarter}},\ }\href@noop {} {\bibfield  {journal} {\bibinfo  {journal}
  {Phys. Rev. B}\ }\textbf {\bibinfo {volume} {\bf 9}},\ \bibinfo {pages}
  {2989} (\bibinfo {year} {1974})}\BibitemShut {NoStop}%
\bibitem [{\citenamefont {von Heimburg}\ and\ \citenamefont
  {Thomas}(1974)}]{vonheimburgthomas-1974}%
  \BibitemOpen
  \bibfield  {author} {\bibinfo {author} {\bibfnamefont {J.}~\bibnamefont {von
  Heimburg}}\ and\ \bibinfo {author} {\bibfnamefont {H.}~\bibnamefont
  {Thomas}},\ }\href@noop {} {\bibfield  {journal} {\bibinfo  {journal} {J.
  Phys. C}\ }\textbf {\bibinfo {volume} {\bf 7}},\ \bibinfo {pages} {3433}
  (\bibinfo {year} {1974})}\BibitemShut {NoStop}%
\bibitem [{\citenamefont {M{\"u}ller-Hartmann}\ and\ \citenamefont
  {Zittartz}(1974)}]{mullerhartmannzittartz-1974}%
  \BibitemOpen
  \bibfield  {author} {\bibinfo {author} {\bibfnamefont {E.}~\bibnamefont
  {M{\"u}ller-Hartmann}}\ and\ \bibinfo {author} {\bibfnamefont
  {J.}~\bibnamefont {Zittartz}},\ }\href@noop {} {\bibfield  {journal}
  {\bibinfo  {journal} {Phys. Rev. Lett.}\ }\textbf {\bibinfo {volume} {\bf
  33}},\ \bibinfo {pages} {893} (\bibinfo {year} {1974})}\BibitemShut {NoStop}%
\bibitem [{\citenamefont {Brout}(1960)}]{brout-1960}%
  \BibitemOpen
  \bibfield  {author} {\bibinfo {author} {\bibfnamefont {R.}~\bibnamefont
  {Brout}},\ }\href@noop {} {\bibfield  {journal} {\bibinfo  {journal} {Phys.
  Rev.}\ }\textbf {\bibinfo {volume} {\bf 118}},\ \bibinfo {pages} {1009}
  (\bibinfo {year} {1960})}\BibitemShut {NoStop}%
\bibitem [{\citenamefont {Kac}(1959)}]{kac-1959}%
  \BibitemOpen
  \bibfield  {author} {\bibinfo {author} {\bibfnamefont {M.}~\bibnamefont
  {Kac}},\ }\href@noop {} {\bibfield  {journal} {\bibinfo  {journal} {Phys.
  Fluids}\ }\textbf {\bibinfo {volume} {\bf 2}},\ \bibinfo {pages} {8}
  (\bibinfo {year} {1959})}\BibitemShut {NoStop}%
\bibitem [{\citenamefont {Lipkin}\ \emph {et~al.}(1965)\citenamefont {Lipkin},
  \citenamefont {Meshkov},\ and\ \citenamefont {Glick}}]{lmg-1965}%
  \BibitemOpen
  \bibfield  {author} {\bibinfo {author} {\bibfnamefont {H.~J.}\ \bibnamefont
  {Lipkin}}, \bibinfo {author} {\bibfnamefont {N.}~\bibnamefont {Meshkov}}, \
  and\ \bibinfo {author} {\bibfnamefont {A.~J.}\ \bibnamefont {Glick}},\
  }\href@noop {} {\bibfield  {journal} {\bibinfo  {journal} {Nucl. Phys.}\
  }\textbf {\bibinfo {volume} {62}},\ \bibinfo {pages} {188} (\bibinfo {year}
  {1965})}\BibitemShut {NoStop}%
\bibitem [{\citenamefont {Dusuel}\ and\ \citenamefont
  {Vidal}(2004)}]{vidal-2004}%
  \BibitemOpen
  \bibfield  {author} {\bibinfo {author} {\bibfnamefont {S.}~\bibnamefont
  {Dusuel}}\ and\ \bibinfo {author} {\bibfnamefont {J.}~\bibnamefont {Vidal}},\
  }\href@noop {} {\bibfield  {journal} {\bibinfo  {journal} {Phys. Rev. Lett.}\
  }\textbf {\bibinfo {volume} {93}},\ \bibinfo {pages} {237204} (\bibinfo
  {year} {2004})}\BibitemShut {NoStop}%
\bibitem [{\citenamefont {Ribiero}\ \emph {et~al.}(2007)\citenamefont
  {Ribiero}, \citenamefont {Vidal},\ and\ \citenamefont
  {Mosseri}}]{vidal-2007}%
  \BibitemOpen
  \bibfield  {author} {\bibinfo {author} {\bibfnamefont {P.}~\bibnamefont
  {Ribiero}}, \bibinfo {author} {\bibfnamefont {J.}~\bibnamefont {Vidal}}, \
  and\ \bibinfo {author} {\bibfnamefont {R.}~\bibnamefont {Mosseri}},\
  }\href@noop {} {\bibfield  {journal} {\bibinfo  {journal} {Phys. Rev. Lett.}\
  }\textbf {\bibinfo {volume} {99}},\ \bibinfo {pages} {050402} (\bibinfo
  {year} {2007})}\BibitemShut {NoStop}%
\bibitem [{\citenamefont {Botet}\ \emph {et~al.}(1982)\citenamefont {Botet},
  \citenamefont {Jullien},\ and\ \citenamefont {Pfeuty}}]{botet-1982}%
  \BibitemOpen
  \bibfield  {author} {\bibinfo {author} {\bibfnamefont {R.}~\bibnamefont
  {Botet}}, \bibinfo {author} {\bibfnamefont {R.}~\bibnamefont {Jullien}}, \
  and\ \bibinfo {author} {\bibfnamefont {P.}~\bibnamefont {Pfeuty}},\
  }\href@noop {} {\bibfield  {journal} {\bibinfo  {journal} {Phys. Rev. Lett.}\
  }\textbf {\bibinfo {volume} {49}},\ \bibinfo {pages} {478} (\bibinfo {year}
  {1982})}\BibitemShut {NoStop}%
\bibitem [{\citenamefont {Weigert}(1994)}]{weigert-1994}%
  \BibitemOpen
  \bibfield  {author} {\bibinfo {author} {\bibfnamefont {S.}~\bibnamefont
  {Weigert}},\ }\href@noop {} {\bibfield  {journal} {\bibinfo  {journal} {Phys.
  Rev. A}\ }\textbf {\bibinfo {volume} {50}},\ \bibinfo {pages} {4572}
  (\bibinfo {year} {1994})}\BibitemShut {NoStop}%
\bibitem [{\citenamefont {van Wezel}\ \emph {et~al.}(2006)\citenamefont {van
  Wezel}, \citenamefont {Zaanen},\ and\ \citenamefont {van~den
  Brink}}]{vanwezel1}%
  \BibitemOpen
  \bibfield  {author} {\bibinfo {author} {\bibfnamefont {J.}~\bibnamefont {van
  Wezel}}, \bibinfo {author} {\bibfnamefont {J.}~\bibnamefont {Zaanen}}, \ and\
  \bibinfo {author} {\bibfnamefont {J.}~\bibnamefont {van~den Brink}},\
  }\href@noop {} {\bibfield  {journal} {\bibinfo  {journal} {Phys. Rev. B}\
  }\textbf {\bibinfo {volume} {\bf 74}},\ \bibinfo {pages} {094430} (\bibinfo
  {year} {2006})}\BibitemShut {NoStop}%
\bibitem [{\citenamefont {van Wezel}\ and\ \citenamefont {van~den
  Brink}(2008)}]{vanwezel2}%
  \BibitemOpen
  \bibfield  {author} {\bibinfo {author} {\bibfnamefont {J.}~\bibnamefont {van
  Wezel}}\ and\ \bibinfo {author} {\bibfnamefont {J.}~\bibnamefont {van~den
  Brink}},\ }\href@noop {} {\bibfield  {journal} {\bibinfo  {journal} {Phys.
  Rev. B}\ }\textbf {\bibinfo {volume} {\bf 77}},\ \bibinfo {pages} {064523}
  (\bibinfo {year} {2008})}\BibitemShut {NoStop}%
\bibitem [{\citenamefont {Sherrington}\ and\ \citenamefont
  {Kirkpatrick}(1975)}]{sherkirk}%
  \BibitemOpen
  \bibfield  {author} {\bibinfo {author} {\bibfnamefont {D.}~\bibnamefont
  {Sherrington}}\ and\ \bibinfo {author} {\bibfnamefont {S.}~\bibnamefont
  {Kirkpatrick}},\ }\href@noop {} {\bibfield  {journal} {\bibinfo  {journal}
  {Phys. Rev. Lett.}\ }\textbf {\bibinfo {volume} {\bf 35}},\ \bibinfo {pages}
  {1792} (\bibinfo {year} {1975})}\BibitemShut {NoStop}%
\bibitem [{\citenamefont {Fradkin}(1989)}]{fradkin-1989}%
  \BibitemOpen
  \bibfield  {author} {\bibinfo {author} {\bibfnamefont {E.}~\bibnamefont
  {Fradkin}},\ }\href@noop {} {\bibfield  {journal} {\bibinfo  {journal} {Phys.
  Rev. Lett.}\ }\textbf {\bibinfo {volume} {63}},\ \bibinfo {pages} {322}
  (\bibinfo {year} {1989})}\BibitemShut {NoStop}%
\bibitem [{\citenamefont {Lopez}\ \emph {et~al.}(1994)\citenamefont {Lopez},
  \citenamefont {Rojo},\ and\ \citenamefont {Fradkin}}]{lopezrojofradkin-1994}%
  \BibitemOpen
  \bibfield  {author} {\bibinfo {author} {\bibfnamefont {A.}~\bibnamefont
  {Lopez}}, \bibinfo {author} {\bibfnamefont {A.~G.}\ \bibnamefont {Rojo}}, \
  and\ \bibinfo {author} {\bibfnamefont {E.}~\bibnamefont {Fradkin}},\
  }\href@noop {} {\bibfield  {journal} {\bibinfo  {journal} {Phys. Rev. B}\
  }\textbf {\bibinfo {volume} {49}},\ \bibinfo {pages} {15139} (\bibinfo {year}
  {1994})}\BibitemShut {NoStop}%
\bibitem [{\citenamefont {Zhang}\ \emph {et~al.}(1989)\citenamefont {Zhang},
  \citenamefont {Hannson},\ and\ \citenamefont
  {Kivelson}}]{zhanghannsonkivelson-1989}%
  \BibitemOpen
  \bibfield  {author} {\bibinfo {author} {\bibfnamefont {S.~C.}\ \bibnamefont
  {Zhang}}, \bibinfo {author} {\bibfnamefont {T.~H.}\ \bibnamefont {Hannson}},
  \ and\ \bibinfo {author} {\bibfnamefont {S.}~\bibnamefont {Kivelson}},\
  }\href@noop {} {\bibfield  {journal} {\bibinfo  {journal} {Phys. Rev. Lett.}\
  }\textbf {\bibinfo {volume} {62}},\ \bibinfo {pages} {82} (\bibinfo {year}
  {1989})}\BibitemShut {NoStop}%
\bibitem [{\citenamefont {Wang}(1991{\natexlab{a}})}]{wang-feb-1991}%
  \BibitemOpen
  \bibfield  {author} {\bibinfo {author} {\bibfnamefont {Y.~R.}\ \bibnamefont
  {Wang}},\ }\href@noop {} {\bibfield  {journal} {\bibinfo  {journal} {Phys.
  Rev. B}\ }\textbf {\bibinfo {volume} {43}},\ \bibinfo {pages} {3786}
  (\bibinfo {year} {1991}{\natexlab{a}})}\BibitemShut {NoStop}%
\bibitem [{\citenamefont {Wang}(1991{\natexlab{b}})}]{wang-june-1991}%
  \BibitemOpen
  \bibfield  {author} {\bibinfo {author} {\bibfnamefont {Y.~R.}\ \bibnamefont
  {Wang}},\ }\href@noop {} {\bibfield  {journal} {\bibinfo  {journal} {Phys.
  Rev. B}\ }\textbf {\bibinfo {volume} {43}},\ \bibinfo {pages} {13774}
  (\bibinfo {year} {1991}{\natexlab{b}})}\BibitemShut {NoStop}%
\bibitem [{\citenamefont {Wang}(1992)}]{wang-1992}%
  \BibitemOpen
  \bibfield  {author} {\bibinfo {author} {\bibfnamefont {Y.~R.}\ \bibnamefont
  {Wang}},\ }\href@noop {} {\bibfield  {journal} {\bibinfo  {journal} {Phys.
  Rev. B}\ }\textbf {\bibinfo {volume} {46}},\ \bibinfo {pages} {151} (\bibinfo
  {year} {1992})}\BibitemShut {NoStop}%
\bibitem [{\citenamefont {Derzhko}\ \emph {et~al.}(2003)\citenamefont
  {Derzhko}, \citenamefont {Verkholyak}, \citenamefont {Schmidt},\ and\
  \citenamefont {Richter}}]{derzhko-2003}%
  \BibitemOpen
  \bibfield  {author} {\bibinfo {author} {\bibfnamefont {O.}~\bibnamefont
  {Derzhko}}, \bibinfo {author} {\bibfnamefont {T.}~\bibnamefont {Verkholyak}},
  \bibinfo {author} {\bibfnamefont {R.}~\bibnamefont {Schmidt}}, \ and\
  \bibinfo {author} {\bibfnamefont {J.}~\bibnamefont {Richter}},\ }\href@noop
  {} {\bibfield  {journal} {\bibinfo  {journal} {Physica A}\ ,\ \bibinfo
  {pages} {407}} (\bibinfo {year} {2003})}\BibitemShut {NoStop}%
\bibitem [{\citenamefont {Yang}\ \emph {et~al.}(1993)\citenamefont {Yang},
  \citenamefont {Warman},\ and\ \citenamefont {Girvin}}]{yang-1993}%
  \BibitemOpen
  \bibfield  {author} {\bibinfo {author} {\bibfnamefont {K.}~\bibnamefont
  {Yang}}, \bibinfo {author} {\bibfnamefont {L.~K.}\ \bibnamefont {Warman}}, \
  and\ \bibinfo {author} {\bibfnamefont {S.~M.}\ \bibnamefont {Girvin}},\
  }\href@noop {} {\bibfield  {journal} {\bibinfo  {journal} {Phys. Rev. Lett.}\
  }\textbf {\bibinfo {volume} {70}},\ \bibinfo {pages} {2641} (\bibinfo {year}
  {1993})}\BibitemShut {NoStop}%
\bibitem [{\citenamefont {Misguich}\ \emph {et~al.}(2001)\citenamefont
  {Misguich}, \citenamefont {Jolicoeur},\ and\ \citenamefont
  {Girvin}}]{misguich-2001}%
  \BibitemOpen
  \bibfield  {author} {\bibinfo {author} {\bibfnamefont {G.}~\bibnamefont
  {Misguich}}, \bibinfo {author} {\bibfnamefont {T.}~\bibnamefont {Jolicoeur}},
  \ and\ \bibinfo {author} {\bibfnamefont {S.~M.}\ \bibnamefont {Girvin}},\
  }\href@noop {} {\bibfield  {journal} {\bibinfo  {journal} {Phys. Rev. Lett.}\
  }\textbf {\bibinfo {volume} {87}},\ \bibinfo {pages} {097203} (\bibinfo
  {year} {2001})}\BibitemShut {NoStop}%
\bibitem [{\citenamefont {Senthil}\ and\ \citenamefont
  {Fisher}(2006)}]{senthilfisher-2006}%
  \BibitemOpen
  \bibfield  {author} {\bibinfo {author} {\bibfnamefont {T.}~\bibnamefont
  {Senthil}}\ and\ \bibinfo {author} {\bibfnamefont {M.~P.~A.}\ \bibnamefont
  {Fisher}},\ }\href@noop {} {\bibfield  {journal} {\bibinfo  {journal} {Phys.
  Rev. B}\ }\textbf {\bibinfo {volume} {74}},\ \bibinfo {pages} {064405}
  (\bibinfo {year} {2006})}\BibitemShut {NoStop}%
\bibitem [{\citenamefont {Fu}\ and\ \citenamefont {Kane}(2007)}]{fukane-2007}%
  \BibitemOpen
  \bibfield  {author} {\bibinfo {author} {\bibfnamefont {L.}~\bibnamefont
  {Fu}}\ and\ \bibinfo {author} {\bibfnamefont {C.~L.}\ \bibnamefont {Kane}},\
  }\href@noop {} {\bibfield  {journal} {\bibinfo  {journal} {Phys. Rev. B}\
  }\textbf {\bibinfo {volume} {76}},\ \bibinfo {pages} {045302} (\bibinfo
  {year} {2007})}\BibitemShut {NoStop}%
\bibitem [{\citenamefont {Polyakov}(1987)}]{polyakov-book}%
  \BibitemOpen
  \bibfield  {author} {\bibinfo {author} {\bibfnamefont {A.~M.}\ \bibnamefont
  {Polyakov}},\ }\href@noop {} {\emph {\bibinfo {title} {Gauge Fields and
  Strings}}}\ (\bibinfo  {publisher} {Harwood Academic Publishers, CH},\
  \bibinfo {year} {1987})\BibitemShut {NoStop}%
\bibitem [{\citenamefont {Wegner}(1971)}]{wegner-1971}%
  \BibitemOpen
  \bibfield  {author} {\bibinfo {author} {\bibfnamefont {F.}~\bibnamefont
  {Wegner}},\ }\href@noop {} {\bibfield  {journal} {\bibinfo  {journal} {J.
  Math. Phys.}\ }\textbf {\bibinfo {volume} {12}},\ \bibinfo {pages} {2259}
  (\bibinfo {year} {1971})}\BibitemShut {NoStop}%
\bibitem [{\citenamefont {Rieger}\ and\ \citenamefont
  {Kawashima}(1999)}]{riegerkawashima-1999}%
  \BibitemOpen
  \bibfield  {author} {\bibinfo {author} {\bibfnamefont {H.}~\bibnamefont
  {Rieger}}\ and\ \bibinfo {author} {\bibfnamefont {N.}~\bibnamefont
  {Kawashima}},\ }\href@noop {} {\bibfield  {journal} {\bibinfo  {journal}
  {Eur. Phys. J. B}\ }\textbf {\bibinfo {volume} {9}},\ \bibinfo {pages} {233}
  (\bibinfo {year} {1999})}\BibitemShut {NoStop}%
\bibitem [{\citenamefont {Bl{\"o}te}\ and\ \citenamefont
  {Deng}(2002)}]{blotedeng-2002}%
  \BibitemOpen
  \bibfield  {author} {\bibinfo {author} {\bibfnamefont {H.~W.~J.}\
  \bibnamefont {Bl{\"o}te}}\ and\ \bibinfo {author} {\bibfnamefont
  {Y.}~\bibnamefont {Deng}},\ }\href@noop {} {\bibfield  {journal} {\bibinfo
  {journal} {Phys. Rev. E}\ }\textbf {\bibinfo {volume} {66}},\ \bibinfo
  {pages} {066110} (\bibinfo {year} {2002})}\BibitemShut {NoStop}%
\bibitem [{\citenamefont {Evenbly}\ and\ \citenamefont
  {Vidal}(2009)}]{evenblyvidal-2009}%
  \BibitemOpen
  \bibfield  {author} {\bibinfo {author} {\bibfnamefont {G.}~\bibnamefont
  {Evenbly}}\ and\ \bibinfo {author} {\bibfnamefont {G.}~\bibnamefont
  {Vidal}},\ }\href@noop {} {\bibfield  {journal} {\bibinfo  {journal} {Phys.
  Rev. Lett.}\ }\textbf {\bibinfo {volume} {102}},\ \bibinfo {pages} {180406}
  (\bibinfo {year} {2009})}\BibitemShut {NoStop}%
\bibitem [{\citenamefont {Mukherjee}\ \emph {et~al.}()\citenamefont
  {Mukherjee}, \citenamefont {Jalal},\ and\ \citenamefont
  {Lal}}]{mukherjeejalallal}%
  \BibitemOpen
  \bibfield  {author} {\bibinfo {author} {\bibfnamefont {A.}~\bibnamefont
  {Mukherjee}}, \bibinfo {author} {\bibfnamefont {S.}~\bibnamefont {Jalal}}, \
  and\ \bibinfo {author} {\bibfnamefont {S.}~\bibnamefont {Lal}},\ }\href@noop
  {} {\bibinfo  {journal} {Manuscript under preparation}\ }\BibitemShut
  {NoStop}%
\end{thebibliography}%
\end{document}